\documentclass[11pt,a4paper]{article}
\pdfoutput=1 

\usepackage[table]{xcolor}
\usepackage{colortbl}
\usepackage{comment}
\RequirePackage{ifpdf} 
\usepackage{amsmath} 
\usepackage{mathtools}
\usepackage{booktabs}
\usepackage{multirow}
\usepackage{fvextra}

\usepackage{jheppub}
\usepackage{pstricks}
\usepackage[final]{pdfpages} 
\usepackage{ifpdf} 
\usepackage{slashed}
\usepackage{pdflscape}
\usepackage[normalem]{ulem}
\usepackage{color}
\usepackage{xcolor}
\definecolor{urlblue}{rgb}{0.2,0.4,0.7} 
\definecolor{citegreen}{rgb}{0,0.6,0.2}
\definecolor{linkred}{rgb}{0.9,0.2,0.1}
\usepackage{hyperref}
\hypersetup{
colorlinks=true, citecolor=citegreen, linkcolor=blue, urlcolor=urlblue}

\usepackage{graphics}
\usepackage{etoolbox} 
\usepackage{fixmath}
\usepackage{psfrag}

\usepackage{notoccite} 

\usepackage{amsfonts}
\usepackage{autobreak}
\usepackage{marginnote}
\usepackage{enumitem}
\usepackage{appendix}
\usepackage{caption}
\usepackage{tabularx}
\usepackage{graphicx}
\newcommand{\NOdisplay}[1]{ }

\def\MSbar{\overline{\mathrm{MS}}}

\def\Xb{\overline{X}}
\def\Yb{\overline{Y}}

\newcommand{\sqrtsnn}{\sqrt{s_{_\text{NN}}}}
\newcommand{\pp}{p-p}

\providecommand{\PbPb}{Pb-Pb}

\newcommand{\eg}{{\it e.g.}}
\newcommand{\ie}{{\it i.e.}}
\newcommand{\derive}{{\mathrm d}}

\usepackage{xcolor}

\def\lblatnlo{\texttt{LbLatNLO}}
\def\phique{\texttt{PHIQUE}}
\def\gammaUPC{\texttt{gamma-UPC}}

\def\helaconia{\texttt{
HELAC-Onia}}
\def\mg{\texttt{
MadGraph5\_aMC@NLO}}
\def\fastgpl{\texttt{FastGPL}}
\def\handyg{\texttt{handyG}}

\definecolor{mypink}{RGB}{219, 48, 122}

\usepackage{tikz}
\usetikzlibrary{positioning,arrows}
\usetikzlibrary{decorations.pathmorphing}
\usetikzlibrary{decorations.markings}
\usetikzlibrary{shapes.geometric}
\tikzset{
    vector/.style={decorate, decoration={snake}, draw},
    provector/.style={decorate, decoration={snake,amplitude=2.5pt}, draw},
    antivector/.style={decorate, decoration={snake,amplitude=-2.5pt}, draw},
    fermion/.style={draw=black,
      postaction={decorate},decoration={markings,mark=at position .55
        with {\arrow[draw=black]{>}}}},
    fermionbar/.style={draw=black, postaction={decorate},
                       decoration={markings,mark=at position .55 with {\arrow[draw=black]{<}}}},
    fermionnoarrow/.style={draw=black},
    gluon/.style={decorate, draw=red,decoration={coil,amplitude=4pt, segment length=6pt}},
    photon/.style={decorate, draw=red,decoration={snake,amplitude=3pt, segment length=6pt}},
    scalar/.style={dashed,draw=black,
      postaction={decorate},decoration={markings,mark=at position .55
        with {\arrow[draw=black]{>}}}},
    scalarbar/.style={dashed,draw=black,
      postaction={decorate},decoration={markings,mark=at position .55
        with {\arrow[draw=black]{<}}}},
    scalarnoarrow/.style={dashed,draw=black},
    electron/.style={draw=black,
      postaction={decorate},decoration={markings,mark=at position .55
        with {\arrow[draw=black]{>}}}},
    bigvector/.style={decorate, decoration={snake,amplitude=4pt}, draw},
}

\title{Light-by-light scattering: asymptotic expansions, Coulomb resummation and NLO corrections}

\author{Ajjath~A~H$^{a,b}$, Ekta Chaubey$^{c}$, and Hua-Sheng Shao$^{d}$}
\emailAdd{ajjath@iisc.ac.in, eekta@uni-bonn.de, huasheng.shao@lpthe.jussieu.fr}
\affiliation{$^a$Institute for Particle Physics Phenomenology, Durham University, South Road, Durham DH1 3LE, U.K.
\\$^b$Centre for High Energy Physics, Indian Institute of Science, C. V. Raman Avenue, Bengaluru-560012, India
\\$^c$Bethe Center for Theoretical Physics, Universit\"at Bonn, D-53115, Germany
\\$^d$Laboratoire de Physique Th\'eorique et Hautes Energies (LPTHE), UMR 7589, Sorbonne Universit\'e et CNRS, 4 place Jussieu, 75252 Paris Cedex 05, France
}

\preprint{BONN-TH-2026-10, IPPP/26/28}

\abstract{Light-by-light (LbL) scattering is one of the earliest predictions of quantum electrodynamics (QED). Interest in this process has been renewed following its experimental observation at the LHC and the prospects of future measurements at free-electron laser facilities. In this paper, we refine theoretical predictions for LbL scattering by improving the full fermion-mass-dependent two-loop QCD and QED helicity amplitudes using high- and low-energy asymptotic expansions, and by performing Coulomb resummation in the threshold region. The $W$-boson contribution, relevant when the di-photon invariant mass exceeds twice the $W$-boson mass, is included at one-loop accuracy. We present state-of-the-art predictions for LbL cross sections in the Standard Model and provide a new event generator, \lblatnlo, for Monte Carlo simulations of LbL scattering.}

\begin{document}
\allowdisplaybreaks[4]
\unitlength1cm
\maketitle
\flushbottom

\section{Introduction}
\label{sec:intro}
Light, as a messenger of electromagnetic interactions, was historically understood primarily as a probe of matter rather than an object capable of interacting with itself. In classical electrodynamics, Maxwell’s equations predict that electromagnetic waves propagate through vacuum without interacting with each other. Two beams of light can cross paths indefinitely without deflection, energy exchange, or distortion. This linearity of classical electromagnetism is one of its most elegant features and underpins much of modern optics and photonics. Quantum field theory radically revises this picture. In quantum electrodynamics (QED), light can indeed interact with light through purely quantum effects. This phenomenon is known as light-by-light (LbL) scattering and constitutes one of the most striking manifestations of vacuum fluctuations and quantum non-linearity in the electromagnetic field. The origin of LbL scattering lies in the quantum structure of the vacuum. In QED, the vacuum is not empty but is populated with transient virtual particles that constantly fluctuate in and out of existence. Two photons can interact by temporarily converting into a virtual charged particle-antiparticle pair, such as an electron-positron loop, which then re-emits photons. This process appears at the one-loop level in QED perturbation theory and represents one of the earliest predictions of non-linear electromagnetic phenomena. 

Historically, the theoretical foundations of LbL scattering were established in the 1930s by Heisenberg and Euler, who derived an effective description of vacuum polarisation effects in strong electromagnetic fields~\cite{Heisenberg:1934pza,Euler:1935zz,Euler:1935qgl,Heisenberg:1936nmg}. Their work revealed that the quantum vacuum behaves as a non-linear optical medium, leading to phenomena such as vacuum birefringence and photon-photon interactions. However, the prospect of observing vacuum birefringence and LbL scattering appeared quite challenging in the laboratory. Early estimates showed that the scattering cross section at optical photon energies is extraordinarily small, being suppressed by the fourth power of the fine-structure constant $\alpha$ and simultaneously by sixth power of the ratio of the photon energy $\omega$ to the electron mass $m_e$. Even with the most intense light sources available throughout much of the twentieth century, the probability of observing two photons interacting in vacuum remained negligible. As a result, LbL scattering long occupied a special place in theoretical physics: a beautiful prediction of QED that seemed experimentally unreachable.

Beyond testing QED and the Standard Model (SM), LbL scattering also provides a sensitive probe of physics beyond the Standard Model (BSM). In particular, it can be used to constrain quartic anomalous gauge couplings~\cite{dEnterria:2013zqi}, axion-like particles~\cite{Knapen:2016moh}, graviton-like states~\cite{dEnterria:2023npy,Atag:2010bh}, the non-linear Born-Infeld extension of QED~\cite{Ellis:2017edi}, photon-photon self-interactions arising in non-commutative QED~\cite{Horvat:2020ycy}, extra-dimension models~\cite{Cheung:1999ja,Davoudiasl:1999di,AitTamlihat:2026unw}, supersymmetric particles~\cite{Greiner:1992fz}, and new exotic charged particles~\cite{Fichet:2014uka}. Moreover, dispersion relations and crossing symmetry in LbL scattering amplitudes can be exploited to derive new sum rules and positivity bounds on higher-dimensional operators~\cite{Henriksson:2021ymi}.

Interest in measuring LbL scattering and observing vacuum birefringence has been periodically revived by technological advances. The development of high-intensity laser facilities has opened the new possibilities for probing non-linear QED effects in strong electromagnetic fields, motivating experimental searches for vacuum birefringence and laser-based LbL scattering (see, \eg, ref.~\cite{Sarri:2025qng}). Recent experimental proposals based on high-intensity lasers and X-ray free-electron laser (XFEL) facilities aim to access these effects through direct measurements of vacuum polarisation phenomena. The BIREF@HIBEF programme~\cite{Ahmadiniaz:2024xob} at the European XFEL proposes to observe vacuum birefringence by scattering polarised XFEL photons off regions of vacuum polarised by an ultra-intense optical laser, providing a direct test of the Euler-Heisenberg effective theory of QED vacuum non-linearities. Complementary strong-field QED studies are planned within the LUXE experiment~\cite{Abramowicz:2021zja} at DESY, where multi-GeV electron and photon beams will be collided with high-power laser pulses to explore non-linear Compton scattering, multi-photon pair production, and higher-order photon-photon interactions in the non-perturbative regime. Additional proposals include XFEL pump-probe configurations and laser-induced vacuum diffraction setups to detect polarisation-flipped or off-axis signal photons generated by photon-photon scattering in strong electromagnetic fields~\cite{Karbstein:2021ldz,Karbstein:2021hwc}, as well as studies exploring the feasibility of dedicated MeV- to hundred-GeV-scale $\gamma\gamma$ colliders~\cite{Telnov:2020gwc,Sangal:2021qeg,Barklow:2023ess,Zhou:2025vgo,Berger:2026obb}.

The self-interaction of photons has been \textit{indirectly} observed in the scattering of $\gamma$ rays in the Coulomb fields of atomic nuclei (so-called Delbr\"uck scattering)~\cite{Jarlskog:1973aui,Akhmadaliev:1998zz}, as well as through precise measurements of the anomalous magnetic moments of the electron and the muon. The virtual Delbr\"uck scattering contribution, which is closely related to the low-energy LbL amplitude, also provides an important contribution to higher-order QED corrections to precision bound-state observables, such as the Lamb shift and the gyromagnetic factor of the bound electron~\cite{Karshenboim:2002jc,Czarnecki:2016lzl,Szafron:2019tho}. These effects are particularly relevant in muonic atoms, where the larger lepton mass enhances the sensitivity to short-distance QED contributions~\cite{Korzinin:2018tnx}. Additional \textit{indirect} evidence for vacuum birefringence comes from neutron stars in astrophysics~\cite{Mignani:2016fwz} and from the modulation of pairs produced via the linear Breit-Wheeler process $\gamma\gamma\to e^+e^-$ in the STAR experiment in heavy-ion collisions~\cite{Brandenburg:2022tna}. In the laboratory, however, vacuum birefringence has not yet been observed \textit{directly}; the current best limits are set by the PVLAS experiment~\cite{Ejlli:2020yhk}.

Nevertheless, the definitive experimental observation of LbL scattering required a radically different approach, one that exploits the electromagnetic fields of relativistic heavy ions. In particular, ultraperipheral collisions (UPCs) of heavy ions at relativistic energies provide an intense flux of quasi-real photons generated by the strong electromagnetic fields surrounding fast-moving ions. These collisions effectively transform heavy ions into photon sources, enabling photon-photon interactions to be studied experimentally. Direct experimental observations of LbL scattering were only achieved recently in the UPC lead-lead (\PbPb) collisions at the LHC~\cite{ATLAS:2017fur,CMS:2018erd,ATLAS:2019azn,ATLAS:2020hii,CMS:2024bnt}. Their feasibility, first suggested in
ref.~\cite{dEnterria:2013zqi}, can be largely attributed to the intense coherent photon flux carried by relativistic nuclei. This enhancement arises from the large charge $Z$ and the Lorentz factor $\gamma_{\mathrm{L}}$, leading to a cross section that scales as $Z^4\approx 4.5\cdot 10^7$ in \PbPb\ collisions relative to
proton-proton (\pp) or lepton-lepton collisions.

The core ingredients of these low- and high-energy experiments are the scattering amplitudes and cross sections for LbL scattering. In the SM, the leading-order (LO) contribution to the cross section is $\mathcal{O}(\alpha^4)$ and arises from one-loop Feynman diagrams with charged particles circulating in the loop, where $\alpha$ denotes the electromagnetic fine-structure constant. In the low-energy (LE) limit, the process can be described effectively by the Euler-Heisenberg Lagrangian, while the first complete QED calculation was performed by Karplus and Neuman~\cite{Karplus:1950zz}. The LO one-loop results in the SM have since been revisited many times in the literature~\cite{Jikia:1993tc,Bohm:1994sf,Yang:1994nu,Jikia:1997yt,Bernicot:2008th,Bardin:2009gq,Harland-Lang:2018iur,Jia:2024hen,Berger:2026obb}, and such computations are now fully automated in modern event generators~\cite{Hirschi:2011pa,Alwall:2014hca,Hirschi:2015iia,Frederix:2018nkq,Shao:2022cly}. Furthermore, the next-to-leading order (NLO) two-loop QCD and QED corrections in the LE and high-energy (HE) limits~\footnote{For fermion loops, a result in the HE limit (\ie, the first term in the HE expansion) is equivalent to carrying out a massless calculation. As we shall see in appendix~\ref{sec:oneloopHE}, this equivalence no longer holds for $W^\pm$ boson loops.} are available in refs.~\cite{Reuter:1996zm,Fliegner:1997ra,Schubert:2001he,Martin:2003gb} and refs.~\cite{Bern:2001dg,Binoth:2002xg}, respectively. The complete two-loop QCD and QED corrections with full fermion-mass dependence have only recently been computed in refs.~\cite{AH:2023kor,AH:2023ewe}. In particular, analytic two-loop helicity amplitudes were presented in ref.~\cite{AH:2023ewe} and cross-checked against numerical results obtained via direct Monte Carlo integration in momentum space~\cite{Capatti:2019ypt,Capatti:2019edf}. Very recently, a three-loop massless QCD and QED calculation of fermionic contributions has appeared in ref.~\cite{Bargiela:2026tcn}.


The main purpose of this paper is multifold. First, we present analytic asymptotic expansions of the one- and two-loop helicity amplitudes for $\gamma\gamma\to\gamma\gamma$ in both the low- and high-energy regimes. These asymptotic expressions are not only important for improving the numerical stability of two-loop amplitude computations, but are also interesting in their own right. The low-energy expansions are useful for determining the low-energy effective Lagrangian, which is relevant for studying vacuum birefringence phenomena, as demonstrated in ref.~\cite{Heinzl:2025xye}, while the high-energy expressions are particularly relevant for understanding subleading-power logarithmic structures in loop-induced processes~\cite{Kotsky:1997rq,Fadin:1997sn,Akhoury:2001mz,Penin:2014msa,Melnikov:2016emg,Penin:2016wiw,Liu:2017vkm,Liu:2018czl,Wang:2019mym,Liu:2019oav,Anastasiou:2020vkr,Liu:2020tzd,Liu:2020wbn,Liu:2021chn,Liu:2022ajh,Bell:2022ott,Hu:2025hfc,Delto:2025epy}. Second, we improve amplitude and cross-section predictions in the threshold region by incorporating Coulomb gluon and photon resummation, which cures the logarithmic Coulomb divergences appearing in the two-loop amplitudes (cf. figure~2 of ref.~\cite{AH:2023kor}). Finally, we release a new event generator, dubbed \lblatnlo. This code not only computes fiducial differential cross sections, but also enables the generation of helicity-dependent unweighted events for Monte Carlo simulations. In addition, we provide technical details that allow interested readers to extract helicity amplitudes and two-loop master integrals within \lblatnlo. Some of these ingredients may be reusable in other processes and related applications.

The paper is organised as follows. In section~\ref{sec:asymexpofamp}, we derive analytic low- and high-energy expansions of the one- and two-loop helicity amplitudes for LbL scattering. To cure the Coulomb singularity observed at NLO, we perform a Coulomb resummation in section~\ref{sec:Coulombres}. Selected results are presented in section~\ref{sec:results}. Conclusions are drawn in section~\ref{sec:conclusions}. Additional details are provided in the appendices. High-energy expansion expressions for one-loop amplitudes can be found in appendix~\ref{sec:oneloopHE}. The treatment of the running of the effective QED coupling $\alpha$ in the on-shell scheme is described in appendix~\ref{sec:alphaRGrun}. Finally, instructions for using \lblatnlo\ are given in appendix~\ref{sec:lblatnlo}.

\section{Asymptotic expansions of helicity amplitudes\label{sec:asymexpofamp}}

We use the same notation and convention as in refs.~\cite{AH:2023kor,AH:2023ewe}. For the LbL scattering process
\begin{equation}
\gamma(p_1,\lambda_1)+\gamma(p_2,\lambda_2)+\gamma(p_3,\lambda_3)+\gamma(p_4,\lambda_4)\to 0\,,
\end{equation}
the helicity amplitude is generically denoted as $\mathcal{M}_{\lambda_1\lambda_2\lambda_3\lambda_4}$, where $p_i$ is the incoming four momentum of $i^\mathrm{th}$ photon and $\lambda_i$ is its helicity. In QCD and QED, up to any loop order, there are five independent helicity amplitudes, namely $\mathcal{M}_{++++}$, $\mathcal{M}_{-+++}$, $\mathcal{M}_{--++}$, $\mathcal{M}_{+-+-}$, and $\mathcal{M}_{+--+}$. All the remaining helicity amplitudes can be expressed in terms of these five ones (cf. eq.~(2.32) of ref.~\cite{AH:2023ewe}). These helicity amplitudes are functions of Mandelstam variables
\begin{equation}
s=(p_1+p_2)^2\,,\quad t=(p_2+p_3)^2\,,\quad u=(p_1+p_3)^2\,,
\end{equation}
and loop-particle masses. We have the relation $s+t+u=0$ from momentum conservation and on-shell condition $p_i^2=0$. Due to crossing symmetry, we have the relation
\begin{equation}
\mathcal{M}_{+-+-}=\left.\mathcal{M}_{--++}\right|_{s\leftrightarrow u}=\left.\mathcal{M}_{+--+}\right|_{t\leftrightarrow u}\,
\end{equation}
among two-minus-two-plus amplitudes.

For the contributions of a given charged fermion $f$ loop with its mass $m_f$, we denote one- and two-loop QCD and QED amplitudes as $\mathcal{M}_{\lambda_1\lambda_2\lambda_3\lambda_4}^{(0,0,f)}$, $\mathcal{M}_{\lambda_1\lambda_2\lambda_3\lambda_4}^{(1,0,f)}$, and $\mathcal{M}_{\lambda_1\lambda_2\lambda_3\lambda_4}^{(0,1,f)}$, respectively. For $W^\pm$ boson, since we only consider the one-loop result, we denote its LO amplitude as $\mathcal{M}_{\lambda_1\lambda_2\lambda_3\lambda_4}^{(0,0,W)}$. The analytic results for these amplitudes can be found in ref.~\cite{AH:2023ewe}. In particular, we have the simple scaling relation
\begin{equation}
\frac{\mathcal{M}_{\lambda_1\lambda_2\lambda_3\lambda_4}^{(1,0,f)}}{\mathcal{M}_{\lambda_1\lambda_2\lambda_3\lambda_4}^{(0,1,f)}}=\frac{\alpha_sC_{F,f}}{\alpha Q_{f}^2}\,,
\end{equation}
where $\alpha_s$ is the strong coupling constant, $C_{F,f}=4/3$ ($C_{F,f}=0$) for $f$ being a quark (a lepton), and $Q_f$ is the electric charge of $f$ in units of the positron charge.

The two-loop helicity amplitudes $\mathcal{M}^{(1,0,f)}_{\lambda_1\lambda_2\lambda_3\lambda_4}$ and
$\mathcal{M}^{(0,1,f)}_{\lambda_1\lambda_2\lambda_3\lambda_4}$
are known in terms of two-loop master integrals, either expressed through Goncharov polylogarithms (GPLs)~\cite{Goncharov:2001,goncharov2011multiple} or in terms of iterated integrals with $d\log$ one-forms. These iterated integrals can be converted into one-fold integrals, which can be evaluated efficiently using the double-exponential (tanh-sinh) quadrature~\cite{Mori:1973}, in analogy with the implementation in\\
\texttt{PentagonFunctions++}~\cite{Chicherin:2020oor}. In asymptotic regions, such as the LE and HE limits, large numerical cancellations occur, typically requiring higher arithmetic precision or a reorganisation of the helicity amplitudes. In this section, we present our analytic results for the LE and HE expansions in sections~\ref{sec:LEexpansion} and \ref{sec:HEexpansion}, respectively, and describe the methods we adopt.

\subsection{Low-energy expansion\label{sec:LEexpansion}}

We first consider the LE region
\begin{equation}
s,|t|,|u|\ll m_f^2\,.\label{eq:LEregion}
\end{equation}
This regime is generally relevant for laser-based LbL scattering and vacuum birefringence experiments. In this region, large cancellations occur in the SM, such that the one- and two-loop amplitudes at LO in the $m_f^{-2}$ expansion start at $\mathcal{O}(m_f^{-4})$ (cf. appendix F of ref.~\cite{AH:2023ewe}). The leading $\mathcal{O}(m_f^{-4})$ expressions for the two-loop fermionic contributions can also be found in ref.~\cite{Martin:2003gb}. However, analytic results for the higher-order terms in the $m_f^{-2}$ expansion at two loops are not yet available. These higher-order terms can become the leading contributions in situations where the $\mathcal{O}(m_f^{-4})$ terms vanish. For instance, in vacuum circular birefringence induced by a circularly polarised laser, the leading contribution arises from the next-to-leading term in the expansion, \ie, $\mathcal{O}(m_f^{-6})$~\cite{Heinzl:2025xye}. The analytic expressions for the helicity amplitudes in the LE region are also potentially useful for constructing the LE effective Lagrangian, as illustrated in the same reference. In addition to these motivations, from a pragmatic point of view, the LE expansion improves numerical stability in the evaluation of the scattering amplitudes. Therefore, in this section, we present the LE expansion of the two-loop LbL amplitudes.

The main difficulty for obtaining the LE-expanded amplitudes from full mass-dependent amplitudes given in ref.~\cite{AH:2023ewe} is master integrals, which are originally expressed in terms of GPLs and iterated integrals. Although it is possible to derive the $m_f^{-2}$ expansions for master integrals with the differential equation method, we adopt another approach which turns out to be more efficient that we are going to describe below. We, however, still use the differential equation approach as a cross check.

Let us consider a generic $d$-dimensional ($d=4-2\epsilon$) two-loop Feynman integral (cf. eqs.~(3.1-3.2) of ref.~\cite{AH:2023ewe})
\begin{align}
\label{eq:define-int-2L}
I_{n_1,\cdots,n_9}^{(2)}(s_{ij},s_{jk},s_{ik}) &= \left(\frac{e^{\epsilon \gamma_E}m_f^{2\epsilon}}{i\pi^{\frac{d}{2}}}\right)^2 \int \derive^d\ell_1 \derive^d\ell_2 \frac{1}{D_1^{n_1} D_2^{n_2} D_3^{n_3}D_4^{n_4}D_5^{n_5}D_6^{n_6}D_7^{n_7}D_8^{n_8} D_9^{n_9}}\,,
\end{align}
where $n_i \in \mathbb{Z}$, $\gamma_E$ is the Euler-Mascheroni constant, and $\ell_1,\ell_2$ are  the loop momenta. The inverse propagators are
\begin{eqnarray}
D_1&=&\ell_1^2-m_f^2+i0^+\,,\quad D_2=(\ell_1+p_i)^2-m_f^2+i0^+\,,\quad D_3=(\ell_1+p_i+p_j)^2-m_f^2+i0^+\,,\nonumber\\
D_4&=&(\ell_1+p_i+p_j+p_k)^2-m_f^2+i0^+\,,\qquad D_5=\ell_2^2-m_f^2+i0^+\,,\nonumber \\  D_6&=&(\ell_2+p_i)^2-m_f^2+i0^+\,,\qquad 
D_7=(\ell_2+p_i+p_j)^2-m_f^2+i0^+\,,\nonumber \\  D_8&=&(\ell_2+p_i+p_j+p_k)^2-m_f^2+i0^+\,,\qquad D_9=(\ell_1-\ell_2)^2+i0^+\,,
\end{eqnarray}
which use the prescription of the Feynman propagators with $i0^+$. Here, we keep the general external momentum dependence with $i,j,k\in\ \left\{1,2,3,4\right\}$ and $i\neq j, i\neq k, j\neq k$. The arguments of $I_{n_1,\cdots,n_9}^{(2)}$ are defined as $s_{ij}=(p_i+p_j)^2,s_{jk}=(p_j+p_k)^2,s_{ik}=(p_i+p_k)^2$, which are simply a permutation of the original Mandelstam variables $s,t,u$.

In the LE region, we find that only the hard momentum region, $\ell_1,\ell_2\sim m_f$, contributes~\cite{Beneke:1997zp,Smirnov:2002pj}. Therefore, in this particular case, the corresponding expansion reduces to a Taylor expansion of the integrand in $m_f^{-2}$ under the integral sign, without the need to separately identify and sum over multiple momentum regions. Hence, the Feynman integral can be expanded as
\begin{eqnarray}
I_{n_1,\cdots,n_9}^{(2)}(s_{ij},s_{jk},s_{ik})&=&\sum_{n=0}^{+\infty}{I_{n_1,\cdots,n_9}^{(2,n)}(s_{ij},s_{jk},s_{ik})}\,,\label{eq:ILEexp0}
\end{eqnarray}
where the leading term in the LE expansion is
\begin{eqnarray}
I_{n_1,\cdots,n_9}^{(2,0)}(s_{ij},s_{jk},s_{ik})&=&I^{V,(2)}_{n_1+n_2+n_3+n_4,n_5+n_6+n_7+n_8,n_9}\,,
\end{eqnarray}
and two-loop vacuum-type integrals are defined as
\begin{eqnarray}
I_{n_1,n_2,n_3}^{V,(2)}&=&\left(\frac{e^{\epsilon \gamma_E}m_f^{2\epsilon}}{i\pi^{\frac{d}{2}}}\right)^2\int{\mathrm{d}^d\ell_1\mathrm{d}^d\ell_2\frac{1}{D_{V,1}^{n_1}D_{V,2}^{n_2}D_{V,3}^{n_3}}},\quad n_i\in\mathbb{Z}\,.\label{eq:VMIs42L}
\end{eqnarray}
The inverse propagators in the vacuum-type integrals are defined as
\begin{eqnarray}
D_{V,1}&=&\ell_1^2-m_f^2+i0^+,\quad D_{V,2}=\ell_2^2-m_f^2+i0^+,\quad D_{V,3}=\left(\ell_1-\ell_2\right)^2+i0^+\,.
\end{eqnarray}
The $n^\mathrm{th}$ expansion term in eq.~\eqref{eq:ILEexp0} is $\mathcal{O}\left(m_f^{-2n}\right)$ with respect to the leading term. Because we have the following identity
\begin{eqnarray}
\frac{1}{(1+x)^n}&=&\sum_{m=0}^{+\infty}{(-)^m\binom{n+m-1}{m}x^m}
\end{eqnarray}
when $|x|<1$, where $\binom{n}{k}$ is the binomial coefficient. In the general case with $n_7\geq 1$, we can obtain, for $n_i\geq 0$,
\begin{eqnarray}
&&\left.I_{n_1,\cdots,n_9}^{(2,n)}(s_{ij},s_{jk},s_{ik})\right|_{n_7\geq 1}=\left(\frac{e^{\epsilon \gamma_E}m_f^{2\epsilon}}{i\pi^{\frac{d}{2}}}\right)^2\int{\mathrm{d}^d\ell_1\mathrm{d}^d\ell_2}\\
&&\left[\sum_{m_2=0}^{2n\Theta(n_2-\frac{1}{2})}{(-)^{m_2}\binom{n_2+m_2-1}{m_2}\left(2\ell_1\cdot p_i\right)^{m_2}}\times\right.\nonumber\\
&&\sum_{m_4=0}^{(2n-m_2)\Theta(n_4-\frac{1}{2})}(-)^{m_4}\binom{n_4+m_4-1}{m_4}\left(2\ell_1\cdot \left(p_i+p_j+p_k\right)\right)^{m_4}\times\nonumber\\
&&\sum_{m_6=0}^{(2n-m_{24})\Theta(n_6-\frac{1}{2})}{(-)^{m_6}\binom{n_6+m_6-1}{m_6}\left(2\ell_2\cdot p_i\right)^{m_6}}\times\nonumber\\
&&\sum_{m_8=0}^{(2n-m_{246})\Theta(n_8-\frac{1}{2})}{(-)^{m_8}\binom{n_8+m_8-1}{m_8}\left(2\ell_2\cdot \left(p_i+p_j+p_k\right)\right)^{m_8}}\times\nonumber\\
&&\sum_{m_3=0}^{(2n-m_{2468})\Theta(n_3-\frac{1}{2})}{(-)^{m_3}\binom{n_3+m_3-1}{m_3}}\times\nonumber\\
&&\sum_{k_3={\rm max}((2m_3+m_{2468}-2n)\Theta(n_3-\frac{1}{2}),0)}^{m_3}{\binom{m_3}{k_3}\left(2\ell_1\cdot \left(p_i+p_j\right)\right)^{k_3}s_{ij}^{m_3-k_3}}\times\nonumber\\
&&\sum_{m_7=n-m_3-\lfloor \frac{m_{2468}-k_3}{2} \rfloor}^{2(n-m_3-\lfloor \frac{m_{2468}-k_3}{2} \rfloor)-\frac{1-(-)^{m_{2468}-k_3}}{2}}{(-)^{m_7}\binom{n_7+m_7-1}{m_7}\binom{m_7}{m_{2468}+2m_{37}-k_3-2n}}\nonumber\\
&&\left.\times\left(2\ell_2\cdot \left(p_i+p_j\right)\right)^{m_{2468}+2m_{37}-k_3-2n}s_{ij}^{2n+k_3-2m_3-m_{24678}}\right]\frac{1}{D_{V,1}^{n_{1234}+m_{234}}D_{V,2}^{n_{5678}+m_{678}}D_{V,3}^{n_9}}\,,\nonumber
\end{eqnarray}
where we have used the shorthand notations $n_{i_1\cdots i_l}=n_{i_1}+\cdots+n_{i_l}$ and $m_{i_1\cdots i_l}=m_{i_1}+\cdots +m_{i_l}$. $\Theta$ is the Heaviside theta function, and $\lfloor x \rfloor$ represents the greatest integer that does not exceed $x$ with $x\in\mathbb{R}$. If $n_3\geq 1$ and $n_7=0$ and $n_i\geq 0$, we have
\begin{eqnarray}
&&\left.I_{n_1,\cdots,n_9}^{(2,n)}(s_{ij},s_{jk},s_{ik})\right|_{n_7=0,n_3\geq 1}=\left(\frac{e^{\epsilon \gamma_E}m_f^{2\epsilon}}{i\pi^{\frac{d}{2}}}\right)^2\int{\mathrm{d}^d\ell_1\mathrm{d}^d\ell_2}\nonumber\\
&&\left[\sum_{m_2=0}^{2n\Theta(n_2-\frac{1}{2})}{(-)^{m_2}\binom{n_2+m_2-1}{m_2}\left(2\ell_1\cdot p_i\right)^{m_2}}\times\right.\nonumber\\
&&\sum_{m_4=0}^{(2n-m_2)\Theta(n_4-\frac{1}{2})}(-)^{m_4}\binom{n_4+m_4-1}{m_4}\left(2\ell_1\cdot \left(p_i+p_j+p_k\right)\right)^{m_4}\times\nonumber\\
&&\sum_{m_6=0}^{(2n-m_{24})\Theta(n_6-\frac{1}{2})}{(-)^{m_6}\binom{n_6+m_6-1}{m_6}\left(2\ell_2\cdot p_i\right)^{m_6}}\times\nonumber\\
&&\sum_{m_8=0}^{(2n-m_{246})\Theta(n_8-\frac{1}{2})}{(-)^{m_8}\binom{n_8+m_8-1}{m_8}\left(2\ell_2\cdot \left(p_i+p_j+p_k\right)\right)^{m_8}}\times\nonumber\\
&&\sum_{m_3=n-\lfloor \frac{m_{2468}}{2} \rfloor}^{2(n-\lfloor \frac{m_{2468}}{2} \rfloor)-\frac{1-(-)^{m_{2468}}}{2}}{(-)^{m_3}\binom{n_3+m_3-1}{m_3}\binom{m_3}{m_{2468}+2m_3-2n}}\nonumber\\
&&\left.\times\left(2\ell_1\cdot \left(p_i+p_j\right)\right)^{m_{2468}+2m_{3}-2n}s_{ij}^{2n-m_3-m_{2468}}\right]\frac{1}{D_{V,1}^{n_{1234}+m_{234}}D_{V,2}^{n_{568}+m_{68}}D_{V,3}^{n_9}}\,.
\end{eqnarray}

To reduce the above integrals, we use the tensor decomposition of two-loop vacuum integrals outlined in ref.~\cite{Davydychev:1995nq}, which has the advantage that we do not need to introduce extra auxiliary propagators as done with the general integration by parts tools and should be able to reduce the overall computation burden. 
After tensor decomposition, we still need to reduce the vacuum-like integrals $I^{V,(2)}_{n_1,n_2,n_3}$ as defined in eq.~\eqref{eq:VMIs42L}. Although it is generally possible to do integration-by-parts reduction for $I_{n_1,n_2,n_3}^{V,(2)}$ with public integral reduction codes, for our special case, we can easily setup the reduction relations for $I^{V,(2)}_{n_1,n_2,n_3}$. When $n_1\geq 1$, we can use the following identity derived from integration by parts
\begin{align}
I_{n_1+1,n_2,n_3}^{V,(2)}&=&\frac{1}{2n_1m_f^2}\left[\left(d-2n_1-n_3\right)I_{n_1,n_2,n_3}^{V,(2)}-n_3 I_{n_1-1,n_2,n_3+1}^{V,(2)}+n_3 I_{n_1,n_2-1,n_3+1}^{V,(2)}\right]\,.
\end{align}
This relation helps to reduce the integrals of $I^{V,(2)}_{n_1,n_2,n_3}$  with $n_1>1$ into integrals $I_{1,n_2,n_3}^{V,(2)}$. Similarly, when $n_2\geq 1$, we can use
\begin{eqnarray}
I_{1,n_2+1,n_3}^{V,(2)}&=&\frac{1}{2n_2m_f^2}\left[\left(d-2n_2-n_3\right)I_{1,n_2,n_3}^{V,(2)}-n_3I_{1,n_2-1,n_3+1}^{V,(2)}\right]
\end{eqnarray}
to further reduce the integrals into the form of $I_{1,1,n_3}^{V,(2)}$. Finally, we can use the recurrence relations
\begin{eqnarray}
I_{1,1,n_3}^{V,(2)}&=&\frac{d-n_3-1}{2\left(d-2n_3-1\right)m_f^2}I_{1,1,n_3-1}^{V,(2)}\,,\quad n_3\geq 2\,,\label{eq:recurrencerelation1}\\
I_{1,1,n_3}^{V,(2)}&=&\frac{2\left(d-2n_3-3\right)m_f^2}{d-n_3-2}I_{1,1,n_3+1}^{V,(2)}\,,\quad n_3\leq -1\label{eq:recurrencerelation2}
\end{eqnarray}
to reduce all integrals expressed in terms of two master integrals $I_{1,1,1}^{V,(2)}$ and $I_{1,1,0}^{V,(2)}$.~\footnote{In fact, the recurrence relation eq.~\eqref{eq:recurrencerelation1} also works when $n_3=1$. Therefore, rigorously speaking, we only need a single master integral $I_{1,1,0}^{V,(2)}$.} The tensor decomposition and integral reduction can be straightforwardly implemented in \texttt{Mathematica} or in any other symbolic mathematics system.

The final master integrals can be easily evaluated with the help of Feynman parametrisation to all orders in the dimensional regulator $\epsilon$. Their explicit expressions are 
\begin{eqnarray}
I_{1,1,0}^{V,(2)}&=&m_f^4 e^{2\epsilon \gamma_E}\left(\Gamma(-1+\epsilon)\right)^2\,,\label{eq:I110exp}\\
I_{1,1,1}^{V,(2)}&=&m_f^2 e^{2\epsilon \gamma_E}2^{1-2\epsilon}\sqrt{\pi}\frac{\Gamma\left(-1+\epsilon\right)\Gamma\left(-1+2\epsilon\right)}{\Gamma\left(\frac{1}{2}+\epsilon\right)}\,,\label{eq:I111exp}
\end{eqnarray}
where $\Gamma(x)$ is Euler's gamma function. As a direct cross check, the vacuum-type integral with Feynman parameterisation gives us
\begin{equation}
I_{1,1,2}^{V,(2)}=-e^{2\epsilon \gamma_E} 2^{-1-2\epsilon}\left(\sqrt{\pi}\right)^3\frac{\Gamma\left(2\epsilon\right)}{\sin{\left(\epsilon\pi\right)}\Gamma\left(2-\epsilon\right)\Gamma\left(\frac{3}{2}+\epsilon\right)}\,.
\end{equation}
One can easily verify that the above expression satisfies eq.~\eqref{eq:recurrencerelation1} after plugging the expression of $I_{1,1,1}^{V,(2)}$ given in eq.~\eqref{eq:I111exp}.

For the fermion $f$, by defining $x_s=s/m_f^2$, $x_t=t/m_f^2$, $x_u=u/m_f^2$, one-loop LE-expanded helicity amplitudes are
\begin{eqnarray}
-i\mathcal{M}_{++++}^{(0,0,f),\mathrm{LE}}&=&N_{c,f}Q_f^4\alpha^2\Bigg[-\frac{1}{15}\left(x_s^2+x_t^2+x_u^2\right)-\frac{2}{63}x_sx_tx_u-\frac{1}{945}\left(x_s^2+x_t^2+x_u^2\right)^2\nonumber\\
&&-\frac{1}{990}x_sx_tx_u\left(x_s^2+x_t^2+x_u^2\right)-\left(\frac{109}{450450}x_s^2x_t^2x_u^2+\frac{1}{48048}\left(x_s^2+x_t^2+x_u^2\right)^3\right)\nonumber\\
&&-\frac{3}{100100}x_sx_tx_u\left(x_s^2+x_t^2+x_u^2\right)^2+\mathcal{O}\left(x_s^8\right)\Bigg]\,,\nonumber\\
-i\mathcal{M}_{-+++}^{(0,0,f),\mathrm{LE}}&=&N_{c,f}Q_f^4\alpha^2\Bigg[-\frac{1}{315}x_sx_tx_u-\frac{1}{41580}x_sx_tx_u\left(x_s^2+x_t^2+x_u^2\right)-\frac{2}{225225}x_s^2x_t^2x_u^2\nonumber\\
&&-\frac{1}{3603600}x_sx_tx_u\left(x_s^2+x_t^2+x_u^2\right)^2+\mathcal{O}\left(x_s^8\right)\Bigg]\,,\nonumber\\
-i\mathcal{M}_{--++}^{(0,0,f),\mathrm{LE}}&=&N_{c,f}Q_f^4\alpha^2\Bigg[\frac{11}{45}x_s^2+\frac{4}{315}x_s^3+\frac{1}{18900}x_s^2\left(91x_s^2-50x_tx_u\right)+\frac{1}{103950}x_s^3\left(48x_s^2+49x_tx_u\right)\nonumber\\
&&+x_s^2\left(\frac{19}{210210}x_s^4+\frac{73}{772200}x_s^2\left(x_t^2+x_u^2\right)+\frac{19}{300300}x_s\left(x_t^3+x_u^3\right)+\frac{19}{600600}\left(x_t^4+x_u^4\right)\right)\nonumber\\
&&+x_s^3\left(\frac{19}{600600}x_s^4-\frac{967}{37837800}x_s^2\left(x_t^2+x_u^2\right)-\frac{1}{38610}x_s\left(x_t^3+x_u^3\right)-\frac{1}{77220}\left(x_t^4+x_u^4\right)\right)\nonumber\\
&&+\mathcal{O}\left(x_s^8\right)\Bigg]\,,\nonumber\\
\mathcal{M}_{+-+-}^{(0,0,f),\mathrm{LE}}&=&\left.\mathcal{M}_{--++}^{(0,0,f),\mathrm{LE}}\right|_{x_s\leftrightarrow x_u}\,,\quad \mathcal{M}_{+--+}^{(0,0,f),\mathrm{LE}}=\left.\mathcal{M}_{--++}^{(0,0,f),\mathrm{LE}}\right|_{x_s\leftrightarrow x_t}\,.
\end{eqnarray}
While the first expansion terms at $\mathcal{O}(m_f^{-4})$ are well known in the literature, the second expansion terms at $\mathcal{O}(m_f^{-6})$ agree with ref.~\cite{Heinzl:2025xye} up to a global phase.
Similarly, the two-loop LE-expanded results are
\begin{eqnarray}
i\mathcal{M}_{++++}^{(1,0,f),\mathrm{LE}}&=&N_{c,f}Q_f^4\alpha^2\frac{\alpha_s}{\pi}C_{F,f}\Bigg[\frac{5}{12}\left(x_s^2+x_t^2+x_u^2\right)+\frac{49}{180}x_sx_tx_u\nonumber\\
&&+\frac{59}{5670}\left(x_s^2+x_t^2+x_u^2\right)^2+\frac{1949}{170100}x_sx_tx_u\left(x_s^2+x_t^2+x_u^2\right)+\mathcal{O}\left(x_s^{6}\right)\Bigg]\,,\nonumber\\
i\mathcal{M}_{-+++}^{(1,0,f),\mathrm{LE}}&=&N_{c,f}Q_f^4\alpha^2\frac{\alpha_s}{\pi}C_{F,f}\Bigg[\frac{53}{2700}x_sx_tx_u+\frac{18971}{95256000}x_sx_tx_u\left(x_s^2+x_t^2+x_u^2\right)+\mathcal{O}\left(x_s^{6}\right)\Bigg]\,,\nonumber\\
i\mathcal{M}_{--++}^{(1,0,f),\mathrm{LE}}&=&N_{c,f}Q_f^4\alpha^2\frac{\alpha_s}{\pi}C_{F,f}\Bigg[-\frac{391}{324}x_s^2-\frac{1849}{16200}x_s^3-x_s^2\left(\frac{178007}{5670000}x_s^2+\frac{118901}{15876000}\left(x_t^2+x_u^2\right)\right)\nonumber\\
&&-\left(\frac{123234401}{17288964000}x_s^5-\frac{3568207}{1920996000}x_s^3\left(x_t^2+x_u^2\right)-\frac{27757}{411642000}x_s^2\left(x_t^3+x_u^3\right)\right.\nonumber\\
&&\left.+\frac{1090403}{2881494000}x_s\left(x_t^4+x_u^4\right)+\frac{475021}{960498000}\left(x_t^5+x_u^5\right)+\frac{475021}{2881494000}\frac{x_t^6+x_u^6}{x_s}\right)+\mathcal{O}\left(x_s^6\right)\Bigg]\,,\nonumber\\
\mathcal{M}_{+-+-}^{(1,0,f),\mathrm{LE}}&=&\left.\mathcal{M}_{--++}^{(1,0,f),\mathrm{LE}}\right|_{x_s\leftrightarrow x_u}\,,\quad \mathcal{M}_{+--+}^{(1,0,f),\mathrm{LE}}=\left.\mathcal{M}_{--++}^{(1,0,f),\mathrm{LE}}\right|_{x_s\leftrightarrow x_t}\,.\label{eq:LEAmp2L}
\end{eqnarray}
The leading $\mathcal{O}(m_f^{-4})$ terms agree with those reported in the literature~\cite{Martin:2003gb}, while the higher-order terms are new. Interestingly, at sufficiently high order in the $m_f^{-2}$ expansion (\ie, $\mathcal{O}(m_f^{-10})$) at the two-loop level, $1/s$, $1/t$, and $1/u$ poles appear due to massless gluon/photon propagation within loop corrections. These pole terms are non-local and infrared-generated, and do not correspond to local operators in the Euler-Heisenberg effective Lagrangian; they must therefore be subtracted when constructing the two-loop Euler-Heisenberg Lagrangian.

For $W^\pm$ boson, if we instead define $x_s=s/m_W^2$, $x_t=t/m_W^2$, $x_u=u/m_W^2$, the one-loop LE-expanded results are
\begin{eqnarray}
-i\mathcal{M}_{++++}^{(0,0,W),\mathrm{LE}}&=&\alpha^2\Bigg[\frac{1}{10}\left(x_s^2+x_t^2+x_u^2\right)+\frac{1}{21}x_sx_tx_u+\frac{1}{630}\left(x_s^2+x_t^2+x_u^2\right)^2\nonumber\\
&&+\frac{1}{660}x_sx_tx_u\left(x_s^2+x_t^2+x_u^2\right)+\left(\frac{109}{300300}x_s^2x_t^2x_u^2+\frac{1}{32032}\left(x_s^2+x_t^2+x_u^2\right)^3\right)\nonumber\\
&&+\frac{9}{200200}x_sx_tx_u\left(x_s^2+x_t^2+x_u^2\right)^2+\mathcal{O}\left(x_s^8\right)\Bigg]\,,\nonumber\\
-i\mathcal{M}_{-+++}^{(0,0,W),\mathrm{LE}}&=&\alpha^2\Bigg[\frac{1}{210}x_sx_tx_u+\frac{1}{27720}x_sx_tx_u\left(x_s^2+x_t^2+x_u^2\right)+\frac{1}{75075}x_s^2x_t^2x_u^2\nonumber\\
&&+\frac{1}{2402400}x_sx_tx_u\left(x_s^2+x_t^2+x_u^2\right)^2+\mathcal{O}\left(x_s^8\right)\Bigg]\,,\nonumber\\
-i\mathcal{M}_{--++}^{(0,0,W),\mathrm{LE}}&=&\alpha^2\Bigg[\frac{14}{5}x_s^2-\frac{47}{630}x_s^3+\frac{1}{6300}x_s^2\left(262x_s^2-345x_tx_u\right)-\frac{1}{103950}x_s^3\left(292x_s^2-1439x_tx_u\right)\nonumber\\
&&+x_s^2\left(-\frac{4211}{7567560}x_s^4+\frac{79}{28600}x_s^2\left(x_t^2+x_u^2\right)+\frac{1244}{675675}x_s\left(x_t^3+x_u^3\right)+\frac{622}{675675}\left(x_t^4+x_u^4\right)\right)\nonumber\\
&&-x_s^3\left(-\frac{1403}{4204200}x_s^4+\frac{947}{1051050}x_s^2\left(x_t^2+x_u^2\right)+\frac{9}{10010}x_s\left(x_t^3+x_u^3\right)+\frac{9}{20020}\left(x_t^4+x_u^4\right)\right)\nonumber\\
&&+\mathcal{O}\left(x_s^8\right)\Bigg]\,,\nonumber\\
\mathcal{M}_{+-+-}^{(0,0,W),\mathrm{LE}}&=&\left.\mathcal{M}_{--++}^{(0,0,W),\mathrm{LE}}\right|_{x_s\leftrightarrow x_u}\,,\quad \mathcal{M}_{+--+}^{(0,0,W),\mathrm{LE}}=\left.\mathcal{M}_{--++}^{(0,0,W),\mathrm{LE}}\right|_{x_s\leftrightarrow x_t}\,.
\end{eqnarray}

\begin{figure}[hbt!]
\includegraphics[width=0.5\columnwidth,draft=false]{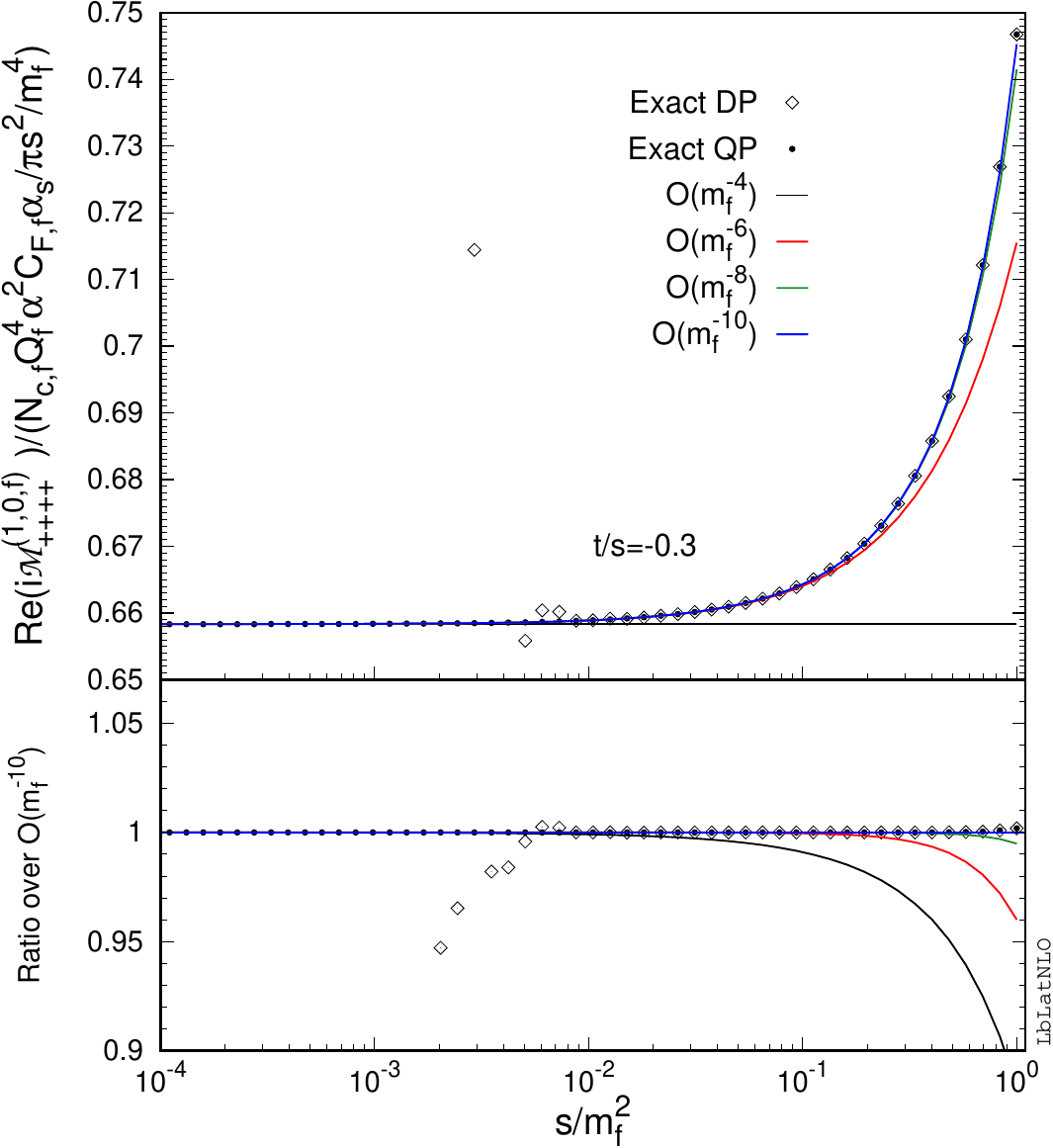}
\includegraphics[width=0.5\columnwidth,draft=false]{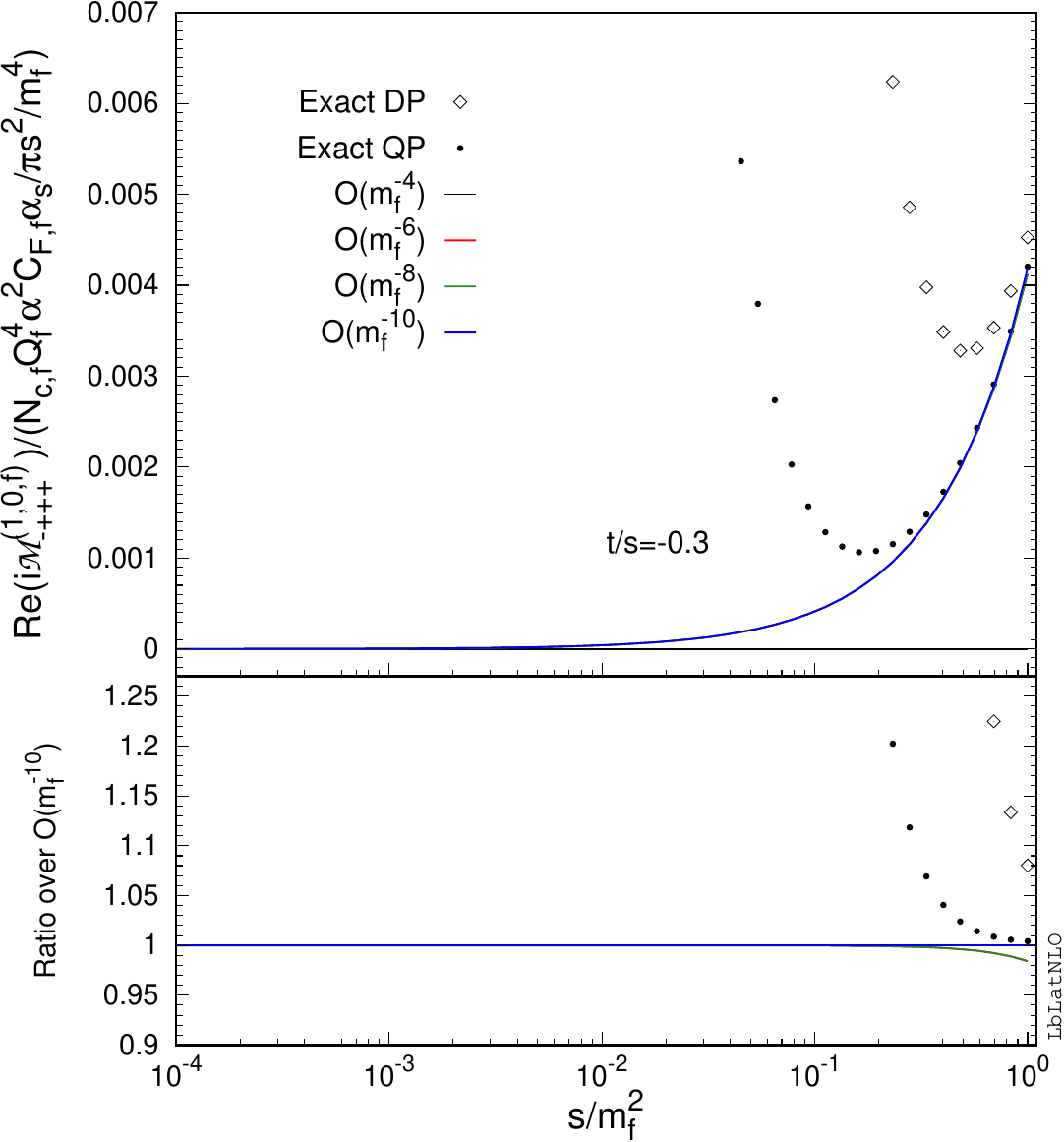}\\
\includegraphics[width=0.5\columnwidth,draft=false]{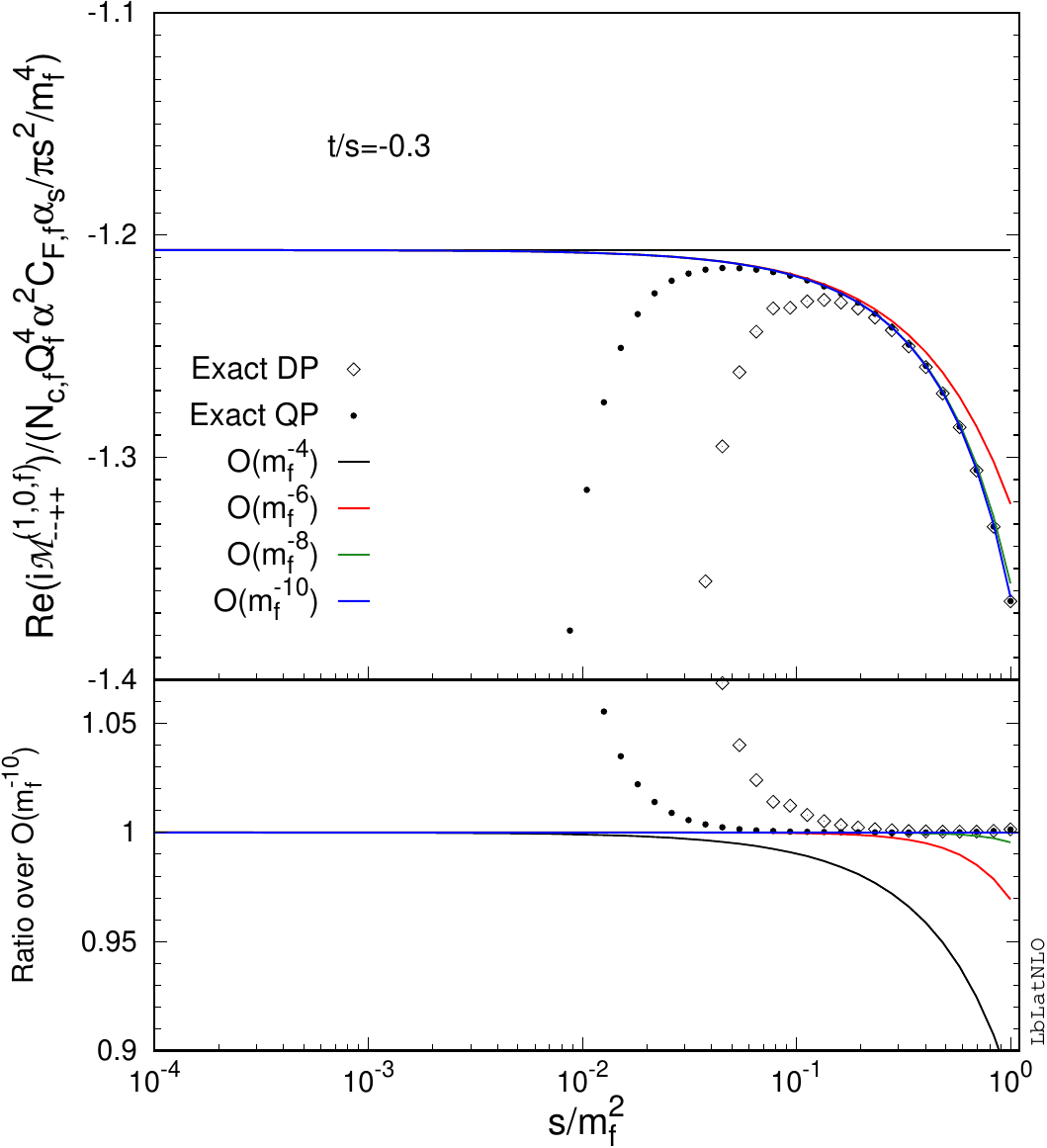}
\includegraphics[width=0.5\columnwidth,draft=false]{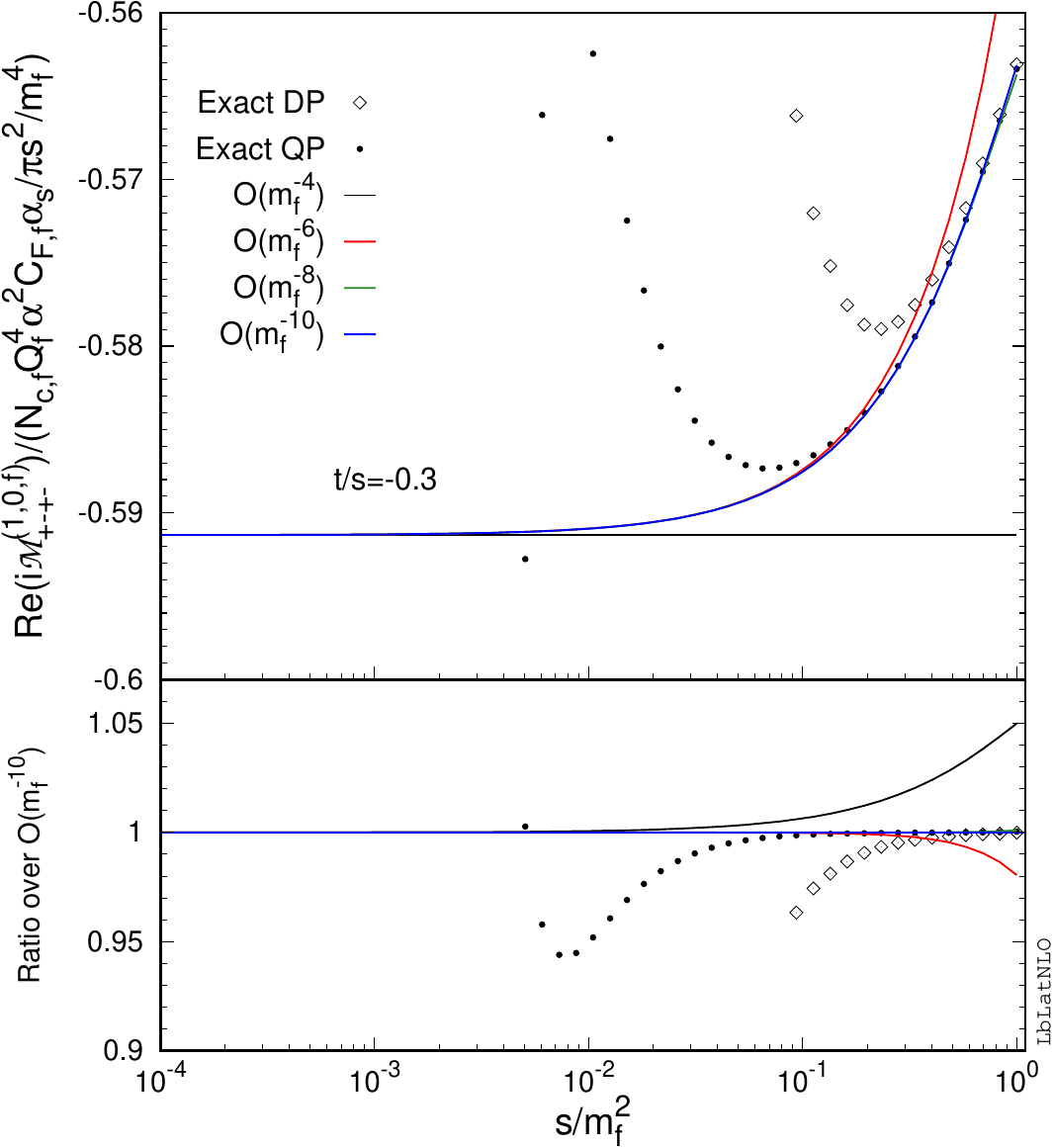}
\caption{The real parts of all-plus (upper left), single-minus (upper right), and two-minus-two-plus (lower) two-loop QCD amplitudes $i\mathcal{M}_{\lambda_1\lambda_2\lambda_3\lambda_4}^{(1,0,f)}$ as functions of $s/m_f^2$ in the LE region. Exact results obtained with double precision (open diamonds) and quadruple precision (filled circles) are compared with the LE expansion truncated at $\mathcal{O}(m_f^{-4})$ (black curves), $\mathcal{O}(m_f^{-6})$ (red curves), $\mathcal{O}(m_f^{-8})$ (green curves), and $\mathcal{O}(m_f^{-10})$ (blue curves). The scattering angle is fixed at $t/s = -0.3$. The lower panels show the ratios with respect to the LE result truncated at $\mathcal{O}(m_f^{-10})$.}
\label{figLE} \vspace*{-0.5cm}
\end{figure}

The LE-expanded expressions for the one- and two-loop master integrals and scattering amplitudes are provided in the ancillary files accompanying this paper.

As an illustration, we present the real parts of four two-loop helicity amplitudes, $\mathcal{M}^{(1,0,f)}_{++++}$ (all-plus), $\mathcal{M}^{(1,0,f)}_{-+++}$ (single-minus), $\mathcal{M}^{(1,0,f)}_{--++}$ (two-minus-two-plus), and $\mathcal{M}^{(1,0,f)}_{+-+-}$ (two-minus-two-plus), in figure~\ref{figLE}. The results are shown for the range $10^{-4} < s/m_f^2 < 1$, with $t/s = -0.3$ fixed. Their imaginary parts vanish for $s < 4m_f^2$. The real parts are normalised by the common prefactor $N_{c,f} Q_f^4 \alpha^2 C_{F,f} \alpha_s/\pi$ and by an additional factor of $s^2/m_f^4$. In each panel, two exact numerical evaluations obtained with different arithmetic precisions are displayed: double precision (exact DP, open diamonds) and quadruple precision (exact QP, filled circles). We also show the results of the LE expansion, including the leading ($\mathcal{O}(m_f^{-4})$, black curves), next-to-leading ($\mathcal{O}(m_f^{-6})$, red curves), next-to-next-to-leading ($\mathcal{O}(m_f^{-8})$, green curves), and terms up to $\mathcal{O}(m_f^{-10})$ (blue curves). The lower panels display the ratios with respect to the LE result truncated at $\mathcal{O}(m_f^{-10})$. The figure clearly demonstrates that the LE expansion substantially improves the numerical stability of the amplitudes. Among the four helicity configurations, the all-plus (single-minus) amplitude is the most (least) stable in the LE region. This behaviour can be readily understood: the exact all-plus two-loop amplitude is structurally the simplest~\cite{AH:2023ewe}, whereas the most severe numerical cancellations occur in the single-minus case because its leading term at $\mathcal{O}(m_f^{-4})$ vanishes. In the all-plus case, the exact quadruple-precision result agrees well with the LE expansion up to $\mathcal{O}(m_f^{-10})$ over the entire range of $s/m_f^2$, while numerical artefacts in the exact double-precision result become visible for $s/m_f^2 < 10^{-2}$. In contrast, the exact quadruple-precision results become unstable for $s/m_f^2 \lesssim 0.06$ in the two-minus-two-plus cases and for $s/m_f^2 \lesssim 0.8$ in the single-minus case. 

\subsection{High-energy expansion\label{sec:HEexpansion}}

In this section, we consider the HE (Sudakov) region, where the Mandelstam variables satisfy
\begin{equation}
s, |t|, |u| \gg m_f^2\,. \label{eq:HEregion}
\end{equation}
In this regime, it is well known that Sudakov logarithms of the form $\log\left(s_{ij}/m_f^2\right)$, with $s_{ij} \in \left\{s,t,u\right\}$, arise. For fermionic contributions to LbL scattering, Sudakov logarithms do not contribute at leading power in $m_f^2$ (\ie, $\mathcal{O}(m_f^0)$)~\cite{Bern:2001dg}. However, they are generically non-vanishing at next-to-leading (subleading) power and beyond in the expansion in $m_f^2$. The resummation of subleading-power Sudakov logarithms has recently attracted significant attention in the literature~\cite{Kotsky:1997rq,Fadin:1997sn,Akhoury:2001mz,Penin:2014msa,Melnikov:2016emg,Penin:2016wiw,Liu:2017vkm,Liu:2018czl,Wang:2019mym,Liu:2019oav,Anastasiou:2020vkr,Liu:2020tzd,Liu:2020wbn,Liu:2021chn,Liu:2022ajh,Bell:2022ott,Hu:2025hfc,Delto:2025epy}. Therefore, knowledge of the HE expansion of the LbL amplitudes may be useful for understanding the subleading-power logarithmic structure of loop-induced $2 \to 2$ scattering processes. The presence of subleading-power Sudakov logarithms also causes the full-$m_f$-dependent two-loop result to converge only slowly towards the HE-limit result with $m_f = 0$ (cf. figure~2 of ref.~\cite{AH:2023kor}). Similar to the LE case, the analytic HE expansion of the amplitudes improves numerical stability in practical computations.

In order to obtain the HE-expanded amplitudes from the known full mass-dependent results~\cite{AH:2023ewe}, we need to perform a series expansion of the master integrals in $m_f^2$. To this end, we adopt the differential-equation method~\cite{Kotikov:1990kg,Remiddi:1997ny,Gehrmann:1999as,Argeri:2007up}.

As explained in section~3 of ref.~\cite{AH:2023ewe}, the full mass-dependent master integrals are expressed in a canonical basis~\cite{Caron-Huot:2014lda}, in which the differential equations take the $\epsilon$-form~\cite{Henn:2013pwa}. Let $\vec{f}^{(2)} = (f_1^{(2)}, \cdots, f_{29}^{(2)})$ denote the full mass-dependent two-loop master integrals in this canonical basis, whose definitions can be found in eq.~(D.1) of ref.~\cite{AH:2023ewe}. The corresponding system of differential equations for $\vec{f}^{(2)}$ reads
\begin{equation}
    \derive \vec{f}^{(2)}= \epsilon\; \derive A^{(2)} \vec{f}^{(2)}\,.\label{eq:DE1}
\end{equation}
Expanding the master integrals in the dimensional regulator,
\begin{equation}
\vec{f}^{(2)}=\sum_{\omega=0}^{+\infty}{\epsilon^\omega \vec{f}^{(2,\omega)}}\,,
\end{equation}
the differential equation in eq.~\eqref{eq:DE1} becomes
\begin{equation}
\derive \vec{f}^{(2,\omega+1)}=\derive A^{(2)}\vec{f}^{(2,\omega)},\quad \omega\in \mathbb{N}\,,\label{eq:DE2}
\end{equation}
where $\omega$ denotes the transcendental weight. The full mass-dependent matrix $\derive A^{(2)}$ is given in eqs.~(3.6, 3.7) of ref.~\cite{AH:2023ewe}.

To obtain HE-expanded expressions for $\vec{f}^{(2,\omega)}$, we assume the following ansatz:
\begin{equation}
\vec{f}^{(2,\omega)}=\sum_{i=0}^{+\infty}{\sum_{j=0}^{\omega}{m_f^{2i}\log^j{\left(m_f^2\right)}\vec{f}^{(2,\omega,i,j)}}}\,,\label{eq:fHEexpansatz}
\end{equation}
where the coefficients $\vec{f}^{(2,\omega,i,j)}$ are independent of $m_f$ and $\epsilon$. Similarly, the series expansion of $\derive A^{(2)}$ in $m_f^2$ is straightforward in eq.~\eqref{eq:DE2}. Unlike the full mass-dependent case, the $m_f^2$-expanded $\derive A^{(2)}$ contains no square roots. This allows us to iteratively solve for $\vec{f}^{(2,\omega,i,j)}$ from lower transcendental weight $\omega$ and smaller powers of $m_f^2$ (\ie, $i$ in eq.~\eqref{eq:fHEexpansatz}) to higher weight and larger $i$, in terms of GPLs, with boundary conditions fixed by the full mass-dependent results. In solving the differential equations, we employ the following recursion relations for GPLs:
\begin{equation}
\begin{aligned}
&\underbrace{\int_0^y{\derive z \frac{1}{(z-a_0)^k}G(a_1,a_{2},\ldots,a_n;z)}}_{\equiv G_k(a_0;a_1,a_2,\ldots,a_n;y)}\\
=&-\frac{1}{k-1}\frac{1}{(y-a_0)^{k-1}}G(a_1,a_{2},\ldots,a_n;y)\\
&+\frac{1}{k-1}\left\{\begin{array}{ll} 
G_{k}\left(a_0;a_2,\ldots,a_n;y\right), & a_1=a_0 \\ 
\frac{1}{(a_1-a_0)^{k-1}}\left[G(a_{1},a_{2},\ldots,a_n;y)-G(a_{0},a_{2},\ldots,a_n;y)\right] &  \\
-\sum_{i=2}^{k-1}{\frac{1}{(a_1-a_0)^{k-i}}G_{i}\left(a_0;a_2,\ldots,a_n;y\right)}, & a_1\neq a_0\\ 
\end{array}\right.,\label{eq:Grecur1}
\end{aligned}
\end{equation}
and 
\begin{equation}
\begin{aligned}
&\int_0^y{\derive z z^k G(a_1,a_{2},\ldots,a_n;z)}\\
=&\left\{\begin{array}{ll}\frac{1}{k+1}\left(y^{k+1}-a_1^{k+1}\right)G(a_1,a_{2},\ldots,a_n;y)-\sum_{i=0}^{k}{\frac{a_1^{k-i}}{k+1}\int_0^y{\derive z z^i G(a_{2},\ldots,a_n;z)}}, & a_1\neq 0\\
\frac{1}{k+1}y^{k+1} G(a_1,a_{2},\ldots,a_n;y)-\frac{1}{k+1}\int_0^{y}{\derive z z^k G(a_2,\ldots,a_n;z)}, & a_1=0\\
\end{array}\right.\,,\label{eq:Grecur2}
\end{aligned}
\end{equation}
where $k\in\mathbb{N}$.
These relations follow from integration by parts (cf. eq.~(3.44) of ref.~\cite{Shao:2025qgv}) and from the definition of GPLs:
\begin{equation}
G(a_1,a_2,\ldots,a_n;z)\equiv\int_0^z{\frac{\derive t}{t-a_1}G(a_2,\ldots,a_n;t)}
\end{equation}
with $G(;z)\equiv 1$ and $G(\underbrace{0,\ldots,0}_{n};z)\equiv\log^n{\left(z\right)}/n!$. The initial conditions for the recursion relations in eqs.~\eqref{eq:Grecur1} and \eqref{eq:Grecur2} are
\begin{equation}
\begin{aligned}
G_k\left(a_0;;y\right)=&-\frac{1}{k-1}\times\left\{\begin{array}{ll}\left(\frac{1}{(y-a_0)^{k-1}}-\frac{1}{(-a_0)^{k-1}}\right), & a_0\neq 0\\
\frac{1}{y^{k-1}}, & a_0=0\\
\end{array}\right.
\,,\\ 
G_1\left(a_0;a_1,\ldots,a_n;y\right)=&G(a_0,a_1,\ldots,a_n;y)\,,
\end{aligned}
\end{equation}
and
\begin{equation}
\int_0^{y}{\derive z z^k G(;z)}=\frac{y^{k+1}}{k+1}\,.
\end{equation}
Using this approach, we have obtained analytic expressions for $\vec{f}^{(2,\omega,i,j)}$ up to $\omega \leq 4$ and $i \leq 10$, which are provided in the ancillary file \texttt{TwoLoopMI\_HE.wl}.

As a cross-check, we have verified that our HE-expanded master integrals can be mapped onto those presented in ref.~\cite{Davies:2018ood} for Higgs boson pair production in the HE limit. Such a linear mapping is non-trivial, since the master integrals are defined in different integral families from ours. With an appropriate correspondence established, we find agreement with their results up to the available orders in the $m_f^2$ expansion.

The final HE-expanded helicity amplitudes can be expressed in terms of the following four dimensionless variables:
\begin{equation}
x\equiv\frac{t}{s}\,,\quad y\equiv\frac{u}{s}\,,\quad \Xb\equiv \log{\left(-x\right)}+i\pi\,,\quad \Yb\equiv \log{\left(-y\right)}+i\pi\,,\label{eq:xyXYdef}
\end{equation}
where $\Xb$ and $\Yb$ depend on $x$ and $y$, respectively. In the HE region, the helicity amplitudes generally take the form of a double series in $m_f^2$ and $\log(m_f^2)$. Since the all-plus $\mathcal{M}_{++++}$, single-minus $\mathcal{M}_{-+++}$, and two-minus-two-plus $\mathcal{M}_{--++}$ helicity amplitudes are symmetric under the exchange $t\leftrightarrow u$, whereas the remaining two-minus-two-plus amplitudes $\mathcal{M}_{+-+-}$ and $\mathcal{M}_{+--+}$ are not, the full mass-dependent amplitudes can be decomposed as
\begin{equation}
i\mathcal{M}_{\lambda_1\lambda_2\lambda_3\lambda_4}^{(l,0,f),\mathrm{HE}}=C^{(l,0,f)}\Bigg[\sum_{p=0}^{+\infty}{\sum_{q=0}^{2+2l}{\left(-\frac{m_f^2}{s}\right)^p\log^q{\left(-\frac{m_f^2}{s}\right)}\left(\mathcal{M}_{\lambda_1\lambda_2\lambda_3\lambda_4}^{(l,0,f,p,q)}(x,y)+\mathcal{M}_{\lambda_1\lambda_2\lambda_3\lambda_4}^{(l,0,f,p,q)}(y,x)\right)}}\Bigg]\label{eq:Ml1HE}
\end{equation}
for $(\lambda_1\lambda_2\lambda_3\lambda_4)\in\left\{(++++),(-+++),(--++)\right\}$, and
\begin{equation}
i\mathcal{M}_{\lambda_1\lambda_2\lambda_3\lambda_4}^{(l,0,f),\mathrm{HE}}=C^{(l,0,f)}\Bigg[\sum_{p=0}^{+\infty}{\sum_{q=0}^{2+2l}{\left(-\frac{m_f^2}{s}\right)^p\log^q{\left(-\frac{m_f^2}{s}\right)}\mathcal{M}_{\lambda_1\lambda_2\lambda_3\lambda_4}^{(l,0,f,p,q)}(x,y)}}\Bigg]\label{eq:Ml2HE}
\end{equation}
for $(\lambda_1\lambda_2\lambda_3\lambda_4)\in\left\{(+-+-),(+--+)\right\}$. From crossing symmetry, we further have
\begin{equation}
\mathcal{M}_{+--+}^{(l,0,f,p,q)}(x,y)=\mathcal{M}_{+-+-}^{(l,0,f,p,q)}(y,x)\,.
\end{equation}
The prefactor in eqs.~\eqref{eq:Ml1HE} and \eqref{eq:Ml2HE} is given by
\begin{equation}
C^{(l,0,f)}=\left\{\begin{array}{ll}-8N_{c,f}Q_f^4\alpha^2\,, & l=0\\
(1+3\delta_{\lambda_1\lambda_2}\delta_{\lambda_2\lambda_3}\delta_{\lambda_3\lambda_4})N_{c,f}Q_f^4\alpha^2\frac{\alpha_s}{\pi}C_{F,f}\,, & l=1\end{array}\right.\,.
\end{equation}

While the one-loop coefficients from both fermionic and $W^\pm$ bosonic contributions are presented in appendix~\ref{sec:oneloopHE}, we restrict ourselves here to the two-loop HE expansion of the fermionic sector. Explicit results are given up to $\mathcal{O}(m_f^2)$, whereas higher-order terms are provided in the ancillary files. For the all-plus amplitude $\mathcal{M}_{++++}^{(1,0,f)}$, the non-vanishing coefficients are given by
\begin{align}
\mathcal{M }^{(1,0,f,0,0)}_{++++}(x,y)=&\dfrac{3}{2}\,, \quad 
\mathcal{M }^{(1,0,f,1,2)}_{++++}(x,y)=\dfrac{1-xy}{xy}\,,\quad
\mathcal{M }^{(1,0,f,1,1)}_{++++}(x,y)=\dfrac{4 \Xb}{x}\,,
\nonumber\\
\mathcal{M }^{(1,0,f,1,0)}_{++++}(x,y)=&-\dfrac{6\Xb^2}{x}-\dfrac{4(1-xy)\Xb^2}{xy}-4 \Xb\ \Yb-\dfrac{12(1-xy)\zeta_2}{xy}\,,
 \end{align}
 where $\zeta_n$ is the Riemann zeta function $\zeta_n\equiv\zeta(n)$. The leading-power term is constant, while the subleading-power contributions contain double logarithms. For the single-minus amplitude $\mathcal{M}_{-+++}^{(1,0,f)}$, the non-zero coefficients read
 \begin{align}
\mathcal{M}^{(1,0,f,0,0)}_{-+++}(x,y)=&
    -\frac{(1+x^2)}{y^2} \left(\Xb^2+\pi^2\right)
    -\frac{1}{2} \left(x^2+y^2\right) \left[\pi^2+\left(\Xb-\Yb\right)^2\right]+\frac{4 \Xb (1-x y)}{y}\,,\nonumber\\

\mathcal{M}^{(1,0,f,1,4)}_{-+++}(x,y)=&\frac{1-x y}{6 x y}\,,\quad \mathcal{M}^{(1,0,f,1,3)}_{-+++}(x,y)=\frac{4 \Xb}{3 x}-\frac{2 (1-x y)}{x y}\,,\nonumber\\

\mathcal{M}^{(1,0,f,1,2)}_{-+++}(x,y)=&\frac{x}{y}\Xb^2+\frac{1+2 x y}{x y} \Xb\Yb-\frac{14 y+x^2}{x y}\Xb+\frac{2 (1+4 \zeta_2) (1-x y)}{x y}\,,\nonumber\\
 
\mathcal{M }^{(1,0,f,1,1)}_{-+++}(x,y)=&
    -\frac{2(y+3)(1+3y)}{3 x y}\Xb^3
    -\frac{4 + 12 x + 2 x^2}{x y}\Xb^2\Yb
    +\frac{2 \left(8 y-5 x^2+3i \pi  \left(y^2-1\right)\right)}{x y}\Xb^2
\nonumber \\ &
    +\frac{1-10 x y
    }{x y}\Xb\Yb
    -\frac{4}{3xy} \left(3 x^2+4 \pi ^2 (x^2+y)
    +9 (x+2) x \text{Li}_2\left(-x\right)\right)\Xb
\nonumber \\ &    
    +\frac{12 (x+2)}{y} \text{Li}_3\left(-x\right)
    +\frac{1}{x y}\left(
    2 \zeta_3 (17-14 x y)
    +4 (1-\pi ^2) (1-x y)
    -3 i \pi ^3 (1+2 x y)\right)\,,\nonumber\\ 
\mathcal{M }^{(1,0,f,1,0)}_{-+++}(x,y)=&
     -\frac{\Xb^4}{6xy}\left(15+20x+2x^2\right)
     + \frac{2\Xb^3\Yb}{3xy}\left(8 + 30 x + 11 x^2\right)
     - \frac{3-22 x y}{x y}\Xb^2\Yb^2
     
\nonumber \\ &
     +\left(\frac{9 x^2-6 y}{x y}-\frac{16 i \pi  \left(2+x\right)}{3 y}\right)\Xb^3
     -\left(\frac{1-7 x^2-30 x}{x y}+\frac{2 i \pi  \left(4-17 x^2-6 x\right)}{xy}\right)\Xb^2\Yb
\nonumber \\ &
     +\left(
     \frac{4 (1-y^2)}{xy} (6 i \pi-\text{Li}_2(-x))
     +\frac{4 \left(x^2+y\right)}{x y}-\frac{20 \pi ^2 y}{3 x}+\frac{5
   \pi ^2 (y-x)}{x y}\right)\Xb^2
\nonumber \\ &
   +\left(
   \frac{2 (1+2 x y)}{x y}-\frac{13 \pi ^2 (1+6 x y)}{3 x y}\right)\Xb\Yb
   -\frac{2 \left(20 x^2+18 x+5\right) }{x y}\Xb\Yb\text{Li}_2(-x)
\nonumber \\ &
   + \frac{1+2 x y}{x y}\pi ^2 \text{Li}_2(-x)
   +\frac{16 (1-y) }{y}
   \Big[
   \Xb\text{Li}_3(-x)
   +3\Xb\text{Li}_2(-x)-3\text{Li}_3(-x)
   \Big]
\nonumber \\ &
   + \left(
   \frac{ 19-4 x^2+6 x}{x y}i \pi ^3
   +\frac{ x}{y}(17 \pi ^2-4)
   +\frac{16 (x+4)\zeta_3}{x}
   +\frac{26 \pi ^2}{x}
   \right)\Xb
\nonumber \\ &
   +\frac{4 \left(3 y^2-8 y-1\right)}{x y}\Xb \text{Li}_3(-y)
   +\frac{40 (x-1) }{x}\text{Li}_4(-x)
   + \frac{20(x-y)}{xy} \text{Li}_4\left(-\frac{x}{y}\right)
\nonumber \\ &
   -\frac{(1-x y)}{x y}(8+4 \zeta_2) +\frac{4  (373+122 x y)\zeta_2^2}{5 x y}-\frac{16
   (4-x y)\zeta_3 }{x y}+\frac{24 i \pi ^3 (1+2 x y)}{x y}\,,
\end{align}
where $\text{Li}_n$ is the classical polylogarithm. For the two-minus-two-plus amplitude $\mathcal{M}_{--++}^{(1,0,f)}$, the non-vanishing coefficients are
\begin{align}
\mathcal{M }^{(1,0,f,0,0)}_{--++}(x,y)=&
    \frac{4x^2}{3}  \Xb^4
    -\frac{16 x^2}{3}\Xb^3 \Yb
    +\frac{2}{3} \left(1+10x+4 i \pi  x^2\right) \Xb^3
    -2(3+2 x y) \Xb \Yb
\nonumber \\  &
    +(1 - 2 x y)
    \left(
    8\Xb^2\Yb^2
    -8(\pi^2-2 \text{Li}_2(-x)) \Xb \Yb 
    +\frac{36}{5} \zeta_2^2
    +16\text{Li}_4(-x) 
    \right)
\nonumber \\  &
    +\left(8 i \pi  \left(x-y-3 y^2\right)-2(1+10 x)\right) \Xb^2 \Yb
     -\left(
    \frac{4}{y}
    +8 x (1-13 \zeta_2)
    -68 \zeta_2
    +\frac{8}{3} i \pi ^3 x^2
    \right)\Xb
\nonumber \\  &
    +\left(10+4 x y+8 \pi ^2 x^2-4 i \pi  (1-2 x)-\frac{2 (3+4 x)}{y^2}
    \right)\Xb^2
\nonumber \\  &
    
    +8 (1-2x) \left(
    \Xb \text{Li}_2(-x)
    -\text{Li}_3(-x)\right)
    -16 x^2\left( 
     \Xb \text{Li}_3(-x)
    +\Xb\text{Li}_3(-y)
    -\zeta_2 \text{Li}_2(-x)
    \right)   
\nonumber \\  &
    + 8 (x - y) \text{Li}_4\left(-\frac{x}{y}\right)
    +16\zeta_3
    +2-8 i \pi^3
    + \frac{2(3-18 x y+19 x^2 y^2+6 x^3 y^3)\zeta_2}{x^2 y^2}\,,
\nonumber \\ 

\mathcal{M }^{(1,0,f,1,4)}_{--++}(x,y)=&\frac{13}{6xy}\,,\quad \mathcal{M }^{(1,0,f,1,3)}_{--++}(x,y)=\frac{12 \Xb}{x}+\frac{2}{3 x y}\,,\nonumber\\

\mathcal{M }^{(1,0,f,1,2)}_{--++}(x,y)=&
    -\frac{2 (4+3 x) \Xb^2}{x}
    -\frac{2 (7+4 x) \Xb}{x y}
    -\frac{6 \zeta_2 (1+3 xy)}{x y}
   +\frac{(x-y)^2}{x y}+6 \Xb \Yb\,,
\nonumber \\ 

\mathcal{M }^{(1,0,f,1,1)}_{--++}(x,y)=&
  \left(
   \frac{24 (1+2 x y)}{x y}-\frac{12 x}{y}\right)\Xb^2
   +\frac{8 \left(3 y^2-8\right) \zeta_2}{x}\Xb
   +4 (x+2) \Xb^3
   +12 y \Xb^2  \Yb\nonumber\\
   &+\frac{2 (1-30 x y)}{x y} \Xb \Yb
   -\frac{4
   (11+23 x) \Xb}{x y}
   +\frac{4 \zeta_2 (22+39 x y)}{x y}
   +\frac{24 \zeta_3}{xy}
   +\frac{6}{x y}\,,
   
\nonumber \\ 
\mathcal{M }^{(1,0,f,1,0)}_{--++}(x,y)=& 
    -\frac{\left(24 x^2+15 x-2\right) \Xb^4}{6 x}
    +2 (1 + 6 x) \Xb^3 \Yb\nonumber\\
    &-\left(\frac{60 x^3-52 x^2-80 x+42}{3 x y}
    +\frac{8i \pi}{3}   (2+3 x)\right)\Xb^3
    +\frac{13}{2}\Xb^2\Yb^2\nonumber \\ &
    +\left(\frac{60 x^3+64 x^2+48 x+40}{x y}
    +8 i \pi  (1+3 x)\right) \Xb^2\Yb
    +\frac{4 \left(6 x^2+3 x+1\right)}{x}\Xb^2 \text{Li}_2(-x)
\nonumber \\ &
    +\left(
    \frac{2 (13 x+25) (x-y)}{x y}
     +\frac{2 \left(30 x^2+7 x+6\right) \zeta_2}{x}
    -\frac{2 i \pi  \left(16 x^2+28 x+15\right)}{xy} \right.\nonumber \\ &   
   \left.+12(x-y) \text{Li}_2(-x)
    +\frac{4}{x}\text{Li}_2(-x)
    \right) \Xb^2
\nonumber \\ &
    -\left(
    12 \text{Li}_2(-x) (x-y)
    +\frac{4 (6-13 x y)}{x y}
    +34 \zeta_2
    \right) \Xb \Yb
    +\frac{4 \left(16 x^2+28 x+15\right)\Xb \text{Li}_2(-x)}{x y}
\nonumber \\ &       
    -\left(
    \frac{16 \text{Li}_3(-x) \left(3x^2-y\right)}{x}
    +16 (1+3 x) \text{Li}_3(-y)
    +\frac{12 \left(10 x^3-8 x^2+3 x+8\right)\zeta_2}{x y}\right.
\nonumber \\ &
     \left.-\frac{4 (23 x+8)}{x y}
    -8 i \pi  (9 x+5) \zeta_2
    +\frac{2}{3} i \pi ^3 (60x+29)
    \right)\Xb
    -\frac{4 \left(16 x^2+28 x+15\right) \text{Li}_3(-x)}{x y}
\nonumber \\ &     
    +\frac{4 \left(6 x^2+x+6\right)\zeta_2 \text{Li}_2(-x)}{x}
    -12 (x-y) \text{Li}_4\left(-\frac{x}{y}\right)
    -\frac{8 (2 x-3)\text{Li}_4(-x)}{x}\nonumber \\ &
    +\frac{6 \zeta_2 (7-38 x y)}{x y}     
    +\frac{3 \zeta_2^2 (23 x y+24)}{5x y}
    +\frac{2 \zeta_3 (35-12 x y)}{x y}
    -\frac{2 i \pi ^3 (9-16 x y)}{x y}-\frac{14}{x y}\,.

\end{align}
Finally, for the $+-+-$ helicity configuration, the non-zero coefficients are
\begin{align}

\mathcal{M }^{(1,0,f,0,0)}_{+-+-}(x,y)=&
   \frac{\left(1+3 x^2\right)}{3 y^2} \Xb^4
   -\frac{8 \left(1+x^2\right)}{3 y^2}\Xb\Yb^3
   +\frac{2 \left(1+x^2\right)}{3 y^2}\Yb^4
\nonumber \\ &
    -\left(\frac{2 (1-9 x)}{3 y}
    +\frac{8 i \pi  x^2}{3 y^2}\right)\Xb^3
    +\frac{8\Xb \Yb^2}{y}\left( 
    1-x
    +\frac{i \pi  \left(1+x^2\right)}{y}\right)
\nonumber \\ &
    +\left(
    2 \left(2+x^2\right)
    +\frac{4 x \left(1-y^2\right)}{y^2}
    -\frac{4 i \pi  (1+3 x)}{y}
    -\frac{2 \pi ^2 \left(1-x\right)}{3 y}
    \right)\Xb^2
\nonumber \\ &
    +\left(
    \frac{8 \pi ^2 \left(1+x^2\right)}{y^2}
    -\frac{16 i \pi  (1-x)}{y}
    -4 (y+x(x-1))
    \right)\Xb\Yb
\nonumber \\ &
    +\left(
    \frac{2 \left(1-2x+x^4\right)}{x^2}+\frac{16}{3} \pi ^2 \left(1-\frac{2 x}{y^2}\right)+4 y-\frac{8 i \pi  (2+y)}{y}
    \right)\Yb^2
\nonumber \\ &
    +\left(
    \frac{16 x^2}{y^2}\left( \text{Li}_3(-x)-\zeta_3\right)
    +\frac{8 (1+3 x)}{y} \text{Li}_2(-x)
    -\frac{8 i \pi ^3 \left(1+2 x^2\right)}{3 y^2}\right.\nonumber\\
    &\left.-\frac{8 x+4y^2}{y}
    -\frac{2 \pi ^2(17-9 x)}{3 y}
    \right)\Xb
    +\left(
    \frac{8 (1+x^2) }{y^2}\left(2\zeta_3-i\pi^3-2\text{Li}_3(-x)\right)\right.\nonumber\\
    &\left.+\frac{16 (1-x)}{y} \text{Li}_2(-x)
    -\frac{8x+4y^2}{x}
   -\frac{8 \pi ^2 (14+9 y)}{3 y}
   \right)\Yb
   +\frac{8 \pi ^2 \left(1-x^2\right) }{3 y^2}\text{Li}_2(-x)\nonumber \\ &
   +\frac{16 \left(1+x^2\right) }{y^2}
   \left(\text{Li}_4\left(-\frac{x}{y}\right)
   - \text{Li}_4(-y)
   \right)
  -\frac{8 (1+3 x)}{y} \text{Li}_3(-x)
\nonumber \\ &
  +\frac{16 (1-x)}{y} 
  \left(\text{Li}_3(-y)
  - \text{Li}_4(-x)
  \right)
  +4 \zeta_2 \left(10-3 (x-3) y+\frac{3-9 x^2}{x^2
   y^2}\right)\nonumber \\ &
   -\frac{\pi ^4 \left(149+183 x^2\right)}{45 y^2}
   +\frac{4 i \pi ^3 (1-5 x)}{y}+\frac{8 (y-2) \zeta_3}{y}+4\,,\nonumber\\

\mathcal{M }^{(1,0,f,1,4)}_{+-+-}(x,y)=&\frac{13 y}{3 x}\,,\quad \mathcal{M }^{(1,0,f,1,3)}_{+-+-}(x,y)=\frac{12 \Xb}{x}-\frac{16\Yb y}{3 x}+\frac{4 y}{3 x}\,,\nonumber\\

\mathcal{M }^{(1,0,f,1,2)}_{+-+-}(x,y)=&
    -\frac{2 \Xb^2 (y-3)}{x y}
    +2 \Xb \left(\frac{4-3 y}{x}
    -\frac{10 \Yb}{x}\right)
    -\frac{2 y\Yb^2}{x}
    +\frac{16 y \Yb}{x}\nonumber\\
    &+\frac{2 \left((1-x)^2-\pi ^2 \left(3 x+y^2\right)\right)}{x y}\,,\nonumber \\
    
 \mathcal{M }^{(1,0,f,1,1)}_{+-+-}(x,y)=&\Xb^2 \left(
    \frac{12 \left(x^2+8 x+2\right)}{x y}
    +\frac{16 \Yb}{x}\right)
    -8\Yb \left(
    \frac{1+x^2}{x y}
    +\frac{\pi ^2 y}{3 x}\right)
    +\frac{4 \pi^2 \left(
    22(1+ x^2)+83 x\right)}{3 x y}
\nonumber \\  &
    -\frac{4 (x+2) \Xb^3}{y^2}
    +4\Xb \left(
    \frac{\Yb^2}{x}
    +\frac{\pi ^2 \left(3-8 y^2\right)}{3 x y^2}
    +\frac{2(2 x+3)\Yb}{x}
   -\frac{(12 x-11)}{x}
   \right)\nonumber\\
   &+\frac{8 y \Yb^3}{3 x}
   +\frac{4 y \Yb^2}{x}
   +\frac{12 y (1+4 \zeta_3)}{x}\,,
\nonumber \\ 

 \mathcal{M }^{(1,0,f,1,0)}_{+-+-}(x,y)=& 
   \Xb^4\left(
   \frac{1}{3 x}+\frac{2 y+5}{y^2}
   \right)
   -\frac{x y+2 (x-y)}{3 x y}\Yb^4
   -\frac{4}{3 x}\Xb^3\Yb
   +4\Xb\Yb^3\left(\frac{x-1}{x}+\frac{2}{3 y}\right)
\nonumber \\  &
   -\left(
   \frac{172 y+60}{3 y^2}
   -\frac{2 (5 x+21)}{3 x}
   +\frac{8 i \pi  (y+3)}{3 y^2}
   \right)\Xb^3
   -\left(
   \frac{16 y}{x}
   +\frac{4 i\pi (x-1)}{x}\right)\Yb^3
\nonumber \\  &
   +\left(
   16-\frac{60}{x}
   -4 i \pi  \left(1-\frac{2}{x}+\frac{2}{y}\right)
   \right)\Xb\Yb^2
   -\left(
   2-\frac{56}{y}
  -\frac{22-12 i \pi }{x}
   \right)\Xb^2\Yb+2\Xb^2
\nonumber \\  &
    \times\left(
    \frac{12 x y-x+25}{x y}
     -\frac{i \pi  (3 x y+x-15)}{x y}
    -\frac{2 \left(1-x+4 x^2\right) }{x y^2}\text{Li}_2(-x)
    -\frac{\pi ^2 (6-11 x y)}{3 x y^2}
    \right)
\nonumber \\  &
    -4\Yb^2\left(
    \frac{2}{y} 
    +\frac{4 y}{x}
     -\frac{6 i \pi  (1-x)}{x}
    +\frac{(1-x)}{x} \text{Li}_2(-x)
    -\frac{\pi ^2 \left(7 y^2+17 y+4\right)}{3 x y}
     \right)
\nonumber \\  &
    +4\Xb\Yb\left(
    \frac{2 (1-6 x)}{x}
    +\frac{12 i \pi  (1-x)}{x}
    +\frac{2 }{x}\text{Li}_2(-x)
    -\frac{ \pi ^2 \left(y^2-2\right)}{x y}
    \right)
\nonumber \\  &
    +\Xb\left(
    60-\frac{32}{x}
    -\frac{4 i \pi ^3
   \left(x^3-x-2\right)}{x y^2}
   +\frac{\pi ^2 \left(-26 x^3+20 x^2+42 x+16\right)}{x y^2}
    +\frac{16}{x} \text{Li}_3(-y)
   \right.
 \nonumber \\  &  \left.
     +\frac{16 \left(3 x^2-y\right)}{x y^2}
    (\text{Li}_3(-x)-\zeta_3) 
    -\frac{4(3 x^2+2 x+15) }{x y}\text{Li}_2(-x)
     \right)
\nonumber \\  &
    +\Yb\left(
    \frac{16y}{x}(1-3\zeta_3)
     +\frac{4 i \pi ^3 \left(2 x+y^2\right)}{x y}
     -\frac{\pi ^2 \left(120-50 y^2-216 y\right)}{3 x y}\right.\nonumber\\
    &\left.+\frac{16  (x-y)}{x y}(\text{Li}_3(-x)-\zeta_3)
   -\frac{16 (1-x) }{x}\left(\text{Li}_3(-y)
    +3 \text{Li}_2(-x)\right)
   \right)\nonumber\\
    &+\frac{16 (5+4 x) (x-y)\zeta_3}{x y}
    -\frac{28 y}{x}
    -\pi ^2 \left(14
    +\frac{14}{x}+\frac{76}{y}
   \right)
\nonumber \\  &
    -\frac{\pi ^4 \left(297 x^3+784 x^2+382 x-54\right)}{45 x y^2}
   -\frac{2 i \pi ^3
   \left(9 x^2-2 x-27\right)}{x y}
   -  \frac{4 \pi ^2\left(1-x^3\right) }{x y^2}\text{Li}_2(-x)
\nonumber \\  &
   -\frac{24+40 x }{x y}\text{Li}_4\left(-\frac{x}{y}\right)
   +\frac{24 (1-x)}{y^2}\text{Li}_4(-x)
   -\left(12+\frac{60}{x}+\frac{64}{y}\right)\text{Li}_3(-x)\nonumber\\ 
   &+\frac{48 (x-1) }{x}\text{Li}_3(-y)
   +\frac{8 (5+3 x)}{y} \text{Li}_4(-y)\,.
\end{align}
In the single-minus and two-minus-two-plus cases, the leading-power terms are kinematically non-trivial, while the leading logarithms at subleading power correspond to quadruple Sudakov logarithms. While the leading terms at $\mathcal{O}(m_f^0)$ agree with the massless calculations in ref.~\cite{Bern:2001dg}, the higher-order terms are new.

\begin{figure}[hbt!]
\includegraphics[width=0.5\columnwidth,draft=false]{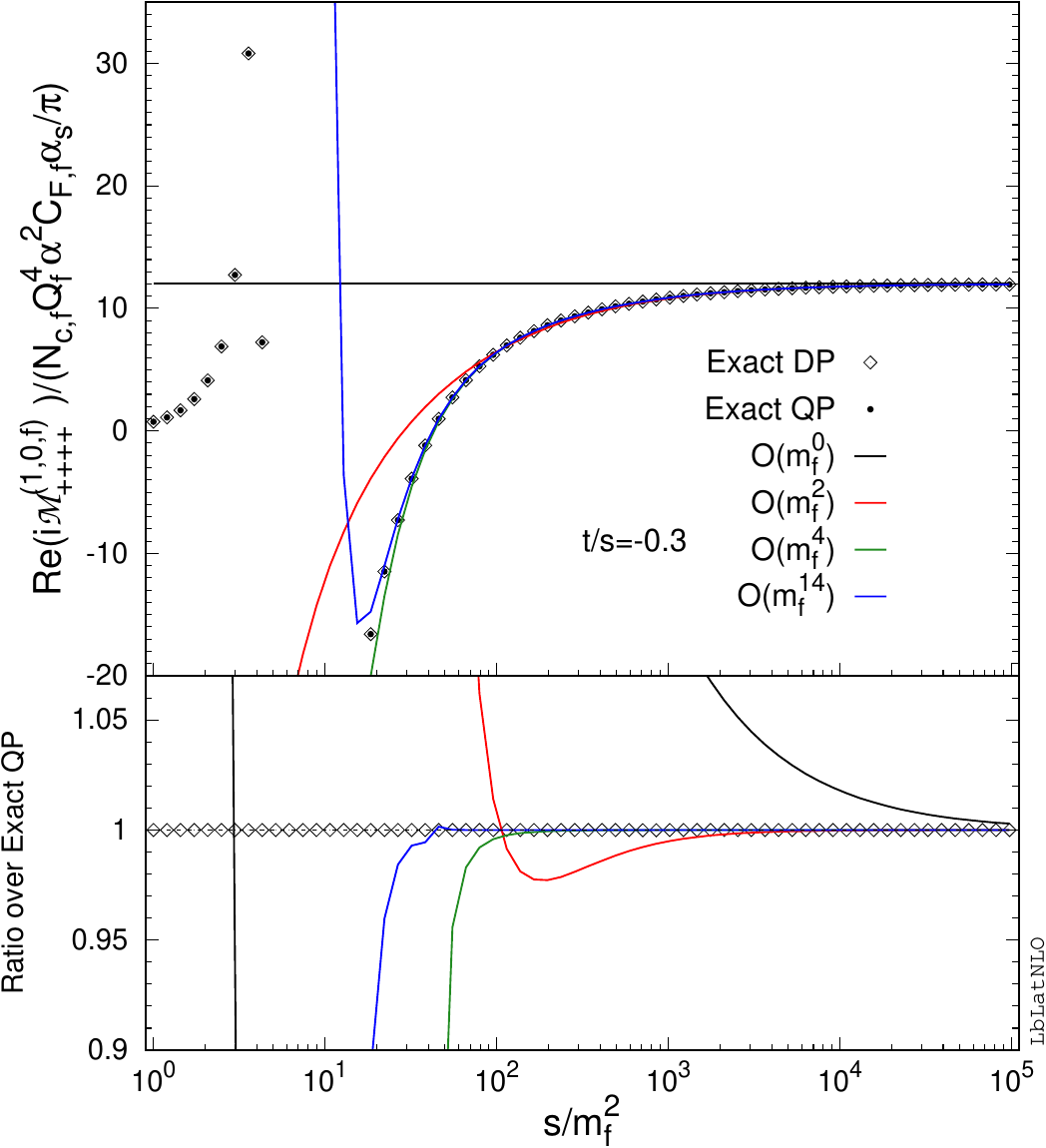}
\includegraphics[width=0.5\columnwidth,draft=false]{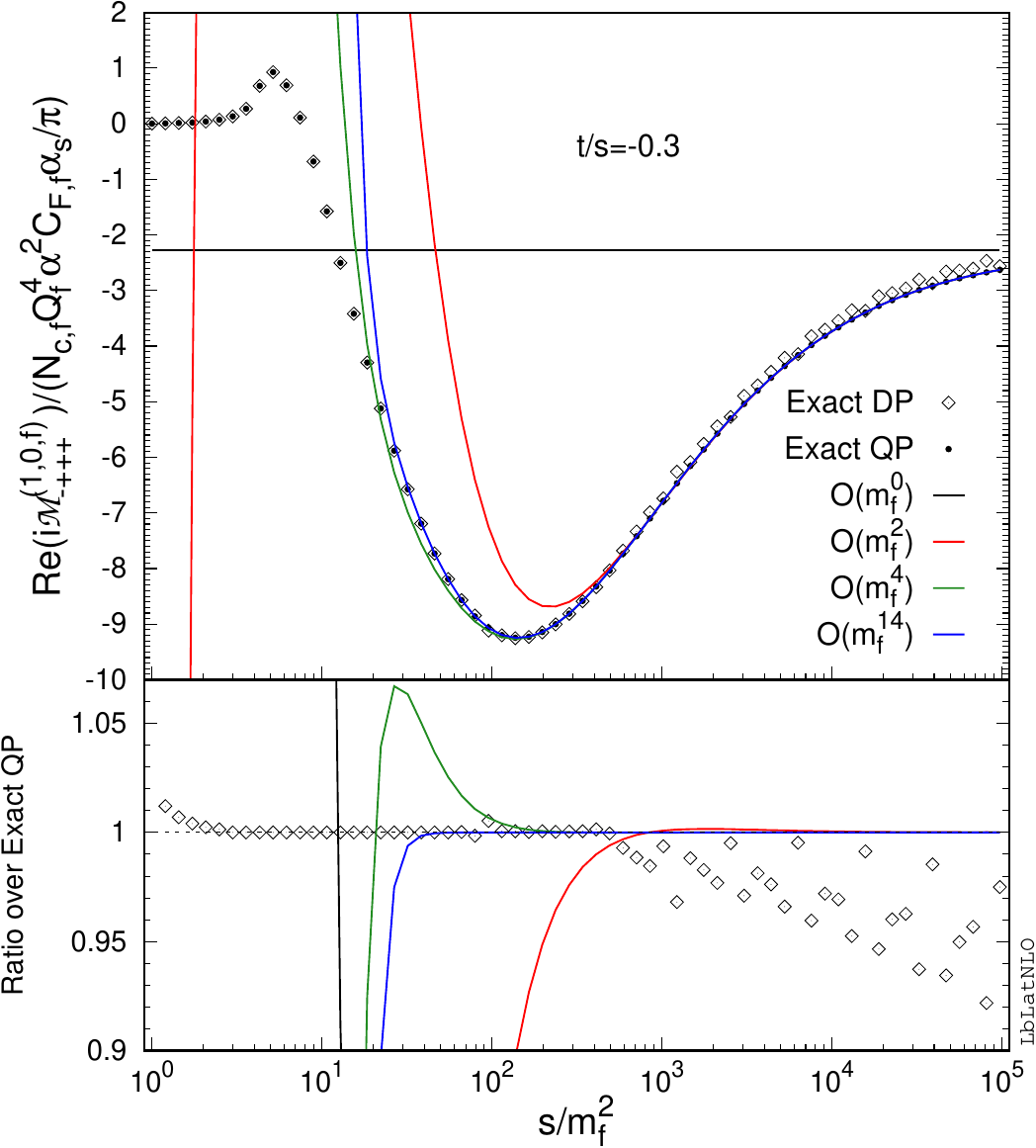}\\
\includegraphics[width=0.5\columnwidth,draft=false]{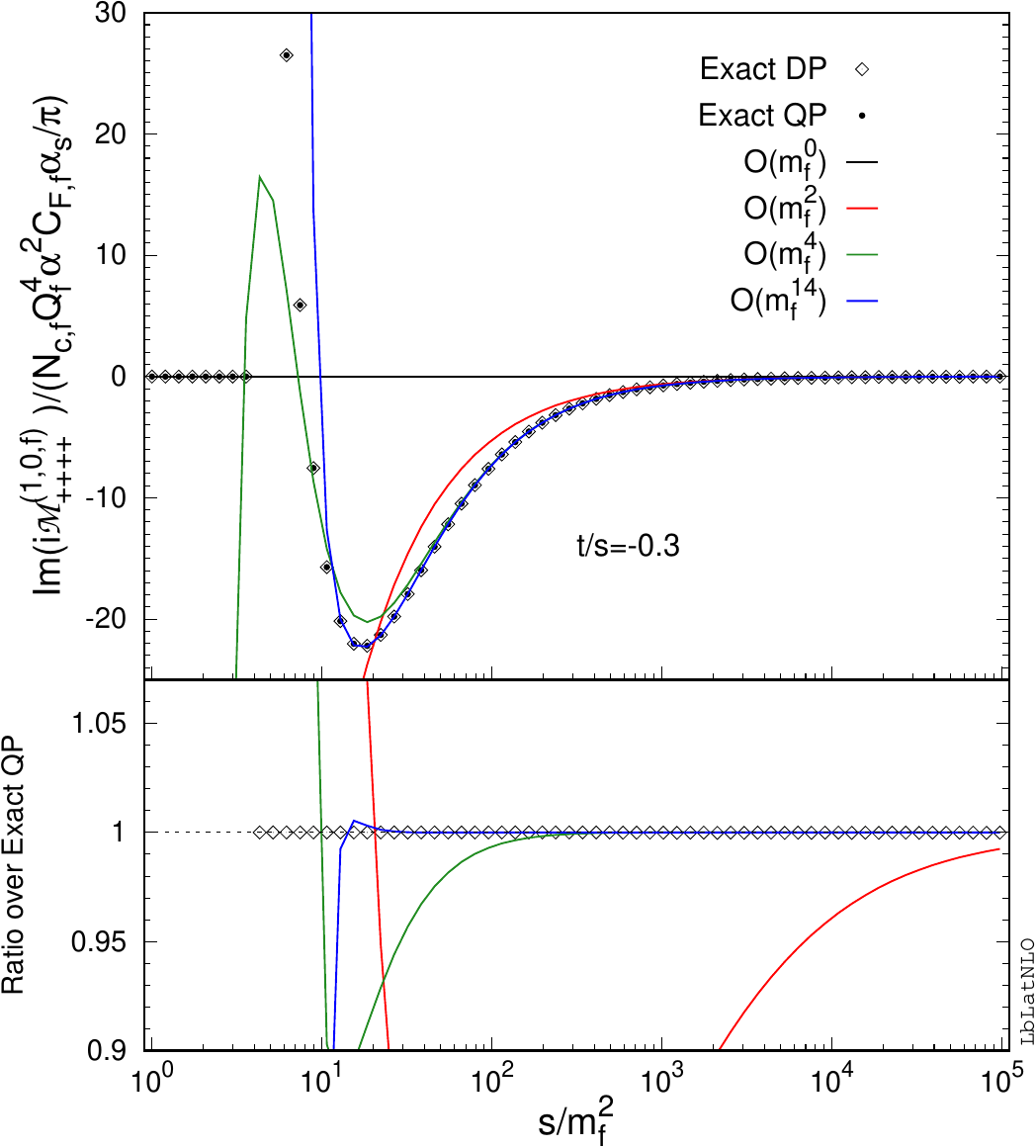}
\includegraphics[width=0.5\columnwidth,draft=false]{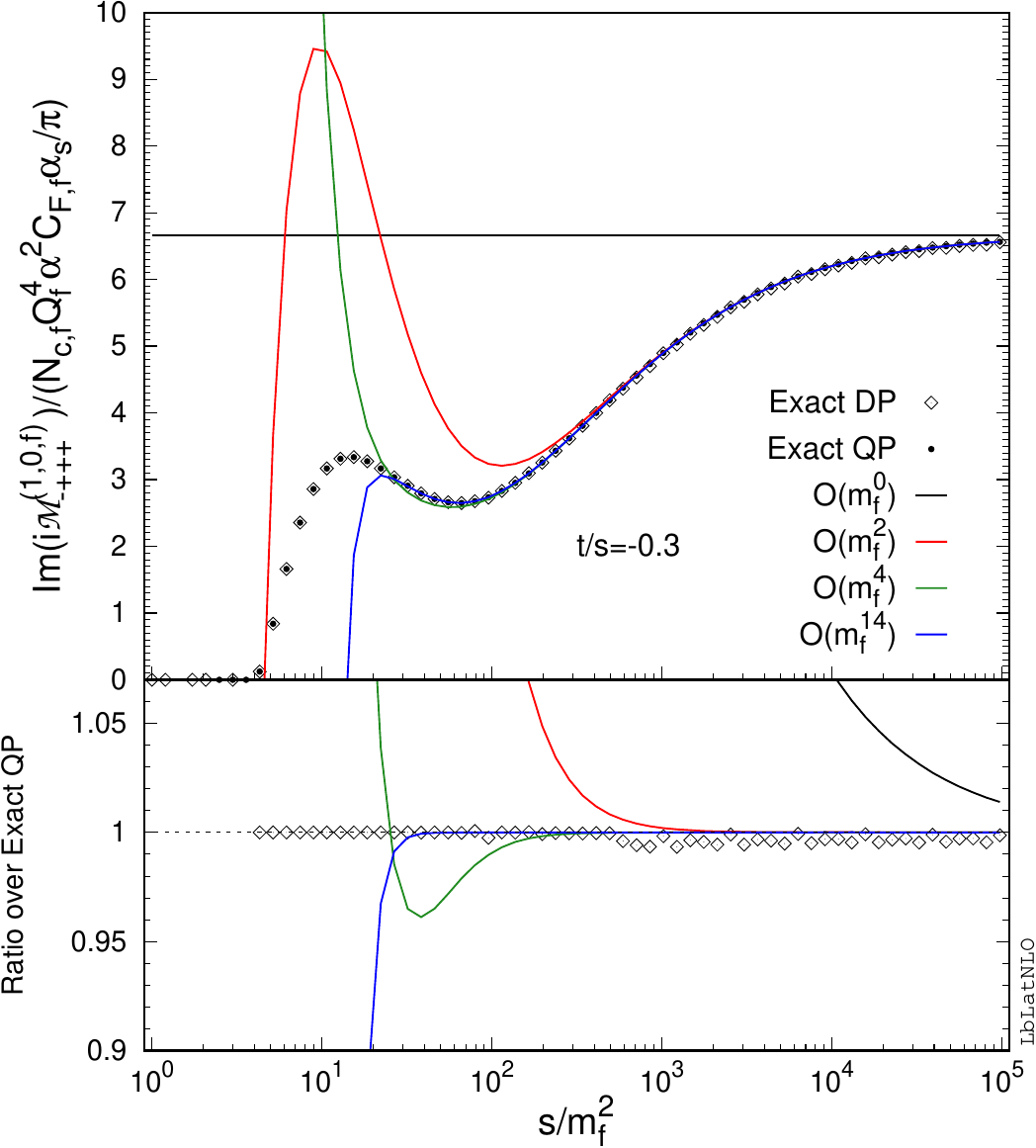}
\caption{All-plus (left) and single-minus (right) two-loop QCD amplitudes $i\mathcal{M}_{\lambda_1\lambda_2\lambda_3\lambda_4}^{(1,0,f)}$ as functions of $s/m_f^2$ in the range $1<s/m_f^2<10^5$. The real (upper panels) and imaginary (lower panels) parts are shown. Exact results obtained in double precision (open diamonds) and quadruple precision (filled circles) are compared with the HE expansion truncated at $\mathcal{O}(m_f^0)$ (black curves), $\mathcal{O}(m_f^2)$ (red curves), $\mathcal{O}(m_f^4)$ (green curves), and $\mathcal{O}(m_f^{14})$ (blue curves). The scattering angle is fixed at $t/s=-0.3$. The bottom panels display the ratios with respect to the exact quadruple-precision result.}
\label{figHE1} \vspace*{-0.5cm}
\end{figure}

\begin{figure}[hbt!]
\includegraphics[width=0.32\columnwidth,draft=false]{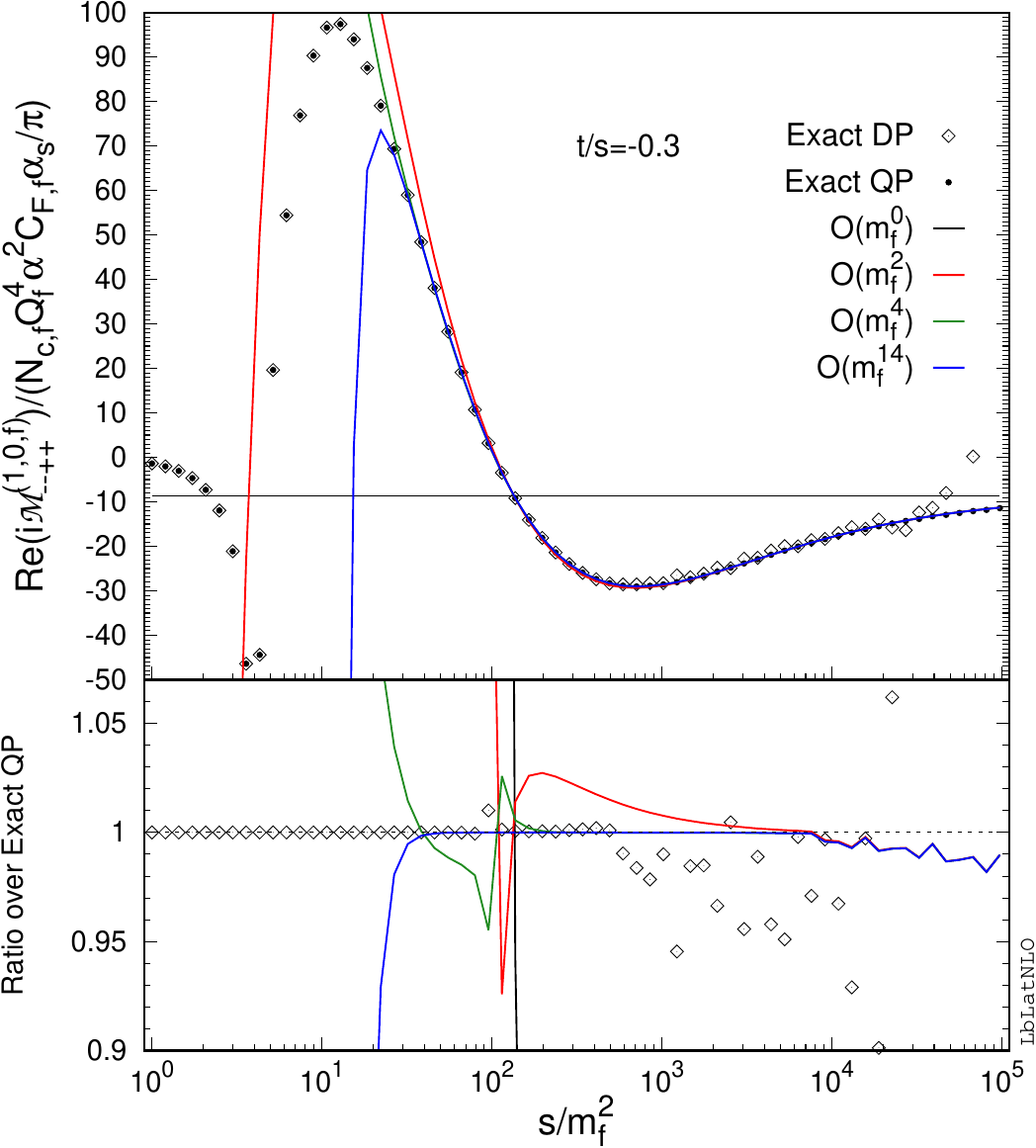}
\includegraphics[width=0.32\columnwidth,draft=false]{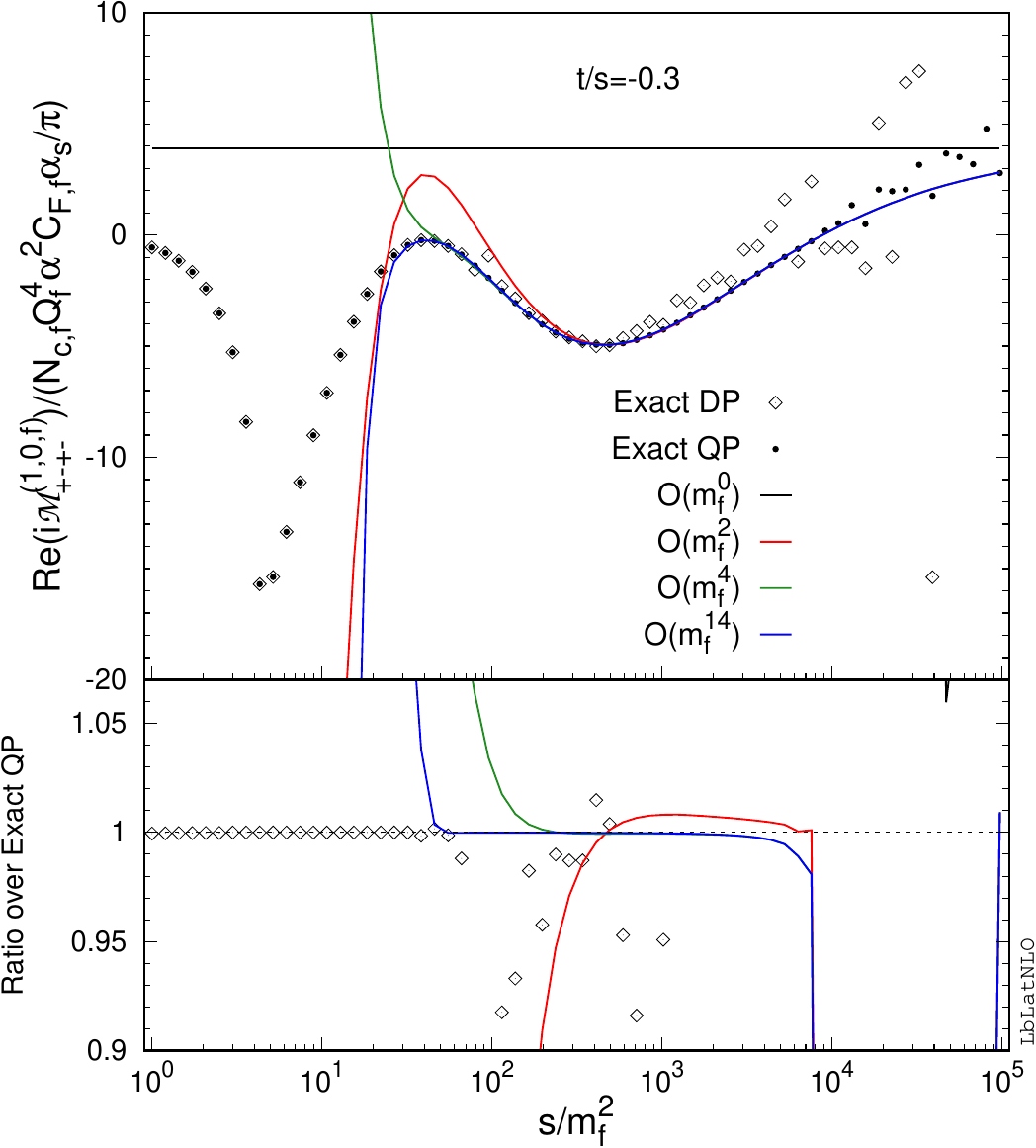}
\includegraphics[width=0.32\columnwidth,draft=false]{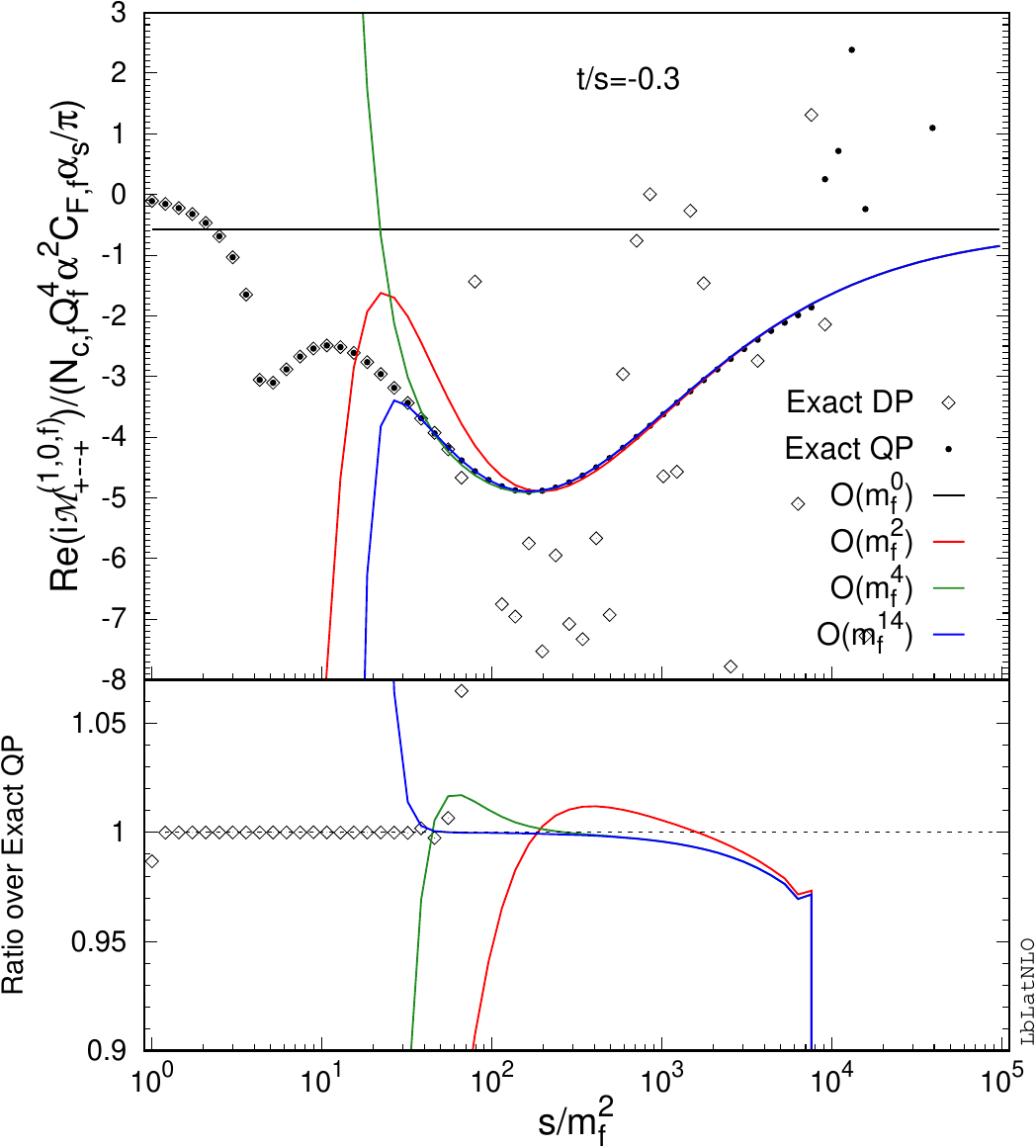}\\
\includegraphics[width=0.32\columnwidth,draft=false]{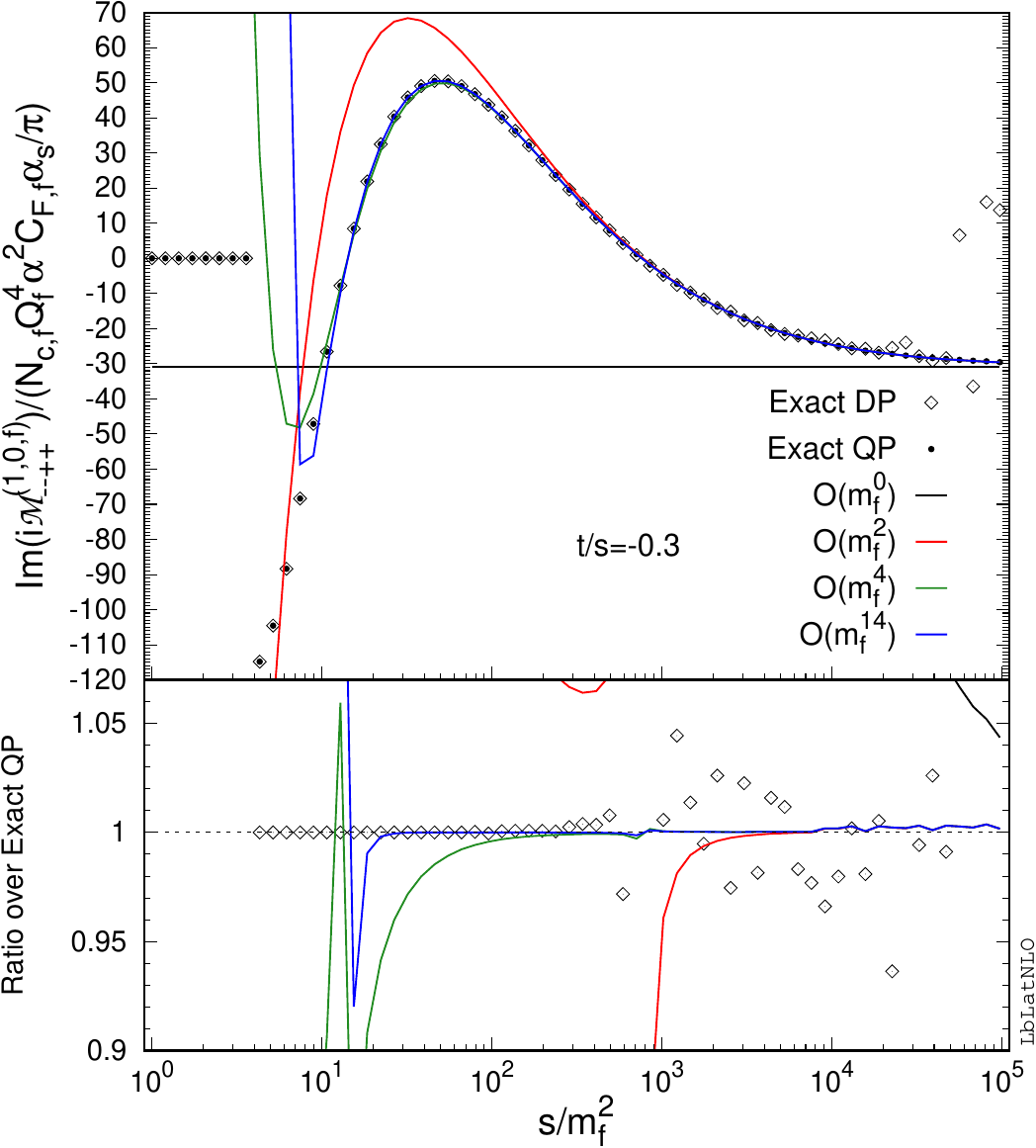}
\includegraphics[width=0.32\columnwidth,draft=false]{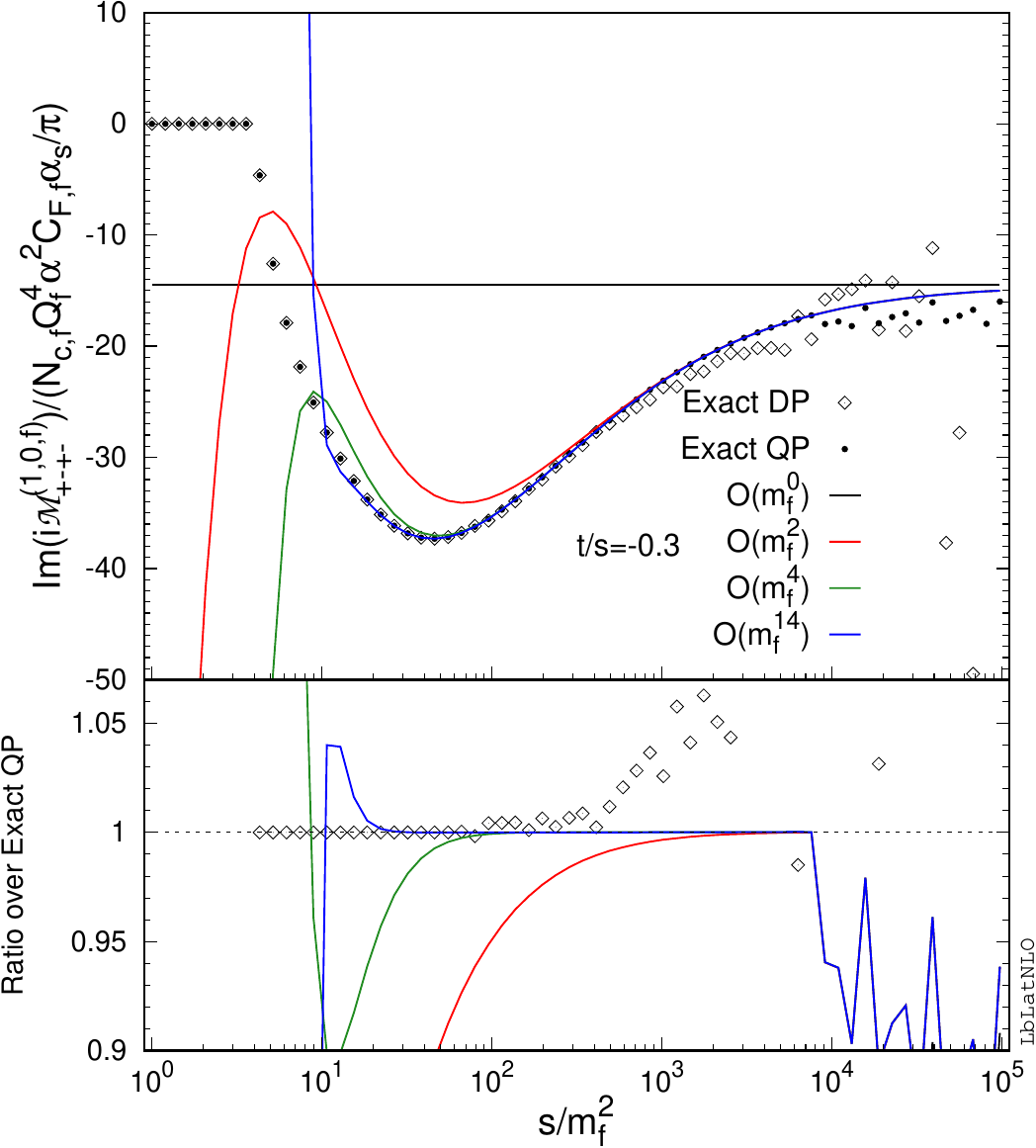}
\includegraphics[width=0.32\columnwidth,draft=false]{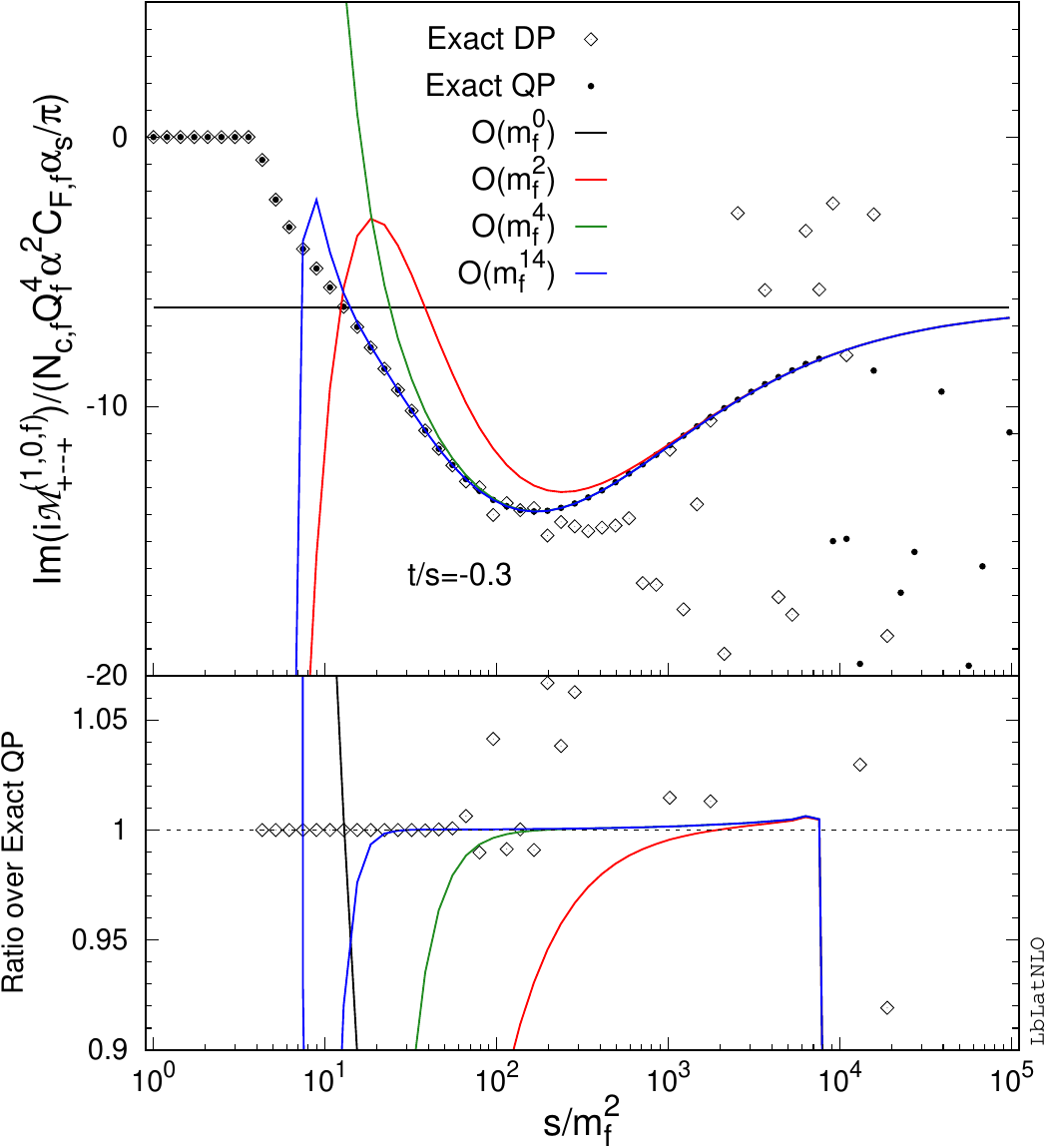}
\caption{Two-minus-two-plus two-loop QCD amplitudes $i\mathcal{M}_{--++}^{(1,0,f)}$ (left),  $i\mathcal{M}_{+-+-}^{(1,0,f)}$ (middle), and $i\mathcal{M}_{+--+}^{(1,0,f)}$ (right) as functions of $s/m_f^2$ in the range $1<s/m_f^2<10^5$. The real (upper panels) and imaginary (lower panels) parts are shown. Exact results obtained in double precision (open diamonds) and quadruple precision (filled circles) are compared with the HE expansion truncated at $\mathcal{O}(m_f^0)$ (black curves), $\mathcal{O}(m_f^2)$ (red curves), $\mathcal{O}(m_f^4)$ (green curves), and $\mathcal{O}(m_f^{14})$ (blue curves). The scattering angle is fixed at $t/s=-0.3$. The bottom panels display the ratios with respect to the exact quadruple-precision result.}
\label{figHE2} \vspace*{-0.5cm}
\end{figure}

Figures~\ref{figHE1} and \ref{figHE2} show the five two-loop QCD helicity amplitudes
$i\mathcal{M}_{\lambda_1\lambda_2\lambda_3\lambda_4}^{(1,0,f)}$, normalised by the common factor $N_{c,f}Q_f^4\alpha^2C_{F,f}\alpha_s/\pi$, in the range $1<s/m_f^2<10^5$ with $t/s=-0.3$ fixed. The real and imaginary parts are displayed in the upper and lower panels, respectively. Among the five helicity configurations, the all-plus amplitude is the most numerically stable. The exact full-$m_f$-dependent results obtained in double precision (exact DP, open diamonds) and quadruple precision (exact QP, filled circles) remain reliable over the entire range of $s/m_f^2$ considered. The single-minus amplitude is the second most stable configuration; however, the exact double-precision result begins to suffer from numerical instabilities for $s/m_f^2\gtrsim 500$. Due to the complexity of the analytic expressions~\cite{AH:2023ewe}, the two-minus-two-plus amplitudes are the least stable. As shown in figure~\ref{figHE2}, the exact double-precision results exhibit fluctuations already at relatively small $s/m_f^2$, while the quadruple-precision results become unstable for $s/m_f^2\gtrsim 8000$ in $i\mathcal{M}_{+-+-}^{(1,0,f)}$ and $i\mathcal{M}_{+--+}^{(1,0,f)}$. The inclusion of higher-order terms in the HE expansion clearly enlarges its range of applicability. In particular, the expansion truncated at $\mathcal{O}(m_f^{14})$ (blue curves) provides a reliable approximation to the full-$m_f$-dependent amplitudes for $s/m_f^2\gtrsim 50$. We emphasise that the massless two-loop amplitudes~\cite{Bern:2001dg} are insufficient over most of the displayed range of $s/m_f^2$, except in the all-plus case, which receives only double Sudakov logarithms at subleading power. This highlights the necessity of employing the HE expansion presented in this section to improve the numerical stability of the amplitude evaluation.

We conclude that, although the numerical stability of exact full mass-dependent amplitudes can be improved by using sufficiently high machine precision whenever this is feasible, this comes at the cost of significantly increased computational resources. Ultimately, a better-organised formulation is required. This clearly highlights the value of the asymptotic expansions presented in this paper.

\section{Coulomb resummation\label{sec:Coulombres}}

It is well known that the NLO cross section for LbL scattering is plagued by Coulomb singularities in the threshold limit $\sqrt{s}\to 2m_f$ (see, \eg, figure 2 of ref.~\cite{AH:2023kor}). This behaviour is analogous to that observed in other loop-induced processes, such as $gg\to\gamma\gamma$~\cite{Kawabata:2016aya,Chen:2019fla} and heavy Higgs boson decay into a photon pair~\cite{Melnikov:1994jb}. In this section, we investigate the Coulomb singularities of $\gamma\gamma\to\gamma\gamma$ in the threshold region (\ie, $\sqrt{s}\to 2m_f$) and improve both the helicity amplitudes and the LbL cross section by means of Coulomb resummation using Green's function approach~\cite{Fadin:1987wz}.

To characterise the kinematics near the fermion-pair production threshold, we define the threshold energy variable $E_f=\sqrt{s}-2m_f$, which measures the deviation of the centre-of-mass energy from the pair-production threshold. In the threshold region ($|E_f|\ll m_f$), the helicity amplitudes can be expanded as
\begin{eqnarray}
i\mathcal{M}_{\lambda_1\lambda_2\lambda_3\lambda_4}^{(0,0,f)}&=&N_{c,f}Q_f^4\alpha^2(0)\Bigg[A_{\lambda_1\lambda_2\lambda_3\lambda_4}^{(0,f)}-8\pi\lambda_1\lambda_3\delta_{\lambda_1\lambda_2}\delta_{\lambda_3\lambda_4}\sqrt{-\frac{E_f}{m_f}}+\mathcal{O}\left(\frac{E_f}{m_f}\right)\Bigg]\,,\\
i\mathcal{M}_{\lambda_1\lambda_2\lambda_3\lambda_4}^{(1,0,f)}&=&N_{c,f}Q_f^4\alpha^2(0)\frac{\alpha_s(\mu_R^2)}{\pi}C_{F,f}\Bigg\{A^{(1,f)}_{\lambda_1\lambda_2\lambda_3\lambda_4}+4\pi\lambda_1\lambda_3\delta_{\lambda_1\lambda_2}\delta_{\lambda_3\lambda_4}\nonumber\\
&&\times\bigg[\pi\left(1-\log{\left(-\frac{4m_fE_f}{\mu_{R}^2}\right)}\right)+2\left(5-\frac{\pi^2}{4}\right)\sqrt{-\frac{E_f}{m_f}}\bigg]+\mathcal{O}\left(\frac{E_f}{m_f}\right)\Bigg\}\nonumber\\
&=&i\mathcal{M}_{\lambda_1\lambda_2\lambda_3\lambda_4}^{(1,0,f),\mathrm{Coul~approx}}+\mathcal{O}\left(\frac{E_f}{m_f}\right)\,,\label{eq:CoulApproxAmp2LQCD}\\
i\mathcal{M}_{\lambda_1\lambda_2\lambda_3\lambda_4}^{(0,1,f)}&=&N_{c,f}Q_f^4\alpha^2(0)\frac{\alpha(\mu_R^2)}{\pi}Q_f^2\Bigg\{A^{(1,f)}_{\lambda_1\lambda_2\lambda_3\lambda_4}+4\pi\lambda_1\lambda_3\delta_{\lambda_1\lambda_2}\delta_{\lambda_3\lambda_4}\nonumber\\
&&\times\bigg[\pi\left(1-\log{\left(-\frac{4m_fE_f}{\mu_{R}^2}\right)}\right)+2\left(5-\frac{\pi^2}{4}\right)\sqrt{-\frac{E_f}{m_f}}\bigg]+\mathcal{O}\left(\frac{E_f}{m_f}\right)\Bigg\}\nonumber\\
&=&i\mathcal{M}_{\lambda_1\lambda_2\lambda_3\lambda_4}^{(0,1,f),\mathrm{Coul~approx}}+\mathcal{O}\left(\frac{E_f}{m_f}\right)\,,\label{eq:CoulApproxAmp2LQED}
\end{eqnarray}
where the analytic continuation $E_f\to E_f+i0^+$ is understood. As it is relevant for the present discussion, we keep the explicit scale dependence of the coupling constants $\alpha$ and $\alpha_s$. The coefficients $A_{\lambda_1\lambda_2\lambda_3\lambda_4}^{(k,f)}$ with $k=0,1$ depend only on the scattering angle (or, equivalently, on the ratio $t/s$) and on the renormalisation scale $\mu_R$, whose dependence cancels in the full two-loop amplitudes. These coefficients can either be derived from a threshold expansion of the helicity amplitudes or determined numerically in the limit $E_f\to 0$ from the full amplitudes. In this work, we adopt the latter approach for simplicity. The terms proportional to $\delta_{\lambda_1\lambda_2}\delta_{\lambda_3\lambda_4}$ are analogous to those appearing in the $gg\to \gamma\gamma$ process discussed in refs.~\cite{Kawabata:2016aya,Chen:2019fla}. These contributions are independent of the scattering angle. This can be understood from the fact that the $f\bar{f}$ pair, at leading order in $E_f/m_f$, can only form a pseudoscalar state with $J^{\mathrm{PC}}=0^{-+}$. We therefore define the corresponding Coulomb contributions as
\begin{eqnarray}
i\mathcal{M}_{\lambda_1\lambda_2\lambda_3\lambda_4}^{(0,0,f),\mathrm{Coul}}&\equiv&N_{c,f}Q_f^4\alpha^2(0)(-8\pi)\lambda_1\lambda_3\delta_{\lambda_1\lambda_2}\delta_{\lambda_3\lambda_4}\sqrt{-\frac{E_f}{m_f}}\,,\label{eq:CoulAmpLO}\\
i\mathcal{M}_{\lambda_1\lambda_2\lambda_3\lambda_4}^{(1,0,f),\mathrm{Coul}}&\equiv&N_{c,f}Q_f^4\alpha^2(0)\frac{\alpha_s(\mu_R^2)}{\pi}C_{F,f}4\pi\lambda_1\lambda_3\delta_{\lambda_1\lambda_2}\delta_{\lambda_3\lambda_4}\nonumber\nonumber\\
&&\times\bigg[\pi\left(1-\log{\left(-\frac{4m_fE_f}{\mu_{R}^2}\right)}\right)+2\left(5-\frac{\pi^2}{4}\right)\sqrt{-\frac{E_f}{m_f}}\bigg]\,,\label{eq:CoulAmpNLOQCD}\\
i\mathcal{M}_{\lambda_1\lambda_2\lambda_3\lambda_4}^{(0,1,f),\mathrm{Coul}}&\equiv&N_{c,f}Q_f^4\alpha^2(0)\frac{\alpha(\mu_R^2)}{\pi}Q_{f}^24\pi\lambda_1\lambda_3\delta_{\lambda_1\lambda_2}\delta_{\lambda_3\lambda_4}\nonumber\nonumber\\
&&\times\bigg[\pi\left(1-\log{\left(-\frac{4m_fE_f}{\mu_{R}^2}\right)}\right)+2\left(5-\frac{\pi^2}{4}\right)\sqrt{-\frac{E_f}{m_f}}\bigg]\,.\label{eq:CoulAmpNLOQED}
\end{eqnarray}
Unlike heavy-quark pair production processes, the Coulomb singularities in loop-induced processes through virtual heavy particles do not produce $1/\sqrt{E_f}$ singularities, but are instead logarithmic, $\log{\left(E_f\right)}$. As a result, Coulomb singularities in loop-induced processes are less pronounced than in heavy-quark pair production.

In order to perform the leading power (LP) Coulomb resummation, we define the following resummed amplitudes:~\footnote{The expansion is organised in powers of the relative velocity $v$ of the fermion-antifermion pair, with $v\ll1$ and $E_f\sim m_fv^2$. In addition, we assume $\alpha_s(m_f^2v^2)\sim v$ ($\alpha(m_f^2v^2)\sim v$) for LP QCD (QED) Coulomb resummation.}
\begin{align}
i\mathcal{M}_{\lambda_1\lambda_2\lambda_3\lambda_4}^{(\mathrm{LP}_{\mathrm{QCD}},f)}=&N_{c,f}Q_f^4\alpha^2(0)\lambda_1\lambda_3\delta_{\lambda_1\lambda_2}\delta_{\lambda_3\lambda_4}\frac{32\pi^2}{m_f^2}\Bigg[1-\frac{\alpha_s(\mu_R^2)}{\pi}C_{F,f}\left(5-\frac{\pi^2}{4}\right)\Bigg]\nonumber\\
&\times G^{(\mathrm{LP}_{\mathrm{QCD}},f)}\left(\vec{0},\vec{0};E_f\right)\,,\\
i\mathcal{M}_{\lambda_1\lambda_2\lambda_3\lambda_4}^{(\mathrm{LP}_{\mathrm{QED}},f)}=&N_{c,f}Q_f^4\alpha^2(0)\lambda_1\lambda_3\delta_{\lambda_1\lambda_2}\delta_{\lambda_3\lambda_4}\frac{32\pi^2}{m_f^2}\Bigg[1-\frac{\alpha(\mu_R^2)}{\pi}Q_{f}^2\left(5-\frac{\pi^2}{4}\right)\Bigg]\nonumber\\
&\times G^{(\mathrm{LP}_{\mathrm{QED}},f)}\left(\vec{0},\vec{0};E_f\right)\,.
\end{align}
The LP QCD and QED Green’s functions at the origin~\cite{Fadin:1987wz,Fadin:1990wx,Beneke:2010da,Beneke:2011mq} are given by
\begin{eqnarray}
G^{(\mathrm{LP}_{\mathrm{QCD}},f)}\left(\vec{0},\vec{0};E_f\right)&=&-\frac{m_f^2}{4\pi}\Bigg[\sqrt{-\frac{E_f}{m_f}}+C_{F,f}\alpha_s(\mu_{C,\mathrm{QCD},f}^{2})\bigg(\frac{1}{2}\log{\left(-\frac{4m_fE_f}{\mu_{C,\mathrm{QCD},f}^2}\right)}-\frac{1}{2}\nonumber\\
&&+\gamma_E+\tilde{\psi}\left(1-\frac{C_{F,f}\alpha_s(\mu_{C,\mathrm{QCD},f}^{2})}{2\sqrt{-E_f/m_f}}\right)\bigg)\Bigg]\,,\label{eq:QCDGLP00}\\
G^{(\mathrm{LP}_{\mathrm{QED}},f)}\left(\vec{0},\vec{0};E_f\right)&=&-\frac{m_f^2}{4\pi}\Bigg[\sqrt{-\frac{E_f}{m_f}}+Q_{f}^2\alpha(\mu_{C,\mathrm{QED},f}^2)\bigg(\frac{1}{2}\log{\left(-\frac{4m_fE_f}{\mu_{C,\mathrm{QED},f}^2}\right)}-\frac{1}{2}\nonumber\\
&&+\gamma_E+\tilde{\psi}\left(1-\frac{Q_{f}^2\alpha(\mu_{C,\mathrm{QED},f}^2)}{2\sqrt{-E_f/m_f}}\right)\bigg)\Bigg]\,,\label{eq:QEDGLP00}
\end{eqnarray}
where $\mu_{C,\mathrm{QCD},f}$ and $\mu_{C,\mathrm{QED},f}$ denote the QCD and QED Coulomb scales, respectively. The function $\tilde{\psi}(1+x)$ is defined as
\begin{equation}
\tilde{\psi}(1+x)=\left\{\begin{array}{rl}\psi(-x), & x\in\mathbb{R}, x<-\frac{1}{2}\\
\psi(1+x), &\mathrm{otherwise}\end{array}\right.\,,\label{eq:tildepsidef}
\end{equation}
where $\psi(x)=\Gamma^\prime(x)/\Gamma(x)$ denotes the digamma function. The seemingly unusual appearance of $\tilde{\psi}(1+x)$ in the Green's functions at the origin in eqs.~\eqref{eq:QCDGLP00} and \eqref{eq:QEDGLP00}, instead of the standard digamma function $\psi(1+x)$, requires some explanation. The definition of $\tilde{\psi}(1+x)$ is motivated by the asymptotic behaviour of the digamma function for $x\to \infty$ in the complex plane ($x\in\mathbb{C}$), since in our computation we neglect the widths of the internal fermions.~\footnote{The formalisms introduced in this section can readily accommodate non-zero widths. However, for simplicity, and in order to remain directly applicable to all fermions, including stable ones such as the electron, we do not consider width effects here.} The digamma function $\psi(1+x)$ has simple poles when $x$ is a negative integer and admits the asymptotic expansion
\begin{equation}
\psi(1+x)=\log{(x)}+\frac{1}{2x}-\sum_{n=1}^{+\infty}{\frac{B_{2n}}{2n x^{2n}}}\,,\quad \mathrm{when} \quad |\arg{(x)}|<\pi-0^+\,,
\end{equation}
where $B_{2n}$ are Bernoulli numbers. If $x$ approaches the negative real axis, one may instead use the reflection identity
\begin{equation}
\psi(1+x)=\psi(-x)-\pi\cot{\left(\pi x\right)}\,.
\end{equation}
However, the presence of the $\cot{\left(\pi x\right)}$ term leads to highly oscillatory behaviour when $x$ is close to the negative real axis. In the context of the Green's functions, this behaviour produces an unexpected -- and, we believe, unphysical -- behaviour of the resummed cross section for $E_f<0$. For this reason, we drop the $\cot{\left(\pi x\right)}$ contribution and instead use
\begin{equation}
\tilde{\psi}(1+x)=\psi(-x)\,,\quad x<-\frac{1}{2}\,,x\in\mathbb{R}\,,
\end{equation}
as defined in eq.~\eqref{eq:tildepsidef}. With this definition, $\tilde{\psi}(1+x)$ admits the asymptotic expansion
\begin{equation}
\tilde{\psi}(1+x)=\log{(|x|)}+\frac{1}{2x}-\sum_{n=1}^{+\infty}{\frac{B_{2n}}{2n x^{2n}}}\,,\quad |x|>1,x\in\mathbb{R}\,.
\end{equation}
The pathological oscillatory behaviour present in the original digamma function $\psi(1+x)$ would be regulated if a non-zero width were introduced, which shifts $x$ away from the real axis. Since we neglect width effects here, we instead employ the modified function $\tilde{\psi}(1+x)$.

To avoid hitting the Landau pole of $\alpha_s$, we define the QCD Coulomb scale as
\begin{equation}
\mu_{C,\mathrm{QCD},f}=\left\{\begin{array}{rl}\mathrm{max}\left(\xi_R\mathrm{max}\left(\sqrt{4m_f|E_f|},\mu_{B,\mathrm{QCD},f}\right),1~\mathrm{GeV}\right), & 4\sqrt{6}-10<\frac{E_f}{m_f}<6-4\sqrt{2}\\
\mathrm{max}\left(\mu_R,1~\mathrm{GeV}\right), & \mathrm{otherwise}\end{array}\right.\,,\label{eq:muCQCDchoice}
\end{equation}
while the QED Coulomb scale is defined as
\begin{equation}
\mu_{C,\mathrm{QED},f}=\left\{\begin{array}{rl}\xi_R\mathrm{max}\left(\sqrt{4m_f|E_f|},\mu_{B,\mathrm{QED},f}\right), & 4\sqrt{6}-10<\frac{E_f}{m_f}<6-4\sqrt{2}\\
\mu_R, & \mathrm{otherwise}\end{array}\right.\,.\label{eq:muCQEDchoice}
\end{equation}
The constants $4\sqrt{6}-10$ and $6-4\sqrt{2}$ in eqs.~\eqref{eq:muCQCDchoice} and \eqref{eq:muCQEDchoice} follow from the condition $\sqrt{4m_f|E_f|}=\sqrt{s}/2$.
The scales associated with the inverse Bohr radii are obtained as solutions of the following equations~\footnote{We include an additional factor $e^{\gamma_E}\approx 1.78$, motivated by the Green's functions in the limit $E_f\to 0$. This definition differs slightly from the corresponding quantity in ref.~\cite{Capatti:2025khs}.}
\begin{eqnarray}
\mu_{B,\mathrm{QCD},f}&=&e^{\gamma_E}C_{F,f}\alpha_s(\mu_{B,\mathrm{QCD},f}^2)m_f\,,\\
\mu_{B,\mathrm{QED},f}&=&e^{\gamma_E}Q_{f}^2\alpha(\mu_{B,\mathrm{QED},f}^2)m_f\,.
\end{eqnarray}
These equations can be solved numerically using the Newton-Raphson method. The renormalisation scale adopted in this work is defined as
\begin{equation}
\mu_R=\xi_R\frac{\sqrt{s}}{2}\,.\label{eq:muRdef}
\end{equation}
In other words, the Coulomb scale $\mu_{C,\mathrm{X},f}$ (with $\mathrm{X}=\mathrm{QCD}$ or $\mathrm{QED}$) interpolates from $\mu_R$ in the relativistic regime ($E_f/m_f>6-4\sqrt{2}$ or $E_f/m_f<4\sqrt{6}-10$) to $\xi_R \mu_{B,\mathrm{X},f}$ in the non-relativistic regime ($|E_f|<\mu_{B,\mathrm{X},f}^2/(4m_f)$). In the intermediate transition region, the Coulomb scale behaves as $\mu_{C,\mathrm{X},f}=\xi_R \sqrt{4m_f|E_f|}$.
All results presented in this work that include Coulomb resummation employ the scale choice defined in eqs.~\eqref{eq:muCQCDchoice} and \eqref{eq:muCQEDchoice}. Moreover, we fully correlate the renormalisation and Coulomb scales through the common variation factor $\xi_R$. Although, in principle, $\mu_{C,\mathrm{X},f}$ and $\mu_R$ could be varied independently, we consider the correlated variation to be a more appropriate and simpler choice. In particular, it guarantees the correct asymptotic limit $\mu_{C,\mathrm{X},f} \to \mu_R$ in the relativistic regime, which is essential for preserving the achieved fixed-order accuracy. If one wishes to decorrelate the $\mu_{C,\mathrm{X},f}$ and $\mu_R$ variations while preserving the asymptotic limit $\mu_{C,\mathrm{X},f}\to\mu_R$, alternative and more complicated prescriptions for $\mu_{C,\mathrm{X},f}$ than those in eqs.~\eqref{eq:muCQCDchoice} and \eqref{eq:muCQEDchoice} would need to be constructed.

The LP Coulomb-resummation-improved two-loop amplitudes are defined as
\begin{align}
i\mathcal{M}_{\lambda_1\lambda_2\lambda_3\lambda_4}^{(1,0,f),\mathrm{LP}}&\equiv&i\mathcal{M}_{\lambda_1\lambda_2\lambda_3\lambda_4}^{(1,0,f)}+c_{\mathrm{damp}}\left(\frac{m_f}{E_f}\right)\left[i\mathcal{M}_{\lambda_1\lambda_2\lambda_3\lambda_4}^{(\mathrm{LP}_{\mathrm{QCD}},f)}-i\mathcal{M}_{\lambda_1\lambda_2\lambda_3\lambda_4}^{(0,0,f),\mathrm{Coul}}-i\mathcal{M}_{\lambda_1\lambda_2\lambda_3\lambda_4}^{(1,0,f),\mathrm{Coul}}\right]\,,\label{eq:QCDAmp2LLP}\\
i\mathcal{M}_{\lambda_1\lambda_2\lambda_3\lambda_4}^{(0,1,f),\mathrm{LP}}&\equiv&i\mathcal{M}_{\lambda_1\lambda_2\lambda_3\lambda_4}^{(0,1,f)}+c_{\mathrm{damp}}\left(\frac{m_f}{E_f}\right)\left[i\mathcal{M}_{\lambda_1\lambda_2\lambda_3\lambda_4}^{(\mathrm{LP}_{\mathrm{QED}},f)}-i\mathcal{M}_{\lambda_1\lambda_2\lambda_3\lambda_4}^{(0,0,f),\mathrm{Coul}}-i\mathcal{M}_{\lambda_1\lambda_2\lambda_3\lambda_4}^{(0,1,f),\mathrm{Coul}}\right]\,,\label{eq:QEDAmp2LLP}
\end{align}
where $c_{\mathrm{damp}}\left(\frac{m_f}{E_f}\right)$ is a damping function that ensures that Coulomb resummation is applied only in the non-relativistic region $E_{f,\mathrm{min}}/m_f<E_f/m_f<E_{f,\mathrm{max}}/m_f$, with $E_{f,\mathrm{min}}=(4\sqrt{6}-10)m_f$ and $E_{f,\mathrm{max}}=(6-4\sqrt{2})m_f$. The damping function is to some extent arbitrary, but it satisfies the following asymptotic limits:
\begin{equation}
\lim_{x\to 0}{c_{\mathrm{damp}}\left(x\right)}=0,\quad \lim_{x\to \pm\infty}{c_{\mathrm{damp}}\left(x\right)}=1\,.
\end{equation}
In this work, we adopt the following ansatz:
\begin{equation}
c_{\mathrm{damp}}(x)=\frac{1-e^{-20|x|}}{1+e^{-10(|x|-x_0)}}\,,
\end{equation}
where
\begin{equation}
x_0=\frac{\Theta(x)}{6-4\sqrt{2}}+\frac{\Theta(-x)}{10-4\sqrt{6}}\,.
\end{equation}
Damping, albeit with some arbitrariness, is common practice in fixed-order and resummation-matched predictions, where resummed contributions can become anomalously large in kinematic regions in which resummation is not necessary. In contrast to the case of $\gamma\gamma\to Q\bar{Q}$ discussed in ref.~\cite{Capatti:2025khs}, the introduction of such a damping function is essential for LbL scattering, where a delicate cancellation occurs between Coulomb and non-Coulomb contributions in the relativistic regions above and below threshold. Without this damping, the extrapolation of Coulomb contributions to higher orders in $\alpha_s$ or $\alpha$ would significantly overshoot the physical amplitudes. For instance, $i\mathcal{M}_{\lambda_1\lambda_2\lambda_3\lambda_4}^{(1,0,f),\mathrm{Coul}}$ in eq.~\eqref{eq:CoulAmpNLOQCD} gives rise to an $\mathcal{O}(m_f^0)$ behaviour in the LE region (where $\sqrt{s}\ll m_f$ and $E_f\to -2m_f$), whereas the physical two-loop LE amplitudes given in eq.~\eqref{eq:LEAmp2L} scale as $\mathcal{O}(m_f^{-4})$.
Similarly, in the HE region (where $\sqrt{s}\gg m_f$ and $E_f\to \sqrt{s}$), $i\mathcal{M}_{\lambda_1\lambda_2\lambda_3\lambda_4}^{(1,0,f),\mathrm{Coul}}$ scales differently from the full amplitude $i\mathcal{M}_{\lambda_1\lambda_2\lambda_3\lambda_4}^{(1,0,f)}$, namely as $\mathcal{O}(m_f^{-1/2})$ rather than $\mathcal{O}(m_f^0)$.

As a marginal comment, we note that, unlike the $\gamma\gamma\to Q\bar{Q}$ case studied in ref.~\cite{Capatti:2025khs}, where only the imaginary parts of the Green's functions enter, the full Green's functions contribute to Coulomb resummation in LbL scattering (as well as in other loop-induced processes such as $gg\to\gamma\gamma$). The explicit scale dependence appearing in the real parts of the Green's functions in eqs.~\eqref{eq:QCDGLP00} and \eqref{eq:QEDGLP00} introduces uncancelled logarithms of the form $\alpha_s\log{\left(\mu_{C,\mathrm{QCD},f}^2/\mu_R^2\right)}$ in $i\mathcal{M}_{\lambda_1\lambda_2\lambda_3\lambda_4}^{(1,0,f),\mathrm{LP}}$ and $\alpha\log{\left(\mu_{C,\mathrm{QED},f}^2/\mu_R^2\right)}$ in $i\mathcal{M}_{\lambda_1\lambda_2\lambda_3\lambda_4}^{(0,1,f),\mathrm{LP}}$ if one attempts to expand $i\mathcal{M}_{\lambda_1\lambda_2\lambda_3\lambda_4}^{(\mathrm{LP}_{\mathrm{QCD}},f)}$ and $i\mathcal{M}_{\lambda_1\lambda_2\lambda_3\lambda_4}^{(\mathrm{LP}_{\mathrm{QED}},f)}$ in perturbative series in $\alpha_s$ and $\alpha$, respectively, in eqs.~\eqref{eq:QCDAmp2LLP} and \eqref{eq:QEDAmp2LLP}. These logarithms cannot be compensated by the renormalisation-group evolution of the couplings. Although we believe that such uncancelled logarithms are phenomenologically irrelevant, particularly in view of our construction of the resummation formula and the choice of the Coulomb scales, we have so far been unable to identify their physical interpretation from a formal perspective. A complete understanding of this issue and its possible resolution would require a dedicated analysis within non-relativistic effective field theories, which is beyond the scope of the present paper.

\begin{figure}[hbt!]
\includegraphics[width=0.32\columnwidth,draft=false]{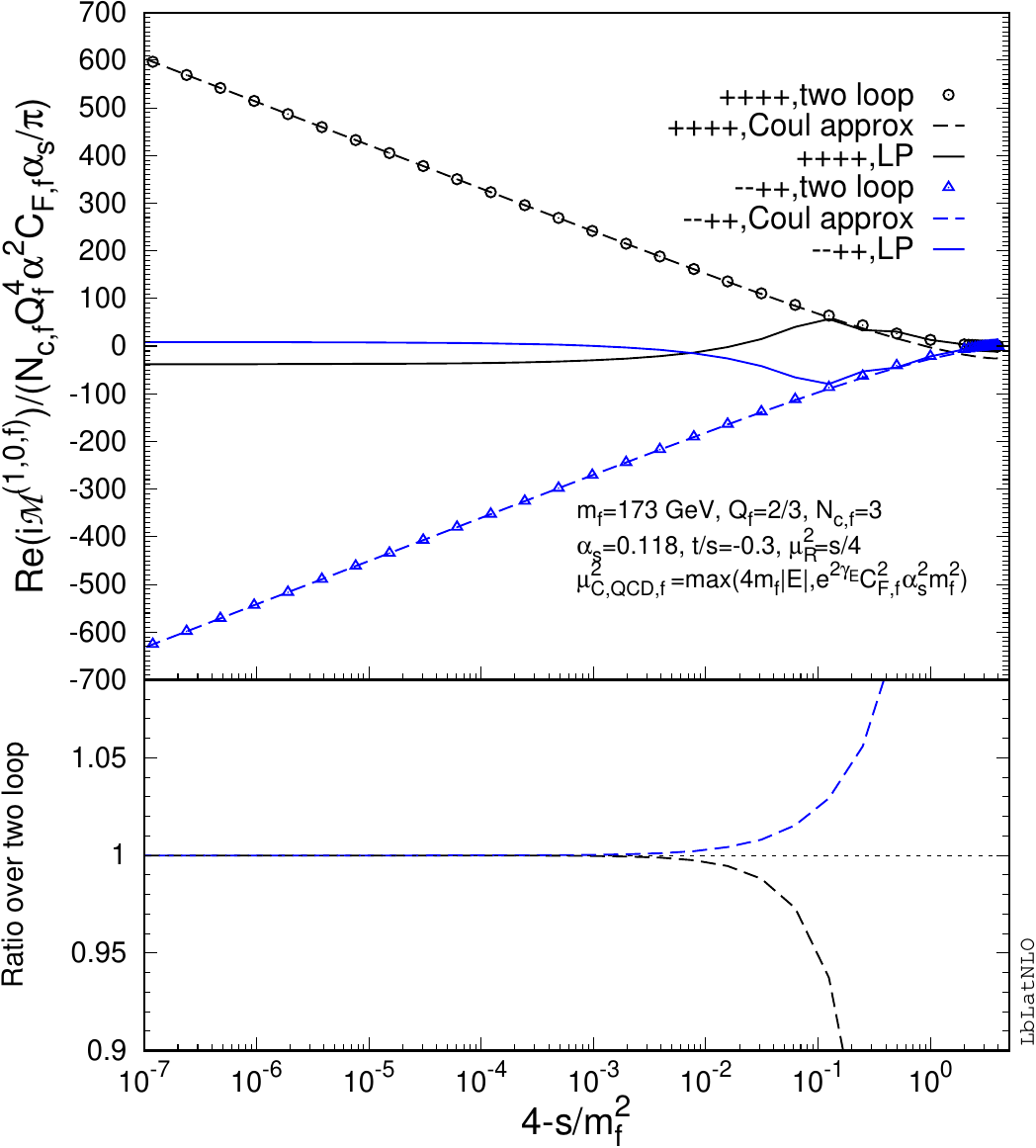}
\includegraphics[width=0.32\columnwidth,draft=false]{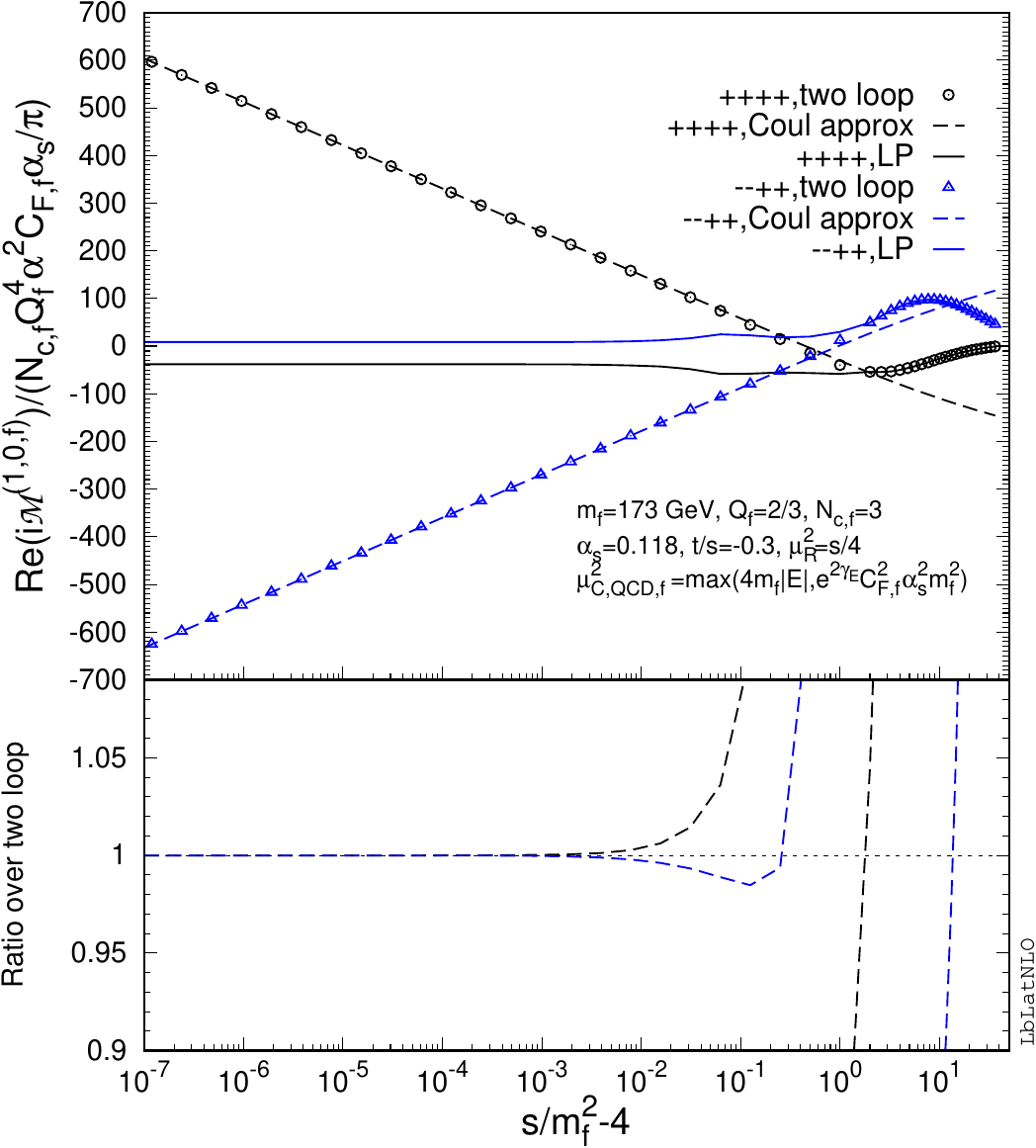}
\includegraphics[width=0.32\columnwidth,draft=false]{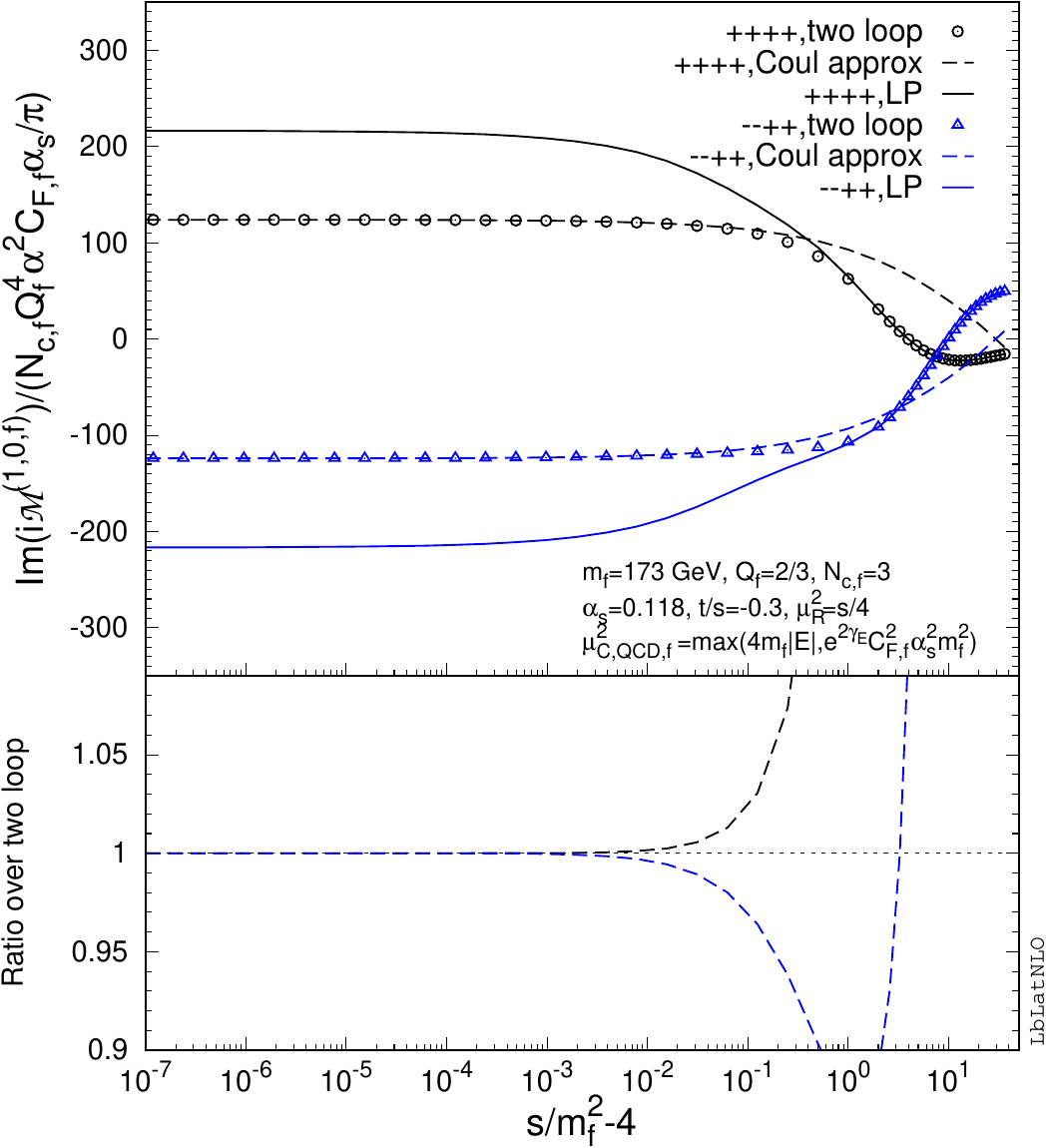}
\caption{All-plus (black) and two-minus–two-plus (blue) two-loop QCD amplitudes $i\mathcal{M}_{\lambda_1\lambda_2\lambda_3\lambda_4}^{(1,0,f)}$ (points), their Coulomb-approximated counterparts $i\mathcal{M}_{\lambda_1\lambda_2\lambda_3\lambda_4}^{(1,0,f),\mathrm{Coul~approx}}$ (long dashed), and the LP Coulomb-resummation-improved amplitudes $i\mathcal{M}_{\lambda_1\lambda_2\lambda_3\lambda_4}^{(1,0,f),\mathrm{LP}}$ (solid), shown as functions of $s/m_f^2$. From left to right, the panels display the real part of the amplitudes for $s<4m_f^2$ (left) and $s>4m_f^2$ (middle), and the imaginary part for $s>4m_f^2$ (right). The scattering angle is fixed at $t/s=-0.3$. The lower panels show the ratio of the Coulomb approximation to the full two-loop amplitude.}
\label{figCoulomb} \vspace*{-0.5cm}
\end{figure}

To demonstrate that the Coulomb approximation $i\mathcal{M}_{\lambda_1\lambda_2\lambda_3\lambda_4}^{(1,0,f),\mathrm{Coul~approx}}$, given in eq.~\eqref{eq:CoulApproxAmp2LQCD}, correctly captures the singular behaviour of the two-loop amplitude $i\mathcal{M}_{\lambda_1\lambda_2\lambda_3\lambda_4}^{(1,0,f)}$, we compare the two as a function of $s/m_f^2$ at fixed scattering angle $t/s=-0.3$ in figure~\ref{figCoulomb}. The full two-loop amplitudes are displayed as open circles (all-plus) and open triangles (two-minus-two-plus), while their Coulomb-approximation counterparts are shown as long-dashed lines. The approximation reproduces the full result asymptotically in the limit $E_f/m_f \to 0$. The relative difference falls below $1\%$ for $|E_f/m_f|\lesssim 0.01$. In the same figure, we also show the LP Coulomb-resummed results (solid lines). The resummation resolves the logarithmic singular behaviour of the two-loop amplitudes in the threshold region. The resummed results approach constant values in the limit $s \to 4m_f^2$.

\begin{figure}[hbt!]
\includegraphics[width=0.47\columnwidth,draft=false]{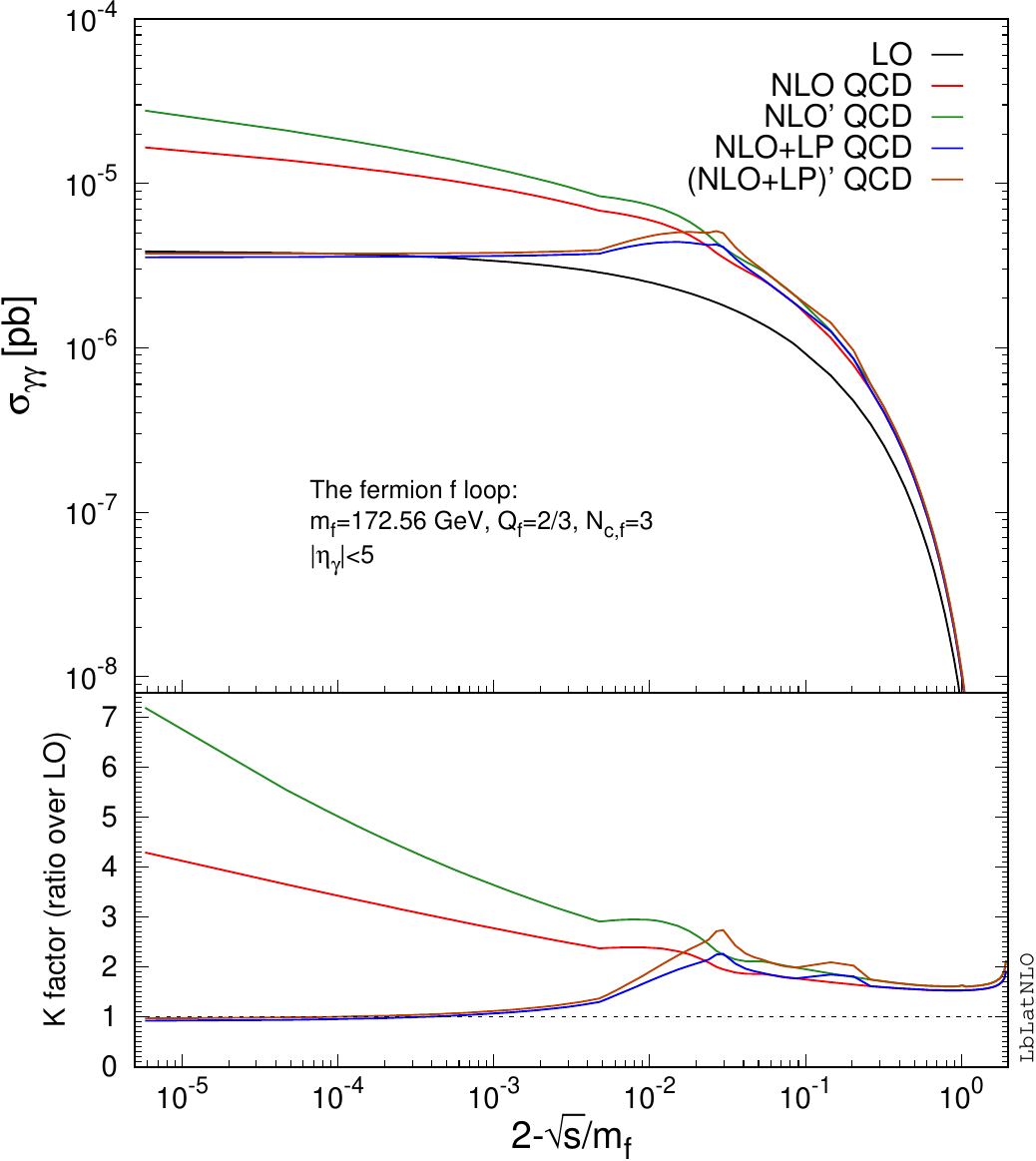}
\includegraphics[width=0.47\columnwidth,draft=false]{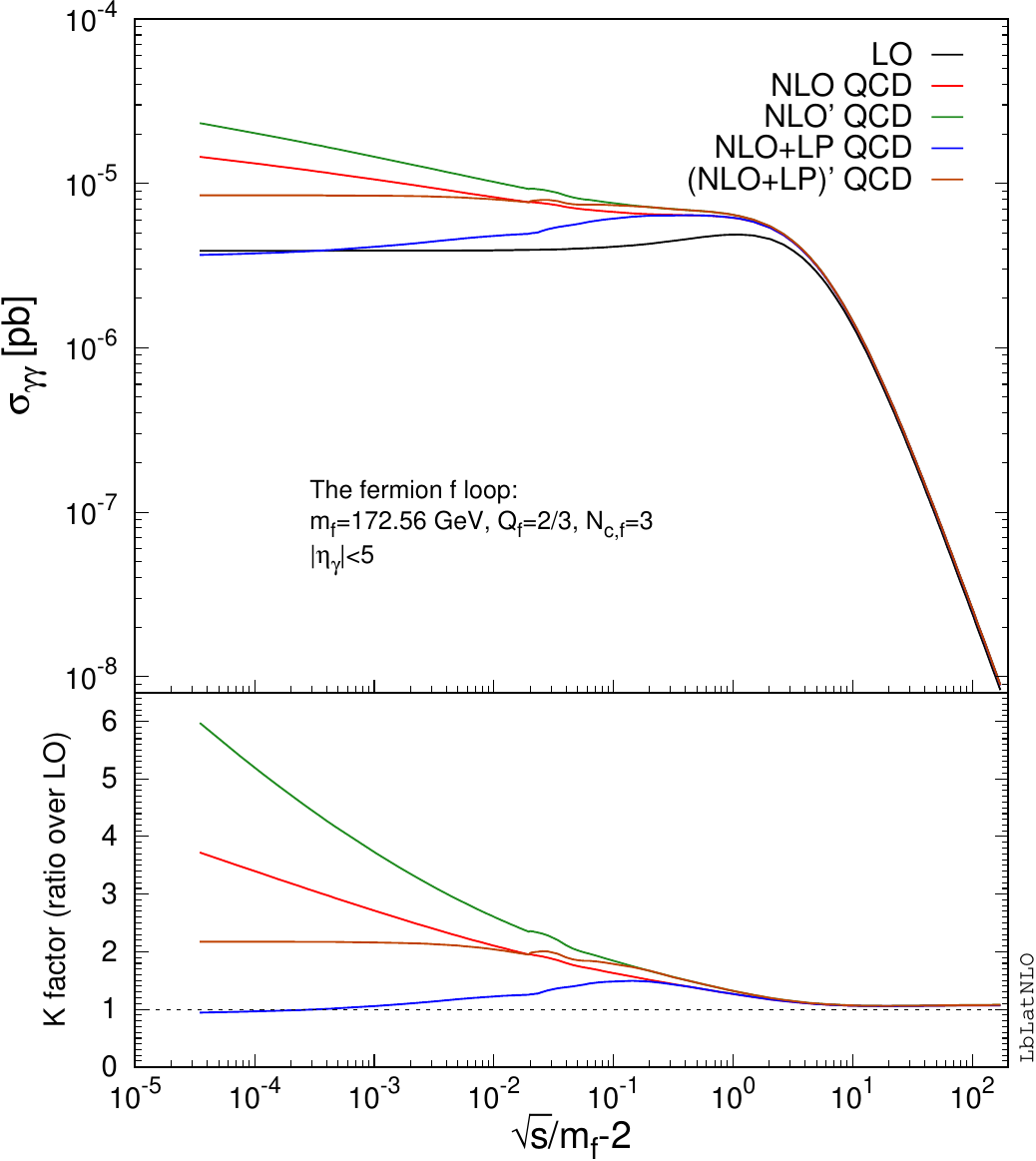}
\caption{The partonic LbL cross sections induced by a top-quark loop at LO (black), NLO QCD (red), NLO$^\prime$ QCD (green), NLO+LP QCD (blue), and (NLO+LP)$^\prime$ QCD (brown). Both the below-threshold region ($\sqrt{s}<2m_f$, left) and the above-threshold region ($\sqrt{s}>2m_f$, right) are shown. The lower panels display the corresponding $K$ factors.}
\label{figCoulombxs} \vspace*{-0.5cm}
\end{figure}

To better understand the impact of Coulomb resummation in the threshold region on the cross section, we consider the top-quark loop (\ie, $f=t$) with QCD corrections only as an illustrative example in figure~\ref{figCoulombxs}. This figure corresponds to a zoomed-in version of figure~2 in ref.~\cite{AH:2023kor}, focusing on the threshold region and using a slightly different setup: the top-quark on-shell mass $m_t=172.56\,\mathrm{GeV}$, the QED fine-structure constant in the Thomson limit $\alpha(0)=1/137.036$, the strong coupling $\alpha_s(\mu_R^2)$ evolved with five-loop QCD renormalisation-group running~\cite{Baikov:2016tgj,Herzog:2017ohr} and $\alpha_s(m_Z^2)=0.118$,~\footnote{We use $m_Z=91.1876$ GeV for the mass of the $Z$ boson.} and a pseudorapidity cut on the final-state photons $|\eta_\gamma|<5$. The calculational setup has been adjusted according to section~\ref{sec:setup}. We compare the LbL cross sections at LO (black), NLO QCD (red), NLO$^\prime$ QCD (green), NLO+LP QCD (blue), and (NLO+LP)$^\prime$ QCD (brown), whose precise definitions will be given in section~\ref{sec:setup}, as $\sqrt{s}$ approaches $2m_f=2m_t$ from both the below- and above-threshold regions. The logarithmically divergent behaviour in the threshold region observed in the NLO and NLO$^\prime$ QCD results is cured by the LP Coulomb resummation. In the NLO$^\prime$ QCD result, a double-logarithmic enhancement appears in the threshold region, stemming from the squared two-loop amplitude. For the same reason, in contrast to NLO+LP QCD, the (NLO+LP)$^\prime$ QCD prediction yields an almost factor-of-two enhancement in the $K$ factor above threshold, $\sqrt{s}\to 2m_f+0^+$. This behaviour arises because the Coulomb-resummed two-loop amplitudes develop an imaginary part proportional to $i\pi$ that is approximately twice as large as that of the non-resummed amplitudes above threshold, as seen in figure~\ref{figCoulomb} (right panel). Consequently, the (NLO+LP)$^\prime$ cross section becomes discontinuous in the limit $\sqrt{s}\to 2m_f\pm 0^+$. In principle, this discontinuity could be removed by introducing an imaginary Coulomb scale, which entails the analytic continuation of $\alpha_s$. However, this procedure cannot be consistently applied to the QED corrections when $\alpha$ is defined in the on-shell renormalisation scheme. Nevertheless, we do not consider this discontinuity to be phenomenologically problematic and therefore do not pursue the use of an imaginary scale in this work.

\section{Selected results\label{sec:results}}

\subsection{Notation and setup\label{sec:setup}}

In this section, we consider the LbL scattering process with two initial photons emitted from two charged sources generically denoted as $\mathrm{A}_1$ and $\mathrm{A}_2$. Our predictions rely on photon collinear factorisation formula, where the physical cross section for the exclusive process $\gamma\gamma\to\gamma\gamma$ can be written as
\begin{equation}
\sigma_{\mathrm{A}_1\mathrm{A}_2}\equiv\sigma\left(\mathrm{A}_1\mathrm{A}_2\overset{\gamma\gamma}{\to}\mathrm{A}_1^{(*)}\gamma\gamma\mathrm{A}_2^{(*)}\right)=\int{\derive x_1 \derive x_2 \mathcal{L}_{\gamma\gamma}^{(\mathrm{A}_1\mathrm{A}_2)}(x_1,x_2)\hat{\sigma}_{\gamma\gamma}(s=x_1x_2 S)}\,,\label{eq:LbLxsinA1A2}
\end{equation}
where $\hat{\sigma}_{\gamma\gamma}$ is the partonic LbL cross section, $\mathcal{L}_{\gamma\gamma}^{(\mathrm{A}_1\mathrm{A}_2)}(x_1,x_2)$ is the photon-photon flux function with $x_1$ and $x_2$ being longitudinal momentum fractions carried by the initial-state photons of initial beam particles, and $\sqrt{S}$ denotes the centre-of-mass energy of the two initial beam particles $\mathrm{A}_1$ and $\mathrm{A}_2$. The symbol $\mathrm{A}_i^{(*)}$ with $i=1,2$ in eq.~\eqref{eq:LbLxsinA1A2} denotes that the particle $\mathrm{A}_i$ can either remain intact or be excited to higher-excited states, where the latter case is only relevant when $\mathrm{A}_i$ is an ion in this study. The precise definition of the flux function $\mathcal{L}_{\gamma\gamma}^{(\mathrm{A}_1\mathrm{A}_2)}(x_1,x_2)$ depends on the species of $\mathrm{A}_1$ and $\mathrm{A}_2$. In this paper, we consider the following three possible scenarios, which are the same as those in ref.~\cite{Capatti:2025khs}.
\begin{itemize}
\item \textbf{Monochromatic photons as beam particles:} if the initial particles $\mathrm{A}_1$ and $\mathrm{A}_2$ are monochromatic photons, the flux function reduces to the product of two Dirac delta functions
\begin{equation}
\mathcal{L}_{\gamma\gamma}^{(\gamma\gamma)}(x_1,x_2)=\delta(1-x_1)\delta(1-x_2)\,.\label{eq:Lggingg}
\end{equation}
In this case, the physical LbL cross section defined in eq.~\eqref{eq:LbLxsinA1A2} is identical to the partonic cross section $\sigma_{\gamma\gamma}=\hat{\sigma}_{\gamma\gamma}$, and $\sqrt{S}=\sqrt{s}$. The final particle set $\left\{\mathrm{A}_1^{(*)},\mathrm{A}_2^{(*)}\right\}$ is empty. The corresponding results in this case will be given in section \ref{sec:LbLxs}.
\item \textbf{UPCs at hadron colliders:} in this case, the beam particles are two charged hadrons, such as protons and ions. The corresponding photon-photon flux function reads~\cite{Shao:2022cly}
\begin{equation}
\mathcal{L}_{\gamma\gamma}^{(\mathrm{A}_1\mathrm{A}_2)}(x_1,x_2)=(x_1x_2)^{-1}\frac{\derive^2N^{(\mathrm{A}_1\mathrm{A}_2)}_{\gamma_1/Z_1,\gamma_2/Z_2}}{\derive E_{\gamma_1}\derive E_{\gamma_2}}\,,\label{eq:LgginUPC}
\end{equation}
where $E_{\gamma_i}$ is the energy of the $i^\mathrm{th}$ initial-state photon, and $Z_i$ is the atomic number of the nucleus $\mathrm{A}_i$. The momentum fractions are defined as
\begin{equation}
x_i=\frac{A_iE_{\gamma_i}}{E_{\mathrm{A}_i}}\,,
\end{equation}
where $A_i$ denotes the atomic mass number of $\mathrm{A}_i$, and $E_{\mathrm{A}_i}$ is the total energy of $\mathrm{A}_i$. Then, $\sqrt{S}$ is the nucleon-nucleon (NN) centre-of-mass energy, 
\begin{equation}
\sqrt{S}=\sqrtsnn\approx \sqrt{\frac{4E_{\mathrm{A}_1}E_{\mathrm{A}_2}}{A_1A_2}}\,.
\end{equation}
The effective photon-photon luminosity can be expressed as a convolution of the two photon number densities, $N_{\gamma_1/Z_1}(E_{\gamma_1},\pmb{b}_1)$ and $N_{\gamma_2/Z_2}(E_{\gamma_2},\pmb{b}_2)$, at impact parameters $\pmb{b}_{1,2}$ with respect to the centres of hadrons $\mathrm{A}_1$ and $\mathrm{A}_2$:
\begin{equation}
\frac{\derive^2N^{(\mathrm{A}_1\mathrm{A}_2)}_{\gamma_1/Z_1,\gamma_2/Z_2}}{\derive E_{\gamma_1}\derive E_{\gamma_2}} =  \int{\derive^2\pmb{b}_1\derive^2\pmb{b}_2\,P_\text{no\,inel}(\pmb{b}_1,\pmb{b}_2)\,N_{\gamma_1/Z_1}(E_{\gamma_1},\pmb{b}_1)N_{\gamma_2/Z_2}(E_{\gamma_2},\pmb{b}_2)}\,,\label{eq:2photonintegral}
\end{equation}
where $P_\text{no\,inel}(\pmb{b}_1,\pmb{b}_2)$ accounts for the probability of no inelastic hadronic interactions between $\mathrm{A}_1$ and $\mathrm{A}_2$. In this work, we use the \gammaUPC\ code~\cite{Shao:2022cly} to evaluate the two-photon flux in UPCs at hadron colliders. In the code, two types of coherent photon fluxes are implemented as functions of the impact parameter, namely those based on the electric-dipole (EDFF) and charge (ChFF) form factors. The latter provides a more reliable description than the former. In fact, the EDFF photon number density diverges as the impact parameter approaches zero (cf. figure 3 of ref.~\cite{Shao:2022cly}), which must therefore be regulated by an arbitrary cutoff (typically $b=|\pmb{b}|\gtrsim R_{\mathrm{A}}$, with $R_{\mathrm{A}}$ the radius of $\mathrm{A}$) in the integral in eq.~\eqref{eq:2photonintegral}. Such a regulation is not required in the ChFF flux. Additionally, cross sections computed with the ChFF flux show better agreement with experimental measurements for dilepton production processes such as $\gamma \gamma\to e^+e^-$~\cite{CMS:2024bnt} and $\gamma \gamma\to \mu^+\mu^-$~\cite{Shao:2024dmk}. Hence, we adopt only the ChFF flux in this work when considering the UPC scenario. The parametric uncertainties in the photon flux modelling have been assessed to be at the percent level, as shown in figure 2 of ref.~\cite{Shao:2024dmk}. We will discuss the data-theory comparisons in lead-lead UPC at the LHC in section \ref{sec:LbLUPC}.
\item \textbf{Electron-positron colliders:} the third case considered in this paper is LbL scattering at electron-positron ($e^-e^+$) colliders. In this scenario, we have $\mathrm{A}_1=\mathrm{A}_1^{(*)}=e^-$ and $\mathrm{A}_2=\mathrm{A}_2^{(*)}=e^+$. The photon-photon flux can be expressed as the product of the photon parton distribution functions (PDFs) of electron and positron,
\begin{equation}
    \mathcal{L}_{\gamma\gamma}^{(e^-e^+)}(x_1,x_2)=f_{\gamma}^{(e^-)}(x_1)f_{\gamma}^{(e^+)}(x_2)\,.\label{eq:Lgginee}
\end{equation}
The photon PDF of electron/positron can be well approximated in the improved Weizs\"acker-Williams (iWW) form~\cite{Frixione:1993yw}:
\begin{equation}
f_\gamma^{(e^{\pm})}(x)=\frac{\alpha(0)}{2\pi}\Bigg[\frac{1+(1-x)^2}{x}\log{\left(\frac{(1-x)Q_{\mathrm{max}}^2}{x^2m_e^2}\right)}+\bigg(\frac{2x m_e^2}{Q_{\mathrm{max}}^{2}}-\frac{2(1-x)}{x}\bigg)\Bigg]\,,\label{eq:WWPDF4lepton}
\end{equation}
where $m_e$ is the electron mass and $Q_{\mathrm{max}}$ is the maximal photon virtuality. In this work we set $Q_{\mathrm{max}}=1$ GeV. The centre-of-mass energy is $\sqrt{S}=\sqrt{s_{e^-e^+}}$. Some predictions for LbL scattering at $e^+e^-$ colliders can be found in section \ref{sec:LbLee}. More accurate determinations of the photon PDF of the electron exist in the literature~\cite{Frixione:2019lga,Bertone:2019hks,Bertone:2022ktl,Stahlhofen:2025hqd,Schnubel:2025ejl}. However, we expect the $K$ factors to be largely insensitive to the precise choice of photon PDF, although this choice may be more relevant for the prediction of absolute cross sections. We leave the quantification of the resulting uncertainties due to the photon PDF for future work.
\end{itemize}

We now turn to the definition of the partonic LbL cross section $\hat{\sigma}_{\gamma\gamma}$. In contrast to inclusive reactions, LbL is an exclusive process, which does not require the inclusion of real emissions to cancel infrared divergences at NLO and beyond.~\footnote{LbL scattering is an exclusive process. Infrared finiteness holds for non-vanishing $s$, $t$, and $u$ kinematics and a non-vanishing $W$-boson mass, to all orders in $\alpha_s$ and $\alpha$. Infrared finiteness may no longer hold when either of these conditions is not satisfied. The relevance of the second condition is evident from our HE expansion of the $W$-loop contribution presented in appendix~\ref{sec:oneloopHE}. For the first condition, consider a massless one-loop fermion $f$ loop in the forward limit $t=0$ ($s=-u>0$). In this case, the LbL scattering amplitude exhibits infrared divergences, which are cancelled by the corresponding $\gamma\gamma\to f\bar{f}$ contribution with $m_f=0$, as required by unitarity and discussed in ref.~\cite{Frye:2018xjj}.} In fact, the single real gluon or photon emission contributions are zero because
of the Furry’s theorem and the conservation of colour and charge-conjugate (C) number in QCD and QED. Hence, two-loop amplitudes for $\gamma\gamma\to\gamma\gamma$ are already infrared finite. We can therefore define the following amplitude combinations~\cite{AH:2023kor} to be used in the cross section definition:
\begin{eqnarray}
\mathcal{M}_{\lambda_1\lambda_2\lambda_3\lambda_4}^{(0,0)}&\equiv&\sum_{l=f,W}{\mathcal{M}_{\lambda_1\lambda_2\lambda_3\lambda_4}^{(0,0,l)}}\,,\\
\mathcal{M}_{\lambda_1\lambda_2\lambda_3\lambda_4}^{(1,0)}&\equiv&\sum_{f}{\mathcal{M}_{\lambda_1\lambda_2\lambda_3\lambda_4}^{(1,0,f)}}\,,\\
\mathcal{M}_{\lambda_1\lambda_2\lambda_3\lambda_4}^{(0,1)}&\equiv&\sum_{f}{\mathcal{M}_{\lambda_1\lambda_2\lambda_3\lambda_4}^{(0,1,f)}}\,,\\
\mathcal{M}_{\lambda_1\lambda_2\lambda_3\lambda_4}^{(1,1,f)}&\equiv&\mathcal{M}_{\lambda_1\lambda_2\lambda_3\lambda_4}^{(1,0,f)}+\mathcal{M}_{\lambda_1\lambda_2\lambda_3\lambda_4}^{(0,1,f)}\,,\\
\mathcal{M}_{\lambda_1\lambda_2\lambda_3\lambda_4}^{(1,1)}&\equiv&\sum_{f}{\mathcal{M}_{\lambda_1\lambda_2\lambda_3\lambda_4}^{(1,1,f)}}\,,
\end{eqnarray}
where the sum of $f$ runs over all SM quarks ($f=u,d,s,c,b,t$) and charged leptons ($f=e^-,\mu^-,\tau^-$). The LO partonic cross-sections can be computed with
\begin{eqnarray}
\hat{\sigma}_{\gamma\gamma}^{(0,0,l)}&=&\frac{1}{2s}\int{\derive \Phi_2\overline{\sum}_{\rm helicity}{\left|\mathcal{M}_{\lambda_1\lambda_2\lambda_3\lambda_4}^{(0,0,l)}\right|^2}}\,,\quad l=f,W,\label{eq:xsdefl0}\\
\hat{\sigma}_{\gamma\gamma}^{(0,0)}&=&\frac{1}{2s}\int{\derive \Phi_2\overline{\sum}_{\rm helicity}{\left|\mathcal{M}_{\lambda_1\lambda_2\lambda_3\lambda_4}^{(0,0)}\right|^2}}\,,\label{eq:xsdef0}
\end{eqnarray}
where $\derive\Phi_2$ is the two-body phase-space measure, and the overlined sum denotes the sum over the helicity configurations of the external photons and the average over the initial-state photon polarisations. We can also define the NLO QCD and/or QED corrections to the partonic cross sections as follows
\begin{eqnarray}
\hat{\sigma}_{\gamma\gamma}^{(i,j,f)}&=&\frac{1}{2s}\int{\derive\Phi_2\overline{\sum}_{\rm helicity}{2\Re{\left\{\mathcal{M}_{\lambda_1\lambda_2\lambda_3\lambda_4}^{(0,0,f)\;\star} \mathcal{M}_{\lambda_1\lambda_2\lambda_3\lambda_4}^{(i,j,f)}\right\}}}}\,,\label{eq:xsdeff1}\\
\hat{\sigma}_{\gamma\gamma}^{(i,j)}&=&\frac{1}{2s}\int{\derive\Phi_2\overline{\sum}_{\rm helicity}{2\Re{\left\{\mathcal{M}_{\lambda_1\lambda_2\lambda_3\lambda_4}^{(0,0)\;\star} \mathcal{M}_{\lambda_1\lambda_2\lambda_3\lambda_4}^{(i,j)}\right\}}}}\label{eq:xsdef1}
\end{eqnarray}
with $(i,j)\in \left\{(1,0),(0,1),(1,1)\right\}$. Thus, the LO, NLO QCD, NLO QED, and NLO QCD+QED cross sections in the SM are 
\begin{eqnarray}
\hat{\sigma}_{\gamma\gamma}^{{\rm LO}}&=&\hat{\sigma}_{\gamma\gamma}^{(0,0)}\,,\\
\hat{\sigma}_{\gamma\gamma}^{{\rm NLO~QCD}}&=&\hat{\sigma}_{\gamma\gamma}^{(0,0)}+\hat{\sigma}_{\gamma\gamma}^{(1,0)}\,,\\
\hat{\sigma}_{\gamma\gamma}^{{\rm NLO~QED}}&=&\hat{\sigma}_{\gamma\gamma}^{(0,0)}+\hat{\sigma}_{\gamma\gamma}^{(0,1)}\,,\\
\hat{\sigma}_{\gamma\gamma}^{{\rm NLO~QCD+QED}}&=&\hat{\sigma}_{\gamma\gamma}^{(0,0)}+\hat{\sigma}_{\gamma\gamma}^{(1,1)}\,.
\end{eqnarray}
Thanks to the exclusiveness of the process, we can also include partial next-to-next-to-leading order (NNLO) contributions by squaring the whole one- and two-loop amplitudes in the partonic cross sections, which we denote as NLO$^\prime$:
\begin{eqnarray}
\hat{\sigma}_{\gamma\gamma}^{{\rm NLO}^\prime~{\rm QCD}}&=&\frac{1}{2s}\int{\derive \Phi_2\overline{\sum}_{\rm helicity}{\left|\mathcal{M}_{\lambda_1\lambda_2\lambda_3\lambda_4}^{(0,0)}+\mathcal{M}_{\lambda_1\lambda_2\lambda_3\lambda_4}^{(1,0)}\right|^2}}\,,\\
\hat{\sigma}_{\gamma\gamma}^{{\rm NLO}^\prime~{\rm QED}}&=&\frac{1}{2s}\int{\derive \Phi_2\overline{\sum}_{\rm helicity}{\left|\mathcal{M}_{\lambda_1\lambda_2\lambda_3\lambda_4}^{(0,0)}+\mathcal{M}_{\lambda_1\lambda_2\lambda_3\lambda_4}^{(0,1)}\right|^2}}\,,\\
\hat{\sigma}_{\gamma\gamma}^{{\rm NLO}^\prime~{\rm QCD+QED}}&=&\frac{1}{2s}\int{\derive \Phi_2\overline{\sum}_{\rm helicity}{\left|\mathcal{M}_{\lambda_1\lambda_2\lambda_3\lambda_4}^{(0,0)}+\mathcal{M}_{\lambda_1\lambda_2\lambda_3\lambda_4}^{(1,1)}\right|^2}}\,.
\end{eqnarray}
We can have the analogous definitions when considering Coulomb resummation. In other words, we can define amplitudes
\begin{eqnarray}
\mathcal{M}_{\lambda_1\lambda_2\lambda_3\lambda_4}^{(1,0),\mathrm{LP}}&\equiv&\sum_{f}{\mathcal{M}_{\lambda_1\lambda_2\lambda_3\lambda_4}^{(1,0,f),\mathrm{LP}}}\,,\\
\mathcal{M}_{\lambda_1\lambda_2\lambda_3\lambda_4}^{(0,1),\mathrm{LP}}&\equiv&\sum_{f}{\mathcal{M}_{\lambda_1\lambda_2\lambda_3\lambda_4}^{(0,1,f),\mathrm{LP}}}\,,\\
\mathcal{M}_{\lambda_1\lambda_2\lambda_3\lambda_4}^{(1,1),\mathrm{LP}}&\equiv&\sum_{f}{\left(\mathcal{M}_{\lambda_1\lambda_2\lambda_3\lambda_4}^{(1,0,f),\mathrm{LP}}+\mathcal{M}_{\lambda_1\lambda_2\lambda_3\lambda_4}^{(0,1,f),\mathrm{LP}}\right)}\,,
\end{eqnarray}
and partonic cross sections
\begin{eqnarray}
\hat{\sigma}_{\gamma\gamma}^{{\rm NLO+LP~QCD}}&=&\hat{\sigma}_{\gamma\gamma}^{(0,0)}+\frac{1}{2s}\int{\derive\Phi_2\overline{\sum}_{\rm helicity}{2\Re{\left\{\mathcal{M}_{\lambda_1\lambda_2\lambda_3\lambda_4}^{(0,0)\;\star} \mathcal{M}_{\lambda_1\lambda_2\lambda_3\lambda_4}^{(1,0),\mathrm{LP}}\right\}}}}\,,\\
\hat{\sigma}_{\gamma\gamma}^{{\rm NLO+LP~QED}}&=&\hat{\sigma}_{\gamma\gamma}^{(0,0)}+\frac{1}{2s}\int{\derive\Phi_2\overline{\sum}_{\rm helicity}{2\Re{\left\{\mathcal{M}_{\lambda_1\lambda_2\lambda_3\lambda_4}^{(0,0)\;\star} \mathcal{M}_{\lambda_1\lambda_2\lambda_3\lambda_4}^{(0,1),\mathrm{LP}}\right\}}}}\,,\\
\hat{\sigma}_{\gamma\gamma}^{{\rm NLO+LP~QCD+QED}}&=&\hat{\sigma}_{\gamma\gamma}^{(0,0)}+\frac{1}{2s}\int{\derive\Phi_2\overline{\sum}_{\rm helicity}{2\Re{\left\{\mathcal{M}_{\lambda_1\lambda_2\lambda_3\lambda_4}^{(0,0)\;\star} \mathcal{M}_{\lambda_1\lambda_2\lambda_3\lambda_4}^{(1,1),\mathrm{LP}}\right\}}}}\,,\\
\hat{\sigma}_{\gamma\gamma}^{{\rm (NLO+LP)}^\prime~{\rm QCD}}&=&\frac{1}{2s}\int{\derive \Phi_2\overline{\sum}_{\rm helicity}{\left|\mathcal{M}_{\lambda_1\lambda_2\lambda_3\lambda_4}^{(0,0)}+\mathcal{M}_{\lambda_1\lambda_2\lambda_3\lambda_4}^{(1,0),\mathrm{LP}}\right|^2}}\,,\label{eq:xsNLOLPpQCD}\\
\hat{\sigma}_{\gamma\gamma}^{{\rm (NLO+LP)}^\prime~{\rm QED}}&=&\frac{1}{2s}\int{\derive \Phi_2\overline{\sum}_{\rm helicity}{\left|\mathcal{M}_{\lambda_1\lambda_2\lambda_3\lambda_4}^{(0,0)}+\mathcal{M}_{\lambda_1\lambda_2\lambda_3\lambda_4}^{(0,1),\mathrm{LP}}\right|^2}}\,,\label{eq:xsNLOLPpQED}\\
\hat{\sigma}_{\gamma\gamma}^{{\rm (NLO+LP)}^\prime~{\rm QCD+QED}}&=&\frac{1}{2s}\int{\derive \Phi_2\overline{\sum}_{\rm helicity}{\left|\mathcal{M}_{\lambda_1\lambda_2\lambda_3\lambda_4}^{(0,0)}+\mathcal{M}_{\lambda_1\lambda_2\lambda_3\lambda_4}^{(1,1),\mathrm{LP}}\right|^2}}\,.\label{eq:xsNLOLPpQCDQED}
\end{eqnarray}

\begin{table}[t!]
\centering
\renewcommand*{\arraystretch}{1.4}
\begin{tabular}[t]{cc|cc}
\toprule
\textbf{parameter}& value & \textbf{parameter}& value \\
\midrule
$m_e$ & $510.99895\,\mathrm{keV}$ & $m_\mu$ & $105.6583755\,\mathrm{MeV}$ \\
$m_u$ & $335\,\mathrm{MeV}$ & $m_d$ & $340\,\mathrm{MeV}$ \\
$m_s$ & $490\,\mathrm{MeV}$ & $m_c$ & $1.5\,\mathrm{GeV}$ \\
$m_\tau$ & $1.77693\,\mathrm{GeV}$ & $m_b$ & $4.75\,\mathrm{GeV}$ \\
$m_t$  & $172.56\,\mathrm{GeV}$ & $m_W$ & $80.3692\,\mathrm{GeV}$ \\
$\alpha_s(m_Z^2)$ & $0.118$ & $m_Z$ & $91.1876\,\mathrm{GeV}$ \\
$\alpha^{-1}(0)$ & $137.036$ & $\alpha^{-1}(m_Z^2)$ & $128.94$ \\
\bottomrule
\end{tabular}
\caption{Summary of the central values of the SM parameters used in all analyses presented in this article. The corresponding parametric uncertainties are discussed in the text.}
\label{tab:setup}
\end{table}

After defining the above notation, we now specify the calculational setup used in this paper. For the on-shell masses of quarks and charged leptons in the SM, we choose the electron mass $m_e=0.51099895$ MeV, the muon mass $m_\mu=105.6583755$ MeV, the tau mass $m_\tau=1.77693$ GeV, the up-quark mass $m_u=335$ MeV, the down-quark mass $m_d=340$ MeV, the strange-quark mass $m_s=490$ MeV, the charm-quark mass $m_c=1.5$ GeV, the bottom-quark mass $m_b=4.75$ GeV, and the top-quark mass $m_t=172.56$ GeV. These values largely follow the latest recommendations of the Particle Data Group~\cite{ParticleDataGroup:2024cfk}. For the light quarks ($u,d,s$), however, we use approximate constituent quark masses instead of current quark masses in order to account for light-hadron effects (see, \eg, table I of ref.~\cite{dEnterria:2022alo}). We estimate the uncertainty associated with constituent quark masses by varying $m_{u,d}$ and $m_s$ within typical ranges used in constituent quark models, $m_{u,d}\in [270,400]~\mathrm{MeV}$ and $m_s\in [400,600]~\mathrm{MeV}$, corresponding to $\pm20\%$ around their central values. The uncertainty associated with the charm-quark mass is estimated by varying the charm pole mass in the range $1.3~\mathrm{GeV}\leq m_c\leq 1.7~\mathrm{GeV}$, around the central value $m_c=1.5~\mathrm{GeV}$. For the $W^\pm$ and $Z$ bosons, their on-shell masses are fixed to $m_W=80.3692$ GeV and $m_Z=91.1876$ GeV. The $m_Z$ dependence in our results enters only through the $\alpha_s$ and $\alpha$ couplings. The widths of all particles are set to zero. In the one- and two-loop helicity amplitudes, we adopt a hybrid scheme: the couplings associated with external on-shell photons use $\alpha(0)=1/137.036$, while those for internal photons and gluons use $\alpha(\mu_R^2)$ in the on-shell scheme and $\alpha_s(\mu_R^2)$ in the modified minimal subtraction ($\MSbar$) scheme, where the renormalisation scale $\mu_R$ follows eq.~\eqref{eq:muRdef}. The two-loop effective QED coupling $\alpha(\mu_R^2)$, obtained from the vacuum-polarisation corrections in the on-shell scheme and discussed further in appendix~\ref{sec:alphaRGrun}, is obtained by imposing the input condition $\alpha(m_Z^2)=1/128.94$. This condition fixes the non-perturbative five-flavour hadronic contribution $\Delta\alpha_{\mathrm{had}}^{(5)}(m_Z^2)$ used in our calculation. The renormalisation-group evolution of the strong coupling $\alpha_s(\mu_R^2)$ follows the standard five-loop QCD beta function~\cite{Baikov:2016tgj,Herzog:2017ohr}, with the initial condition $\alpha_s(m_Z^2)=0.118$. The four-loop decoupling relations for the strong coupling~\cite{Chetyrkin:1997un,Schroder:2005hy,Chetyrkin:2005ia,Gerlach:2018hen} are employed with the on-shell top-quark (bottom-quark, charm-quark) mass thresholds at $172.56$ GeV ($4.75$ GeV, $1.5$ GeV). We vary $\alpha^{-1}(m_Z^2)\in [128.90,128.98]$ and $\alpha_s(m_Z^2)\in [0.117,0.119]$ to estimate the corresponding parametric uncertainties. We additionally turn off NLO QCD corrections when $\sqrt{s}<2$ GeV, which is equivalent to setting $\alpha_s=0$ in $\mathcal{M}_{\lambda_1\lambda_2\lambda_3\lambda_4}^{(1,0,f)}$ and $\mathcal{M}_{\lambda_1\lambda_2\lambda_3\lambda_4}^{(1,0,f),\mathrm{LP}}$. The central values of the SM parameters used in this paper are summarised in table~\ref{tab:setup}. Finally, we stress that thanks to the improvements of our two-loop amplitudes reported in this paper, we are able to use the same particle-mass setup in the \lblatnlo\ code (cf. appendix~\ref{sec:lblatnlo}) regardless of the value of $\sqrt{s}$, which significantly simplifies the use of the code.

\subsection{Light-by-light scattering cross sections\label{sec:LbLxs}}

In this section, we consider the simplest case without folding in any photon flux. In other words, we use the flux function as given in eq.~\eqref{eq:Lggingg}. The LbL cross section, $\sigma_{\gamma\gamma}$, is a fundamental quantity that enters any LbL-like scattering experiment or its theoretical interpretation. Therefore, our first goal is to understand the quantum corrections to $\sigma_{\gamma\gamma}$.

\begin{figure}[hbt!]
\includegraphics[width=0.95\columnwidth,draft=false]{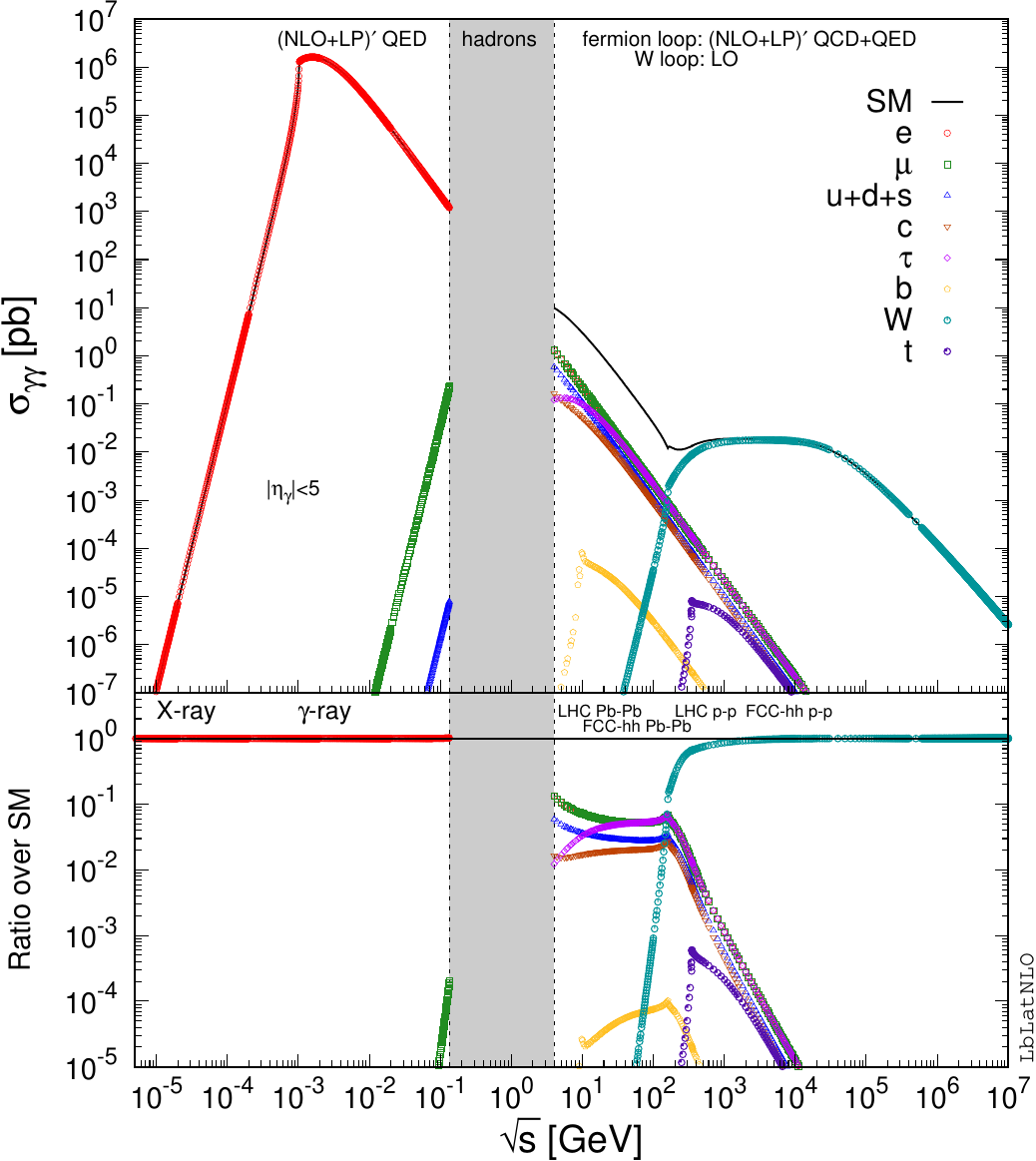}
\caption{Breakdown of the LbL cross section $\sigma_{\gamma\gamma}$ into the various SM components (upper panel), together with the fractional contribution of each component (lower panel). The region $134~\mathrm{MeV} \le \sqrt{s} \le 4~\mathrm{GeV}$ is dark shaded, where the hadronic contribution cannot be reliably computed using quark loops. The $W$-loop contribution is included only at one-loop (LO) accuracy.}
\label{figLbLxs1} \vspace*{-0.5cm}
\end{figure}

Figure~\ref{figLbLxs1} shows the state-of-the-art LbL cross section $\sigma_{\gamma\gamma}$ in the SM (black curve) for $\sqrt{s}\in [2\cdot10^{-6},0.134)\bigcup (4,10^7]$ GeV, where a pseudorapidity cut $|\eta_\gamma|<5$ is imposed on the final-state photons in the centre-of-mass frame of the two initial-state photons. In our calculations, we assume the validity of quark-hadron duality. This assumption is expected to break down in the region $134~\mathrm{MeV}\lesssim\sqrt{s}\lesssim4~\mathrm{GeV}$, which is affected by significant hadronic resonances (see, \eg, figure~13 of ref.~\cite{dEnterria:2025ecx}). Therefore, this region, shown as a dark-shaded area in figure~\ref{figLbLxs1}, has been excluded. As mentioned in the previous section, we achieve (NLO+LP)$^\prime$ QCD+QED accuracy (cf. eq.~\eqref{eq:xsNLOLPpQCDQED}) for $2~\mathrm{GeV}\leq \sqrt{s}\lesssim 156$ GeV and (NLO+LP)$^\prime$ QED accuracy (cf. eq.~\eqref{eq:xsNLOLPpQED}) for $\sqrt{s}<2$ GeV. The SM cross section spans more than 13 orders of magnitude over the considered range, reaching a maximum value of $1.61~\mu\mathrm{b}$ at $\sqrt{s}\simeq1.56~\mathrm{MeV}$. The bumps in the black curve reflect the onset of different particle component contributions, which are more clearly visible from the decomposition of the SM contributions shown in the figure. The electron contribution is the dominant one up to $\sqrt{s}\simeq156$ GeV, above which the $W^\pm$ boson loop becomes dominant. The kinks occur at $\sqrt{s}\simeq2m_f$ and $\sqrt{s}\simeq2m_W$. In the lower panel of figure~\ref{figLbLxs1}, the fractional contributions of the different SM components are displayed. From the LE and HE scattering amplitudes presented in section~\ref{sec:asymexpofamp}, the cross section $\sigma_{\gamma\gamma}$ for each particle species follows the expected scaling behaviour. In the LE limit, all components exhibit the same $\mathcal{O}(s^3)$ scaling, while in the HE limit the cross section scales as $\mathcal{O}(s^{-1})$, with different logarithmic enhancements for fermions and the $W^\pm$ boson. The bottom- and top-quark fractions never exceed $10^{-3}$. In the region $4~\mathrm{GeV} \lesssim \sqrt{s} \lesssim 150~\mathrm{GeV}$, all charged fermions except the bottom and top quarks contribute more than $1\%$, and the interference contribution is significant. For instance, more than $70\%$ of the cross section arises from interference in the range $4~\mathrm{GeV} \lesssim \sqrt{s} \lesssim 140~\mathrm{GeV}$.


\begin{figure}[hbt!]
\includegraphics[width=0.50\columnwidth,draft=false]{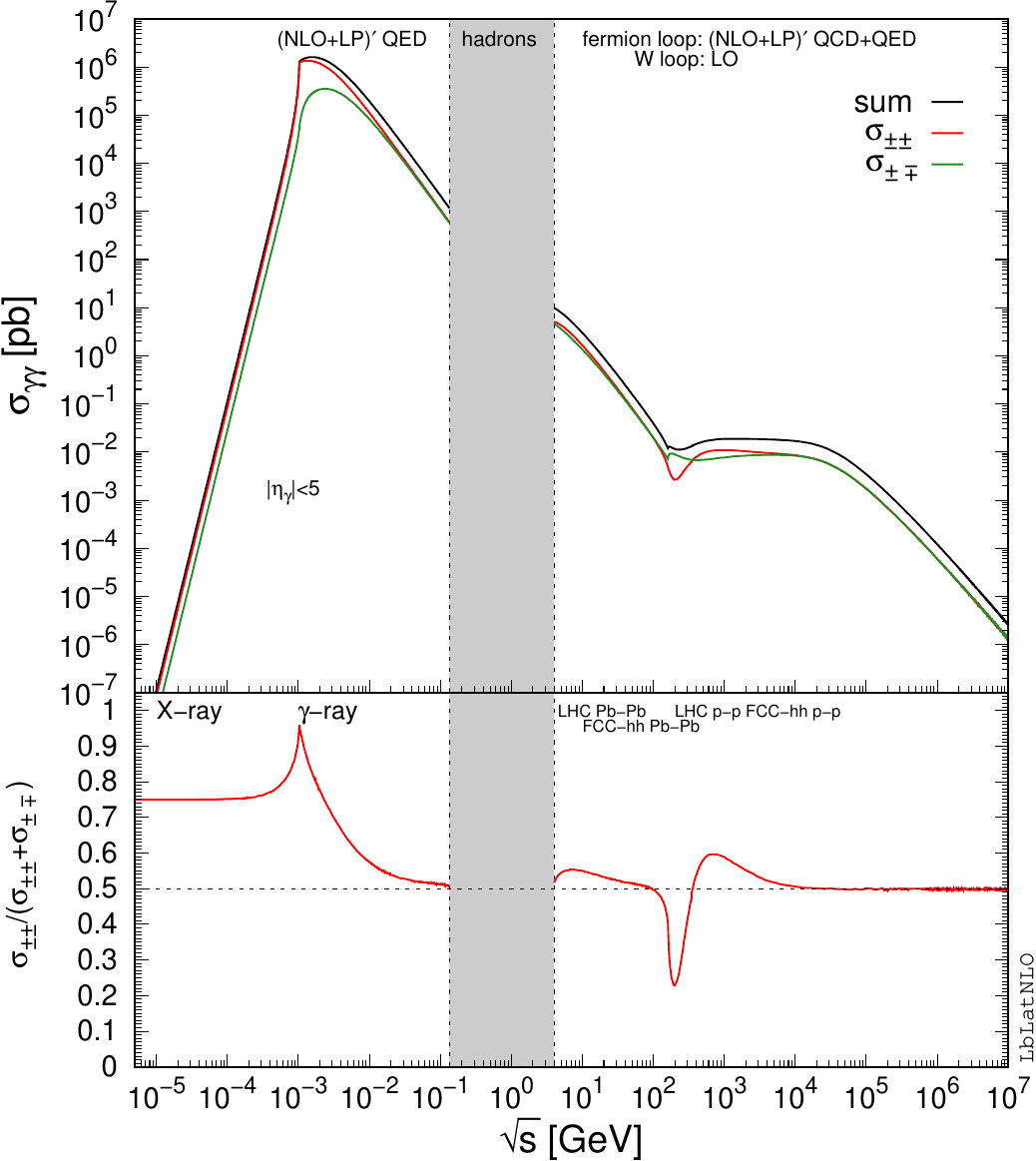}
\includegraphics[width=0.50\columnwidth,draft=false]{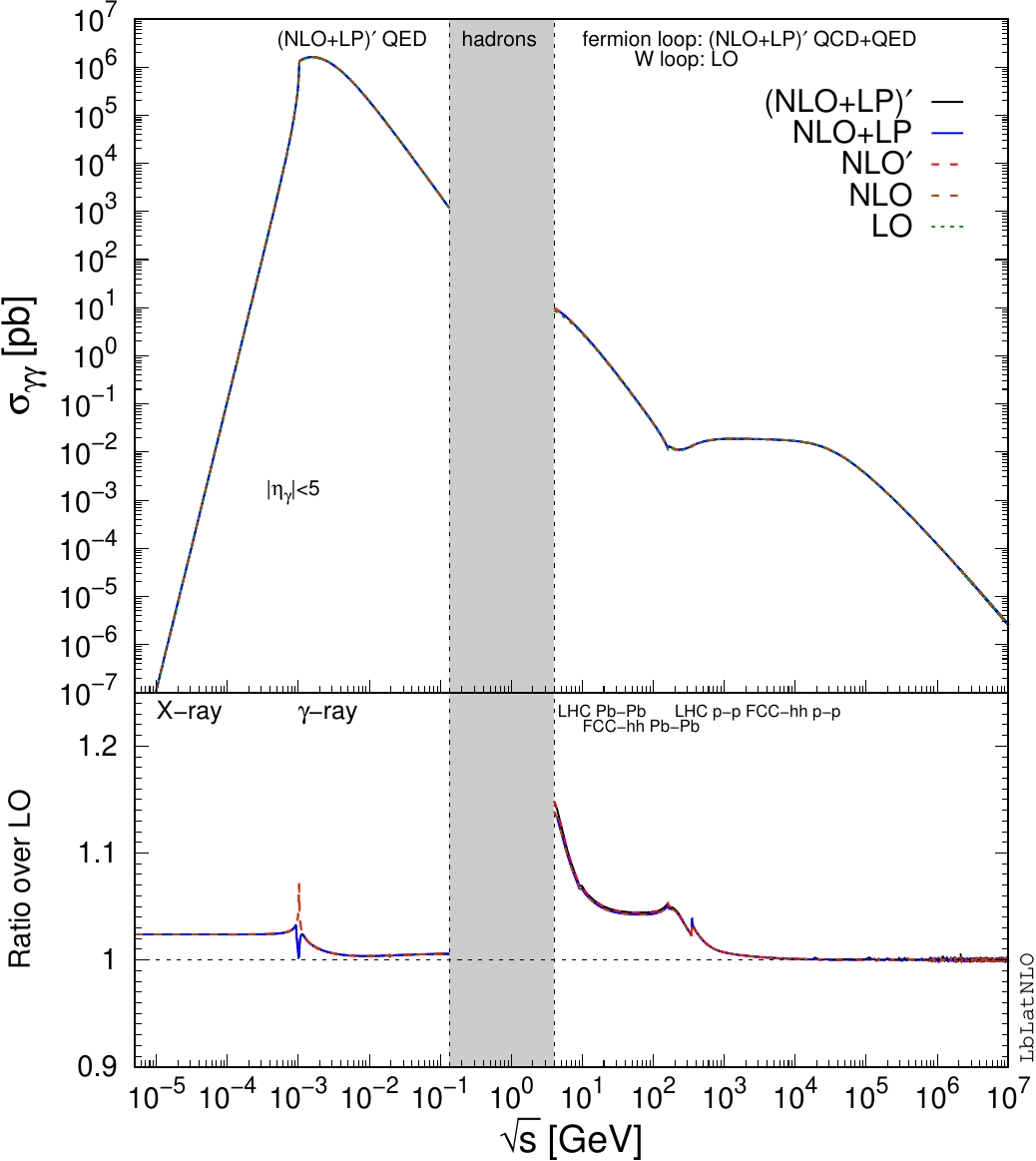}
\caption{The final-state photon helicity-dependent $\sigma_{\gamma\gamma}$ (left panel) and the LbL cross section at various levels of accuracy (right panel). The region $134~\mathrm{MeV} \le \sqrt{s} \le 4~\mathrm{GeV}$ is dark shaded, where the hadronic contribution cannot be reliably computed using quark loops. The $W$-loop contribution is included only at one-loop (LO) accuracy.}
\label{figLbLxs2} \vspace*{-0.5cm}
\end{figure}

The knowledge of the LbL helicity amplitudes also enables us to simulate the polarised $\gamma\gamma\to\gamma\gamma$ process using the \lblatnlo\ code. As a showcase, we consider the cross sections with the same final-state photon helicities, denoted by $\sigma_{\pm\pm}$, and with opposite photon helicities, $\sigma_{\pm\mp}$, shown in the left panel of figure~\ref{figLbLxs2}. In the lower panel, we display the ratio $\sigma_{\pm\pm}/(\sigma_{\pm\pm}+\sigma_{\pm\mp})$ as a function of $\sqrt{s}$. The black curve in the figure is identical to the black curve in figure~\ref{figLbLxs1}, \ie, the state-of-the-art SM prediction. Near the electron-positron threshold, $\sqrt{s}\simeq2m_e$, the ratio $\sigma_{\pm\pm}/(\sigma_{\pm\pm}+\sigma_{\pm\mp})$ approaches unity, which can be readily understood from the results in section~\ref{sec:Coulombres}. In this region, the virtual electron-positron pair forms a $J^{\mathrm{PC}}=0^{-+}$ pseudoscalar state, which predominantly selects the same-helicity configurations. On the other hand, dips appear in the ratio near other fermion-antifermion thresholds and around $\sqrt{s}\simeq2m_W$. This occurs because the scattering amplitudes have opposite signs in the threshold and HE regions, leading to destructive interference in the same-helicity contributions. In the X-ray region, where $\sqrt{s}\ll m_e$, the ratio $\sigma_{\pm\pm}/(\sigma_{\pm\pm}+\sigma_{\pm\mp})$ approaches approximately $0.75$. In contrast, in the HE region where $\sqrt{s}\gg m_W$, there is roughly a $50\%$ probability that the two final-state photons are produced with the same helicity.

We finally discuss the quantum corrections to the LbL cross section as a function of $\sqrt{s}$. The corresponding results at LO, NLO, NLO$^\prime$, NLO+LP, and (NLO+LP)$^\prime$ accuracies are shown as green short-dashed, orange long-dashed, red long-dashed, blue, and black curves, respectively, in the right panel of figure~\ref{figLbLxs2}. In the region $\sqrt{s}<2$ GeV, where only QED corrections are included, the quantum corrections lead to a few-percent enhancement of the LO LbL cross section. In particular, when the centre-of-mass energy is below $1$ MeV, the QED corrections amount to about $2.4\%$. Once again, we observe that the Coulomb resummation removes the logarithmic enhancement observed at NLO and NLO$^\prime$ in the threshold region. Additionally, for $\sqrt{s}>4$ GeV, where our calculations remain reliable, the QCD and QED corrections can increase the LO cross section by as much as $15\%$. As $\sqrt{s}$ increases, the $K$ factor decreases from $1.15$ to a minimum of $1.044$ at around $\sqrt{s}\simeq58$ GeV. Due to the onset of the $W^\pm$ boson contribution, the $K$ factor increases mildly thereafter until $\sqrt{s}\simeq 2m_W$. When $\sqrt{s}\gtrsim2m_t$, the $K$ factor asymptotically approaches unity, since we include only the one-loop amplitudes for the $W^\pm$ boson. We also notice that the partial NNLO corrections, obtained by comparing NLO (NLO+LP) with NLO$^\prime$ ((NLO+LP)$^\prime$), are quite small in view of the size of the NLO corrections. For instance, at $\sqrt{s}=4$ GeV, the NLO and partial NNLO corrections amount to about $14\%$ and $1\%$ of the LO contribution, respectively.

\subsection{Light-by-light scattering at hadron colliders\label{sec:LbLUPC}}

A real phenomenological application of the LbL cross sections, as well as our \lblatnlo\ event generator (cf. appendix \ref{sec:lblatnlo}), is the study of the exclusive di-photon final state produced via initial quasi-real two-photon collisions in UPCs at hadron colliders.~\footnote{Other two-photon processes produced in proton and/or nuclear UPCs that feature non-zero tree-level helicity amplitudes in the SM, such as $\gamma\gamma\to \ell^+\ell^-$ ($\ell=e,\mu,\tau$), have been automated at NLO in the \mg\ framework~\cite{Alwall:2014hca,Frederix:2018nkq} thanks to the recent development~\cite{Shao:2025bma}.} As mentioned in the introduction, the LbL scattering process was first directly observed in ultraperipheral lead-lead (\PbPb) collisions at the LHC~\cite{ATLAS:2017fur,CMS:2018erd,ATLAS:2019azn,ATLAS:2020hii,CMS:2024bnt}.
In this section, we compare our theoretical calculations with the latest ATLAS~\cite{ATLAS:2020hii} and CMS~\cite{CMS:2024bnt} measurements in \PbPb\ collisions at $\sqrtsnn=5.02$ TeV. The fiducial cuts applied to the final-state photons in these two experiments are summarised in table~\ref{tab:LbLUPCcuts}. In particular, the requirement that the di-photon invariant mass $m_{\gamma\gamma}$ be larger than $5$ GeV in both experiments justifies the validity of our computations based on quarks rather than hadrons under the quark-hadron duality assumption. The small di-photon acoplanarity cut $A_{\phi}^{\gamma\gamma}\equiv 1-|\Delta \phi_{\gamma\gamma}|/\pi<0.01$ is important for reducing backgrounds, notably those related to gluon-induced central exclusive production, where $\Delta \phi_{\gamma\gamma}$ is the di-photon azimuthal angle difference. The main difference in the kinematic phase space between the two experiments is the photon transverse momentum cut $p_T^\gamma$, while the impact of the difference in the photon pseudorapidity cut $\eta_\gamma$ on the LbL cross section is secondary. To account for the non-zero initial-state photon virtualities, a $k_\perp$ smearing of the four-momenta of the external photons has been applied, as explained in appendix A of ref.~\cite{Shao:2024dmk}. This effectively allows us to generate non-zero di-photon transverse momentum $p_T^{\gamma\gamma}$ and acoplanarity $A_\phi^{\gamma\gamma}$ in the simulated events.

\begin{table}[htbp!]
\tabcolsep=3.mm
\vspace{0.2cm}
\centering
\begin{tabular}{|l|c|}\hline
System, experiment & $\gamma\gamma\to\gamma\gamma$ fiducial phase space\\\hline
\multirow{2}{*}{\PbPb\ at 5.02 TeV, ATLAS~\cite{ATLAS:2020hii}} & $p_T^{\gamma}>2.5$ GeV, $|\eta_{\gamma}|<2.37$, \\
& $m_{\gamma\gamma}>5$ GeV, $p_T^{\gamma\gamma}<1$ GeV, $A_{\phi}^{\gamma\gamma}<0.01$\\\hline
\multirow{2}{*}{\PbPb\ at 5.02 TeV, CMS~\cite{CMS:2024bnt}} & $p_T^{\gamma}>2$ GeV, $|\eta_{\gamma}|<2.2$, \\
& $m_{\gamma\gamma}>5$ GeV, $p_T^{\gamma\gamma}<1$ GeV, $A_{\phi}^{\gamma\gamma}<0.01$\\
\hline
\end{tabular}
\caption{Exclusive LbL measurements~\cite{ATLAS:2020hii,CMS:2024bnt} in \PbPb\ UPCs at the LHC with their kinematic fiducial phase space.\label{tab:LbLUPCcuts}}
\end{table}

\begin{table}[htpb!]
\centering
\tabcolsep=3.5mm
\vspace{0.2cm}
\begin{tabular}{c|c|c|c} \hline
Process: $\gamma\gamma\to \gamma\gamma$ & \multicolumn{3}{c}{\gammaUPC+\lblatnlo} \\
Experiment & $\sigma_{\mathrm{PbPb}}^{\mathrm{LO}}$ & $\sigma_{\mathrm{PbPb}}^{\mathrm{NLO}\ \mathrm{QCD}}$ & $\sigma_{\mathrm{PbPb}}^{(\mathrm{NLO+LP})^\prime\ \mathrm{QCD+QED}}$ \\ \hline
$\sigma_{\mathrm{PbPb}}^{\mathrm{ATLAS}}=120\pm22$ nb &  $76.3^{+1.8}_{-1.9}$ nb &  $80.1^{+1.2+1.3}_{-1.5-0.7}$ nb & $80.5^{+1.2+1.4}_{-1.5-0.7}$ nb \\
$\sigma_{\mathrm{PbPb}}^{\mathrm{CMS}}=107\pm27$ nb &  $92.0^{+2.6}_{-2.6}$ nb &  $97.3^{+1.8+1.9}_{-2.2-1.0}$ nb & $97.8^{+1.8+2.0}_{-2.2-1.1}$ nb \\
\hline
 $K$ factor & & {\large $\sigma_{\mathrm{PbPb}}^{\mathrm{NLO}\ \mathrm{QCD}}/\sigma_{\mathrm{PbPb}}^{\mathrm{LO}}$} & $\sigma_{\mathrm{PbPb}}^{(\mathrm{NLO+LP})^\prime\ \mathrm{QCD+QED}}/\sigma_{\mathrm{PbPb}}^{\mathrm{LO}}$ \\ \hline
ATLAS fiducial vol. & & $1.049_{-0.009-0.009}^{+0.007+0.017}$ & $1.055_{-0.009-0.010}^{+0.007+0.018}$ \\
CMS fiducial vol. &  &  $1.058_{-0.009-0.011}^{+0.007+0.020}$ & $1.063_{-0.009-0.012}^{+0.007+0.022}$ \\
\hline
\end{tabular}
\caption{Fiducial exclusive LbL cross sections measured in \PbPb\ UPCs at the LHC (within the phase space defined in table~\ref{tab:LbLUPCcuts}), compared with the theoretical LO, NLO QCD, and (NLO+LP)$^\prime$ QCD+QED predictions obtained with \gammaUPC+\lblatnlo\ using the ChFF $\gamma$ flux. The first (second) quoted theoretical uncertainties stem from variations of the charm-quark mass (scale).\label{tab:PbPbUPCxs}
}
\end{table}

\begin{figure}[hbt!]
\includegraphics[width=0.50\columnwidth,draft=false]{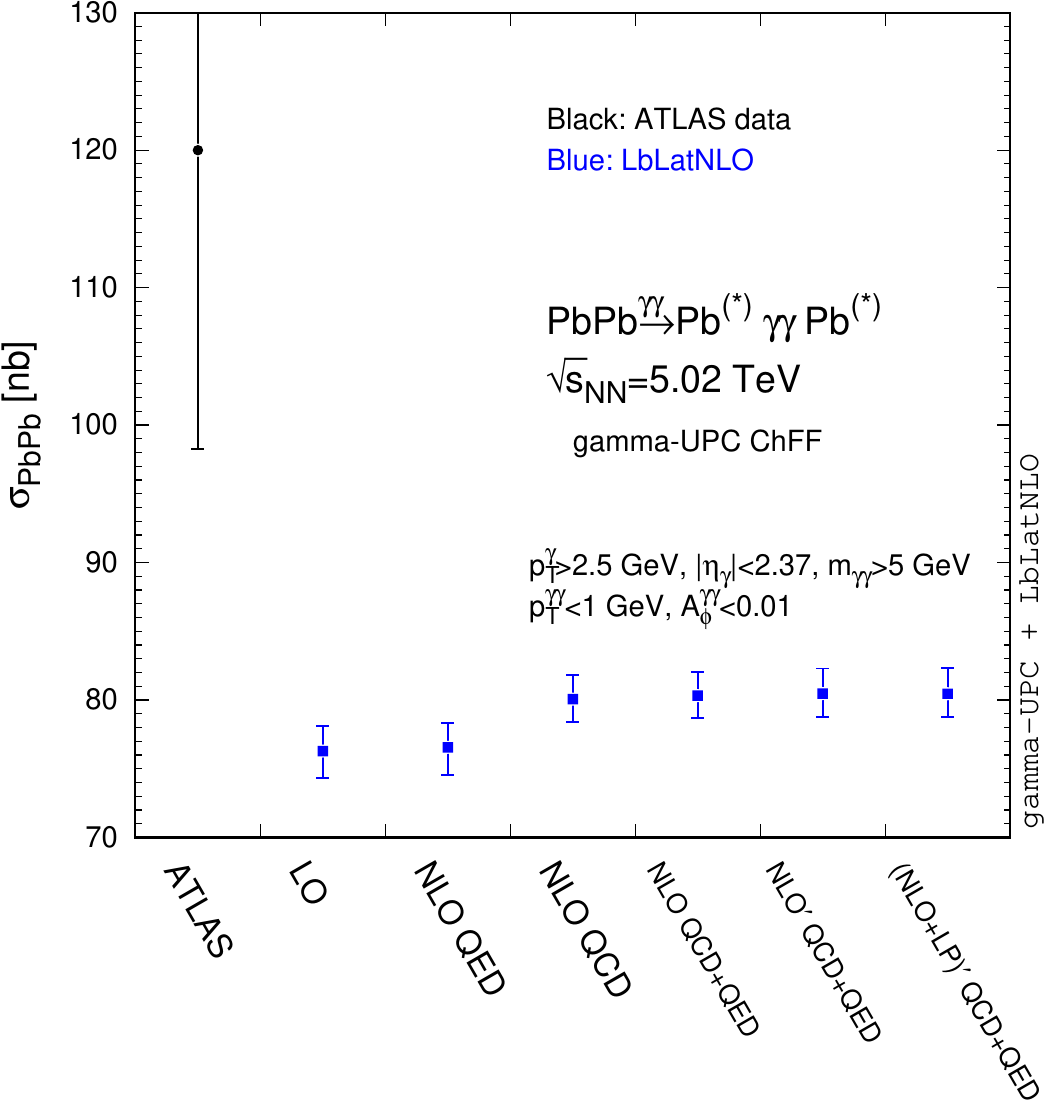}
\includegraphics[width=0.50\columnwidth,draft=false]{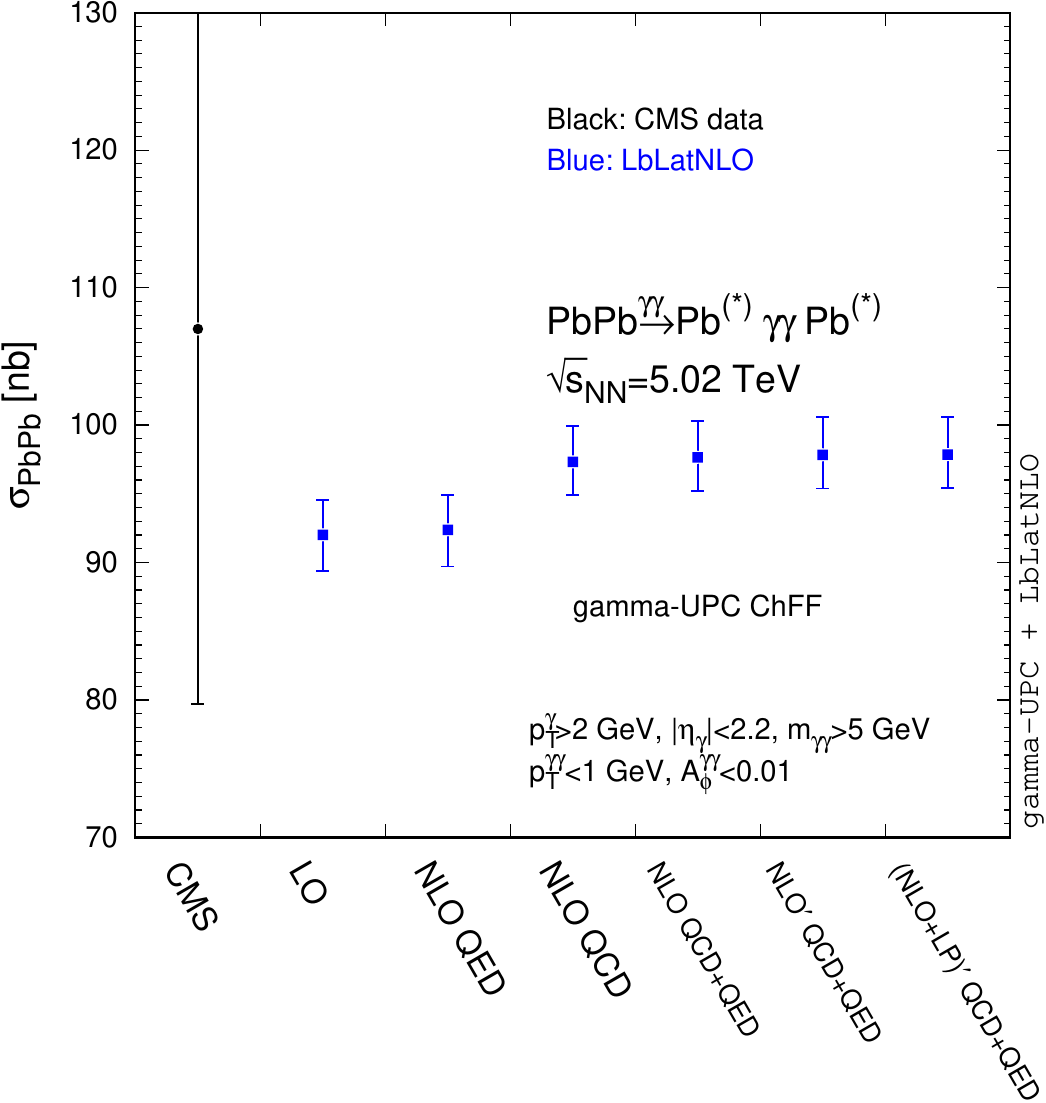}
\caption{Phase-space-integrated fiducial LbL cross sections measured in \PbPb\ UPCs by ATLAS (left) and CMS (right) at the LHC, within the phase space defined in table~\ref{tab:LbLUPCcuts}, compared with theoretical predictions at various perturbative accuracies obtained with \gammaUPC+\lblatnlo\ using the ChFF $\gamma$ flux. The quoted theoretical uncertainties stem from charm-quark mass and scale variations and are combined in quadrature.}
\label{figLbLPbPbfidxs} \vspace*{-0.5cm}
\end{figure}

The theoretical predictions for the phase-space-integrated fiducial cross sections, together with the corresponding experimental measurements, are reported in table~\ref{tab:PbPbUPCxs} and figure~\ref{figLbLPbPbfidxs}. The experimental uncertainties combine statistical and systematic components in quadrature, whereas the quoted theoretical uncertainties stem from charm-quark mass and scale variations, respectively. There is no scale dependence at LO. Other parametric uncertainties are negligible. In particular, variations of the constituent quark masses $m_{u,d,s}$ lead to uncertainties below $0.3\%$. Variations of $\alpha_s(m_Z^2)$ and $\alpha(m_Z^2)$ result in uncertainties below $0.2\%$ and $0.02\%$, respectively. The uncertainty due to the bottom-quark mass is also negligible, given the small size of the bottom-quark contribution (cf. figure~\ref{figLbLxs1}). Since the top-quark and $W$-boson loops are too heavy to contribute appreciably within our fiducial volumes, finite-width effects and the missing full electroweak corrections are also expected to be negligible. Therefore, in the following, we consider only the charm-quark mass and scale uncertainties.

The ATLAS measurement, $120\pm22$ nb, corresponds to a deviation of about $2\sigma$ above the LO prediction $\sigma^{\mathrm{LO}}_{\mathrm{PbPb}}=76.3^{+1.8}_{-1.9}$ nb. NLO QCD corrections increase the LO prediction by $4.9\%$, while NLO QED corrections raise the LO cross section by approximately $0.3\%$, which is a simple consequence of $\alpha_s\gg\alpha$. Partial NNLO effects, estimated by comparing the NLO$^\prime$ QCD+QED and NLO QCD+QED results, yield an additional $+0.2\%$ correction, which is of the same order as the NLO QED contribution following the na\"ive coupling scaling $\alpha_s^2\sim\alpha$. The LP Coulomb resummation contribution is well below one per mille and can therefore be safely neglected. This is unsurprising since the logarithmic Coulomb enhancement in the amplitudes is strongly suppressed by the phase-space measure near threshold. Our state-of-the-art prediction, based on a (NLO+LP)$^\prime$ QCD+QED computation, still lies $1.8\sigma$ below the ATLAS value.~\footnote{
The slightly lower $K$ factor we obtain compared with ref.~\cite{AH:2023kor} is due to a different calculational setup.} 

On the other hand, although the CMS fiducial phase space is slightly larger, mainly driven by the $p_T^\gamma$ cut, the CMS collaboration reports a smaller central value for the fiducial LbL cross section, $107$ nb. This value nevertheless remains compatible with the ATLAS measurement given the large experimental uncertainties. In contrast to the ATLAS case, theoretical predictions at all perturbative orders agree well with the CMS result. The perturbative corrections in the CMS fiducial phase space are slightly larger than those in the ATLAS phase space, as the former includes more events at small scales. Specifically, NLO QCD (QED) corrections amount to $+5.8\%$ ($+0.4\%$) of the LO cross section $92.0$ nb. Partial NNLO corrections are again around $+0.2\%$, with a negligible Coulomb resummation contribution (well below $0.1\%$). 

The residual scale uncertainties in our best predictions are $^{+1.7\%}_{-0.9\%}$ and $^{+2.1\%}_{-1.1\%}$ in the ATLAS and CMS fiducial volumes, respectively, which are comparable to the corresponding charm-quark mass uncertainties of $^{+1.5\%}_{-1.9\%}$ and $^{+1.8\%}_{-2.2\%}$. The former uncertainties are traditionally used to estimate missing higher-order perturbative contributions and are expected to decrease as higher perturbative orders are included. However, we stress that, in analogy with the $\gamma\gamma\to Q\bar{Q}$ case studied in ref.~\cite{Capatti:2025khs}, such scale uncertainties may underestimate the true NNLO QCD corrections due to the (over)simplicity of the $\mu_R$ dependence at NLO.~\footnote{The two-loop amplitudes depend on 
$\mu_R$ only implicitly through the couplings $\alpha_s(\mu_R^2)$ and $\alpha(\mu_R^2)$.} In particular, it has been pointed out in ref.~\cite{Bargiela:2026tcn} that, in massless QCD, singlet three-loop Feynman-diagram contributions (cf. figure 3d therein)~\footnote{The ``singlet" contributions feature two closed quark
loops, each coupled to two photons and two gluons. This terminology follows ref.~\cite{Capatti:2025khs}, where the NNLO QCD corrections to $\gamma\gamma\to Q\bar{Q}$ are computed using the newly proposed Local Unitarity approach~\cite{Capatti:2020xjc,Capatti:2021bsm,Capatti:2022tit}.} give rise to sizeable NNLO corrections due to the enhancement factor $(\sum_q{Q_q^2})^2 / (\sum_q{Q_q^4})$. Therefore, a full quark-mass-dependent calculation of the NNLO QCD corrections to LbL scattering at the three-loop level would be highly desirable for achieving percent-level theoretical precision, although it is clearly beyond the scope of the present work.~\footnote{The massless three-loop QCD calculation of ref.~\cite{Bargiela:2026tcn} shows that the NNLO QCD corrections enhance the LO ATLAS fiducial cross section by less than $2\%$, which is consistent with our scale-uncertainty estimate. However, as we showed in ref.~\cite{AH:2023kor}, the two-loop calculation based on massless quarks significantly underestimates the true $K$ factor obtained from the full mass-dependent two-loop calculation. We expect a similar behaviour at three loops. In particular, the higher powers of subleading Sudakov logarithms that arise at three loops are expected to lead to an even slower convergence of the full mass-dependent three-loop result towards the corresponding massless result. Therefore, it is unclear whether the known massless three-loop QCD result can be used to reliably estimate the missing full mass-dependent three-loop QCD corrections.} On the other hand, the charm-quark-mass uncertainties in the $K$ factors are subdominant compared to the scale uncertainties because much of the charm-mass dependence cancels in the cross-section ratios.

\begin{figure}[hbt!]
\includegraphics[width=0.50\columnwidth,draft=false]{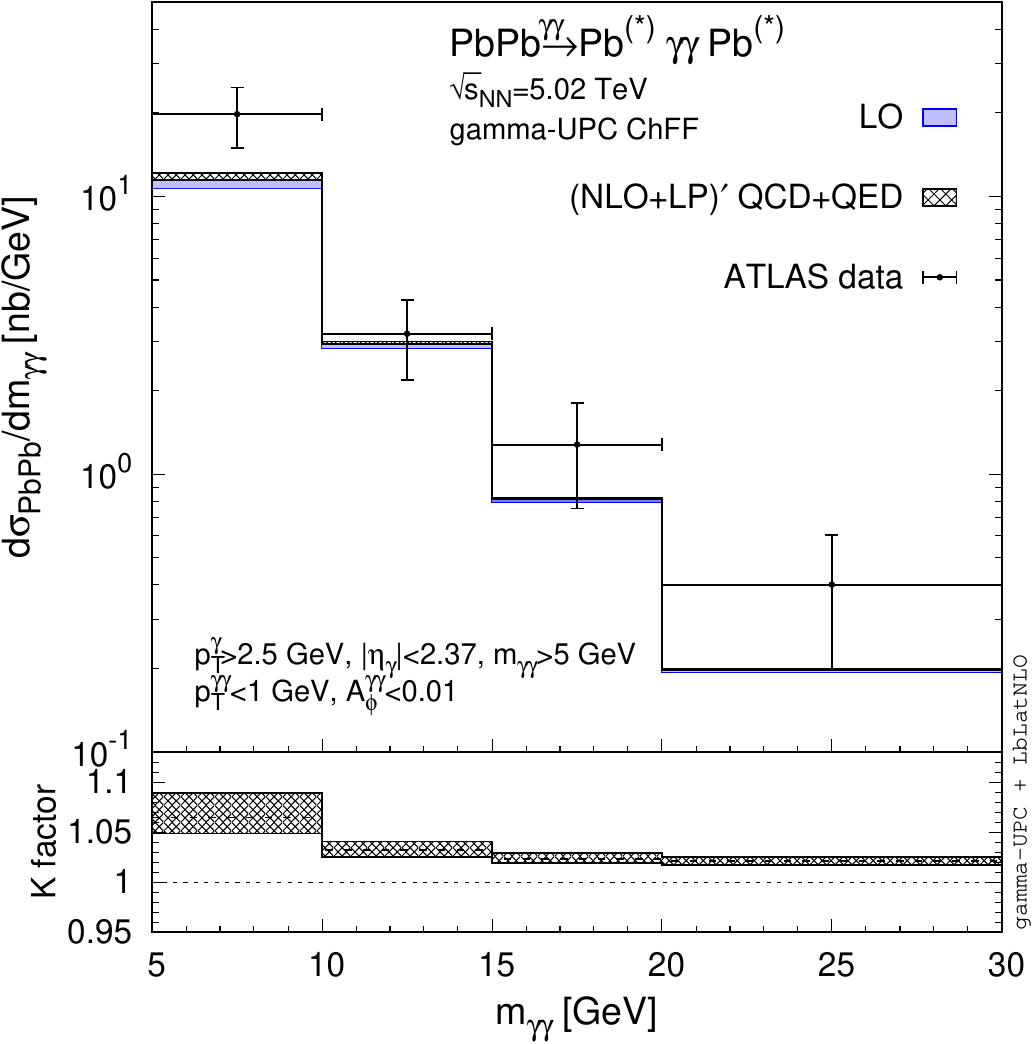}
\includegraphics[width=0.50\columnwidth,draft=false]{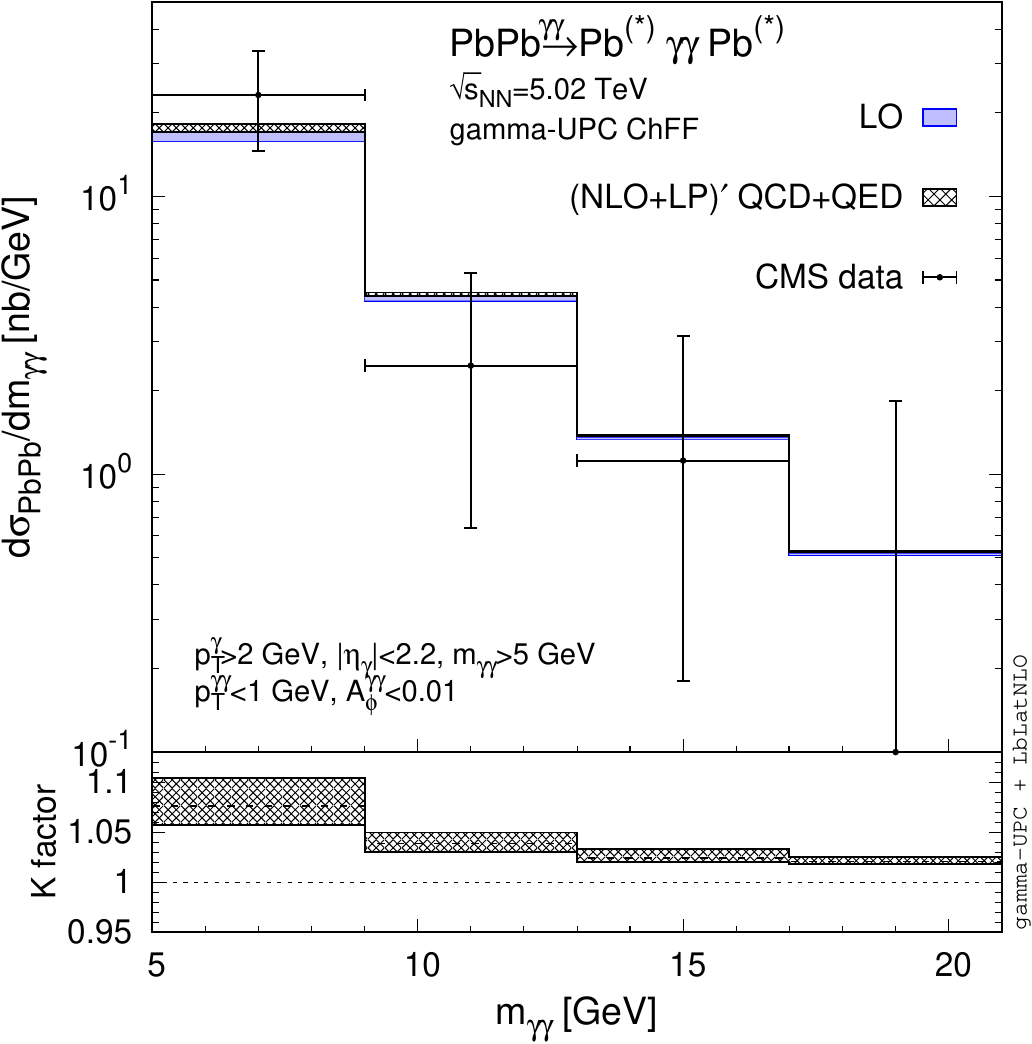}
\caption{Di-photon invariant mass distributions in the ATLAS (left panel) and CMS (right panel) fiducial phase space (see table~\ref{tab:LbLUPCcuts}). The experimental data (error bars) are compared with the LO (blue band) and (NLO+LP)$^\prime$ QCD+QED (black hatched) predictions. The theoretical error bands represent the combined charm-quark mass and scale uncertainties in quadrature. The differential $K$ factors are displayed in the lower panels.}
\label{figLbLPbPbdsigmadM} \vspace*{-0.5cm}
\end{figure}

We now discuss differential distributions. The first observable of particular interest is the di-photon invariant mass $m_{\gamma\gamma}$, whose distributions in the ATLAS and CMS fiducial regions are shown in figure~\ref{figLbLPbPbdsigmadM}. Potential resonances, such as C-even bottomonia, the $X(6900)$~\cite{LHCb:2020bwg,ATLAS:2023bft,CMS:2023owd}, or even more interesting new BSM particles, could manifest themselves as Breit-Wigner-like structures in the $m_{\gamma\gamma}$ spectrum. In fact, a small excess observed in the ATLAS measurement with respect to the LO theoretical prediction in the bin $m_{\gamma\gamma}\in [5,10]$ GeV, as shown in the left panel of figure~\ref{figLbLPbPbdsigmadM}, has motivated studies of the impact of C-even bottomonia and the $X(6900)$ fully-charmed tetraquark state in the literature~\cite{Krintiras:2023axs,Biloshytskyi:2022dmo}. However, using the \helaconia\ event generator~\cite{Shao:2012iz,Shao:2015vga}, it has been shown that the contribution from these hadronic resonances is negligible compared to the continuum LbL process, as illustrated in figure~3 of ref.~\cite{AH:2023kor}. A caveat is that the partial widths of some hadrons used in that analysis are taken from model calculations rather than experimental measurements. The inclusion of higher-order QCD and QED corrections reduces the discrepancy slightly, but not sufficiently to eliminate it. On the other hand, the comparison between theoretical predictions and the CMS measurement, shown in the right panel of figure~\ref{figLbLPbPbdsigmadM}, does not reveal any tension. The $K$ factors, defined as the ratios of (NLO+LP)$^\prime$ QCD+QED to LO results and displayed in the lower panels, indicate that the size of the quantum corrections (mainly NLO QCD corrections) decreases with increasing $m_{\gamma\gamma}$. The relative correction decreases from $+6.4\%^{+2.4\%}_{-1.6\%}$ ($+7.6\%^{+2.9\%}_{-1.8\%}$) in the lowest $m_{\gamma\gamma}$ bin to $+2.1\%^{+0.4\%}_{-0.4\%}$ ($+2.1\%^{+0.4\%}_{-0.3\%}$) in the highest bin for ATLAS (CMS), where the charm-quark-mass and scale uncertainties are summed in quadrature. This trend is qualitatively consistent with the scale evolution of $\alpha_s$, the decreasing importance of the charm-quark mass, and the observed behaviour of the amplitudes.

\begin{figure}[hbt!]
\includegraphics[width=0.50\columnwidth,draft=false]{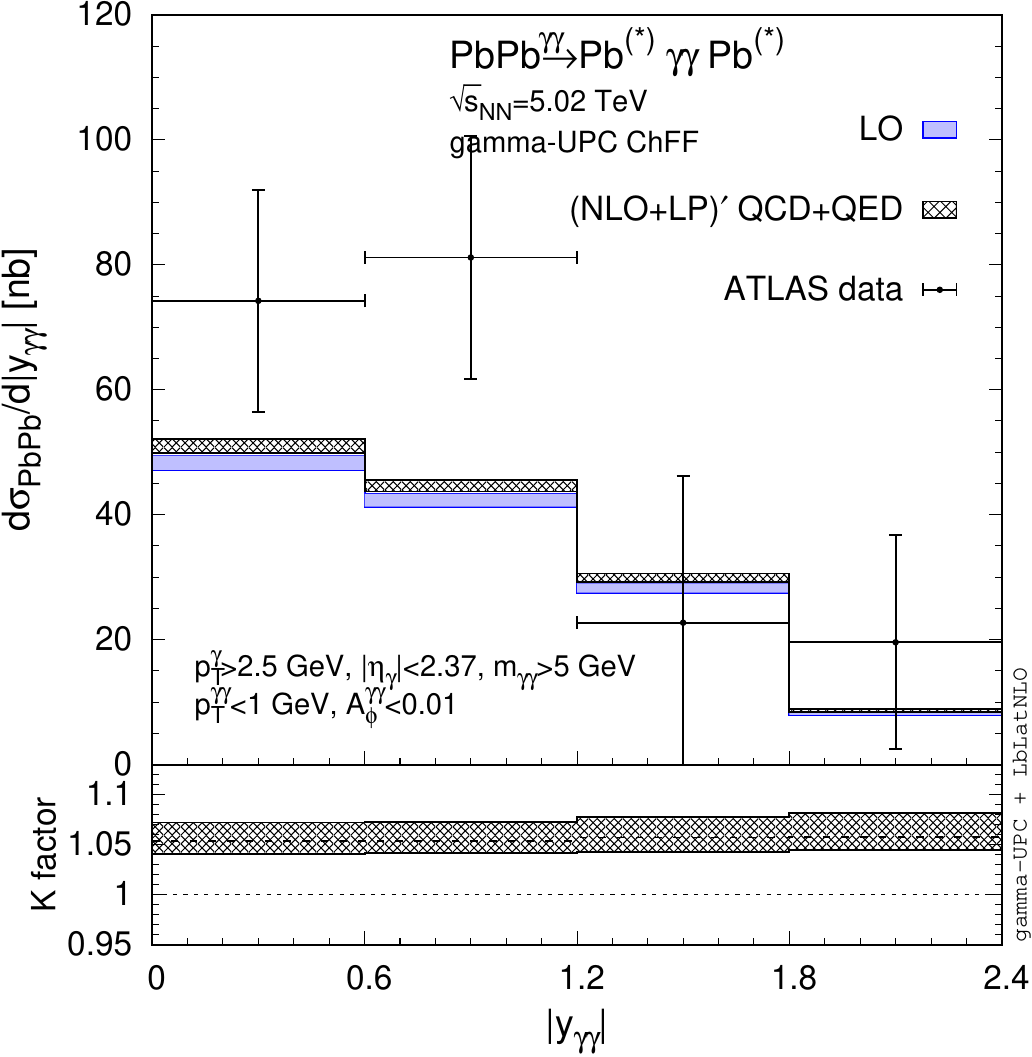}
\includegraphics[width=0.50\columnwidth,draft=false]{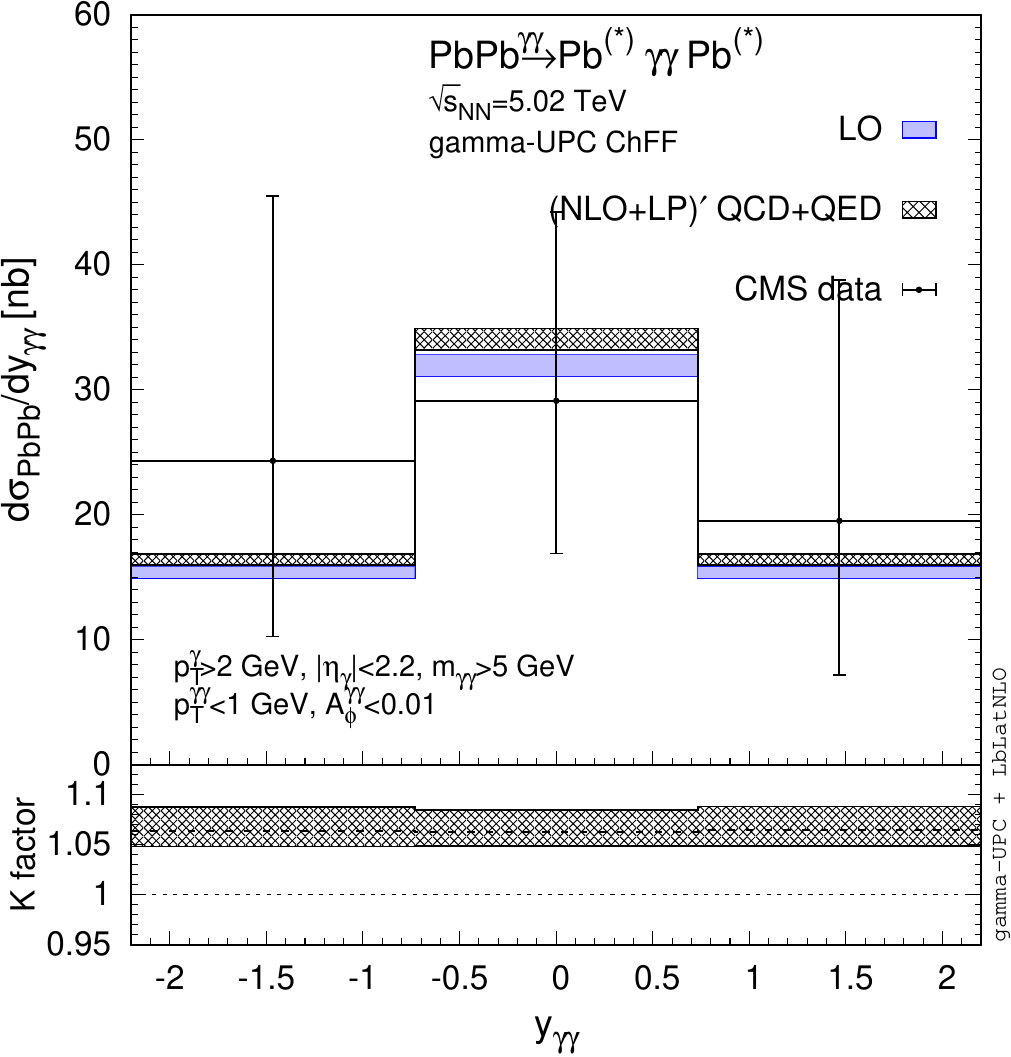}
\caption{Di-photon rapidity distributions in the ATLAS (left panel) and CMS (right panel) fiducial phase space (see table~\ref{tab:LbLUPCcuts}). The experimental data (error bars) are compared with the LO (blue band) and (NLO+LP)$^\prime$ QCD+QED (black hatched) predictions. The theoretical error bands represent the combined charm-quark mass and scale uncertainties in quadrature. The differential $K$ factors are displayed in the lower panels.}
\label{figLbLPbPbdsigmady} \vspace*{-0.5cm}
\end{figure}

The di-photon rapidity $y_{\gamma\gamma}$ distributions shown in figure~\ref{figLbLPbPbdsigmady} are closely related to the photon-photon flux function $\mathcal{L}_{\gamma\gamma}^{(\mathrm{Pb}\mathrm{Pb})}(x_1,x_2)$ defined in eq.~\eqref{eq:LgginUPC}. In the strict collinear factorisation limit, without performing initial $k_\perp$ smearing, one simply has $y_{\gamma\gamma}=\log{(x_1/x_2)}/2$. Therefore, the theory-data comparison for this observable can, to some extent, test our photon-flux modelling and potentially indicate where improvements are needed. While our calculation agrees well with the CMS measurement (right panel of figure~\ref{figLbLPbPbdsigmady}), a small tension is observed when comparing our theoretical predictions with the ATLAS data for $|y_{\gamma\gamma}|<1.2$, as can be seen in the left panel. The $K$ factors show only a very weak dependence on $y_{\gamma\gamma}$ with a very small increasing trend at larger $|y_{\gamma\gamma}|$.

\begin{figure}[hbt!]
\includegraphics[width=0.50\columnwidth,draft=false]{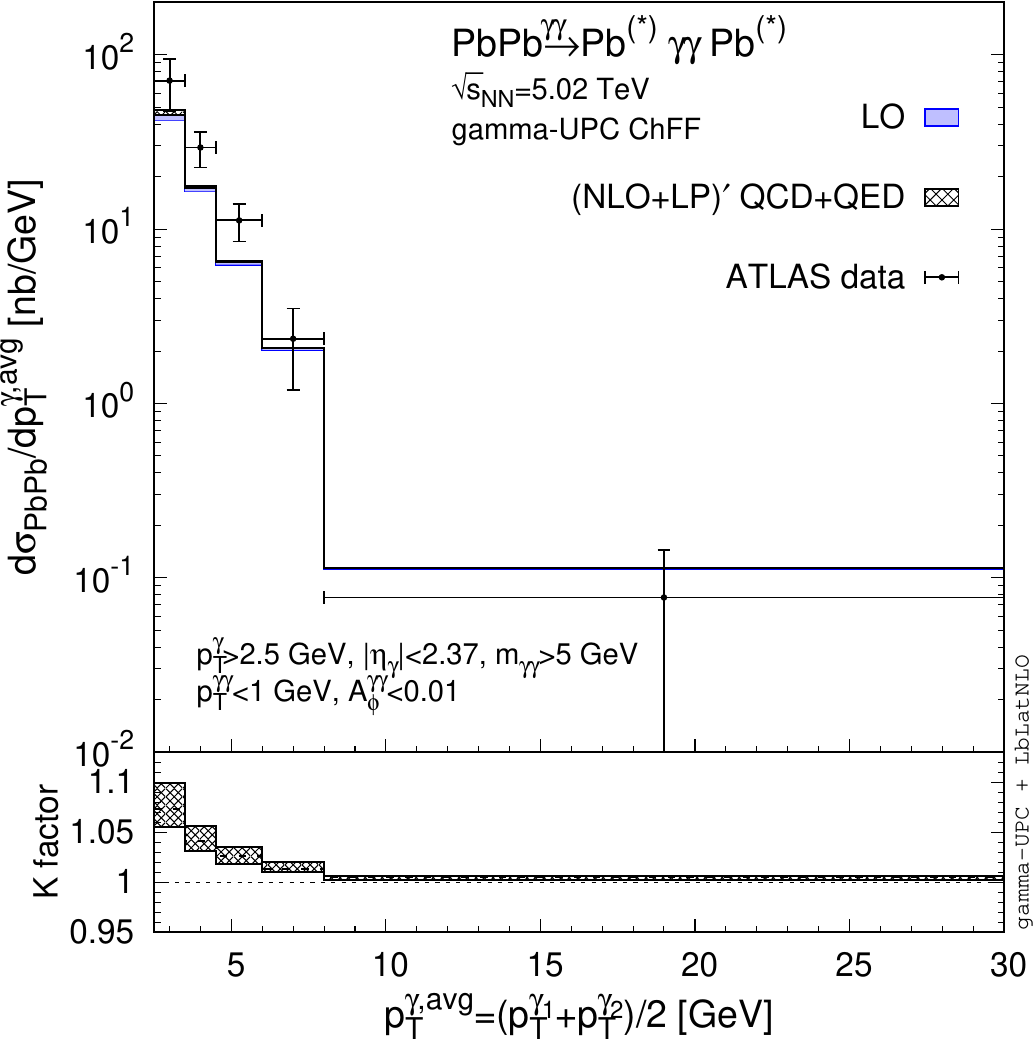}
\includegraphics[width=0.50\columnwidth,draft=false]{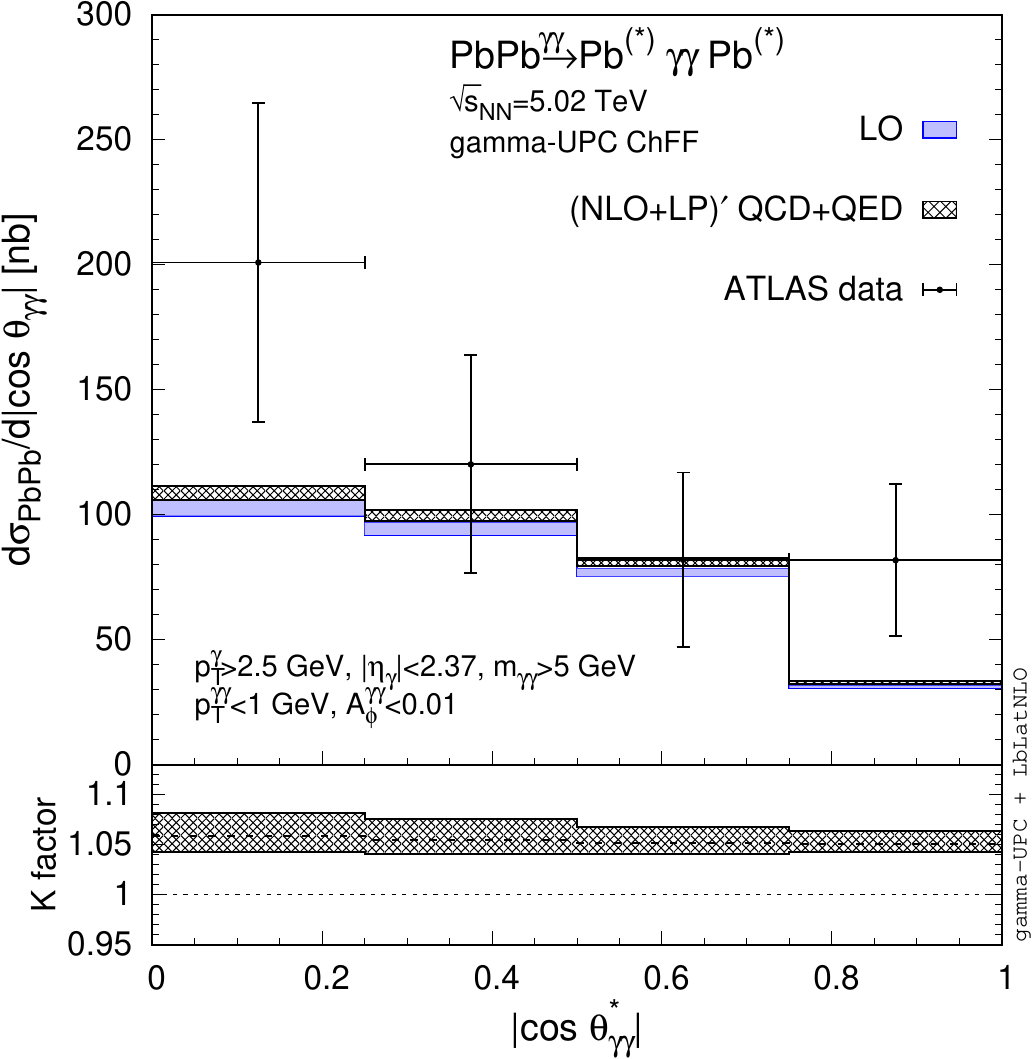}
\caption{Average photon transverse momentum (left panel) and the absolute cosine of the di-photon scattering angle, $|\cos{\theta^\star_{\gamma\gamma}}|$ (right panel), distributions in the ATLAS fiducial phase space (see table~\ref{tab:LbLUPCcuts}). The experimental data (error bars) are compared with the LO (blue band) and (NLO+LP)$^\prime$ QCD+QED (black hatched) predictions. The theoretical error bands represent the combined charm-quark mass and scale uncertainties in quadrature. The differential $K$ factors are shown in the lower panels.}
\label{figLbLPbPbdsigmadpTdcosth} \vspace*{-0.5cm}
\end{figure}

The ATLAS experiment~\cite{ATLAS:2020hii} additionally measures two particle-level observables: the average photon transverse momentum $p_T^{\gamma,\mathrm{avg}}\equiv (p_T^{\gamma_1}+p_T^{\gamma_2})/2$ (left panel of figure~\ref{figLbLPbPbdsigmadpTdcosth}) and the absolute cosine of the di-photon scattering angle in the $\gamma\gamma$ centre-of-mass frame, $|\cos{\theta^\star_{\gamma\gamma}}|$ (right panel of figure~\ref{figLbLPbPbdsigmadpTdcosth}). The latter can also be expressed in terms of the rapidity gap between the final-state photons, $\Delta y_{\gamma_1,\gamma_2}\equiv y_{\gamma_1}-y_{\gamma_2}$, as
\begin{equation}
\cos{\theta^\star_{\gamma\gamma}}=\tanh{\left(\frac{\Delta y_{\gamma_1,\gamma_2}}{2}\right)}\,.
\end{equation}
In these two observables, we again observe small excesses in the range $3.5~\mathrm{GeV}<p_T^{\gamma,\mathrm{avg}}<6~\mathrm{GeV}$ and in the first and last bins of $|\cos{\theta^\star_{\gamma\gamma}}|$. It will be interesting to determine, with more accumulated data in the future, whether these discrepancies arise simply from statistical fluctuations or reflect a genuine effect. The latter possibility would certainly be more intriguing, as it could potentially point to new physics or missing ingredients in the theoretical description. Regarding the $K$ factors, they show a strong (weak) dependence on $p_T^{\gamma,\mathrm{avg}}$ ($|\cos{\theta^\star_{\gamma\gamma}}|$) and decrease with increasing $p_T^{\gamma,\mathrm{avg}}$ and $|\cos{\theta^\star_{\gamma\gamma}}|$. In the $p_T^{\gamma,\mathrm{avg}}$ distribution, the $K$ factor varies from $1.073_{-0.011-0.014}^{+0.008+0.026}$ in the lowest bin to $1.005^{+0.001+0.001}_{-0.002-0.001}$ in the highest bin, where the quoted theoretical uncertainties stem from charm-quark mass and scale variations. For $|\cos\theta^\star_{\gamma\gamma}|$, it ranges from $1.059^{+0.009+0.021}_{-0.012-0.011}$ to $1.050^{+0.004+0.013}_{-0.001-0.008}$.

\subsection{Light-by-light scattering at lepton colliders\label{sec:LbLee}}

It was proposed as early as the 1970s in refs.~\cite{Baier:1975qqg,Baier:1980kx} to study LbL scattering at $e^-e^+$ colliders. However, direct experimental measurements of LbL scattering have essentially not been performed at $e^-e^+$ colliders, although related two-photon processes have been studied extensively~\cite{Jikia:1993tc,Gounaris:1998qk,ECFADESYPhotonColliderWorkingGroup:2001ikq,Ellis:2022uxv,dEnterria:2022ysg,dEnterria:2023yao}. From a theoretical point of view, it is largely immaterial whether the LbL scattering process is studied at hadron or lepton colliders; however, the experimental environments, particularly the background processes, can be radically different. Some challenges associated with backgrounds in measuring LbL scattering at $e^-e^+$ colliders have been discussed recently in ref.~\cite{Beloborodov:2023ajj}. Compared to LbL measurements in UPCs at hadron colliders, backgrounds from processes such as $e^-e^+\to \gamma\gamma (\gamma)$ and misidentified events are much larger at $e^-e^+$ colliders. In particular, in UPCs the signal can be selected by tagging intact protons or ions (modulo ion excitation), since many background processes involve projectile dissociation. No analogous tagging handle exists for the forward electrons and positrons at $e^-e^+$ colliders. In addition, the relatively low luminosity of LEP was insufficient for a clean measurement. Nevertheless, given that a high-energy electron-positron collider has been prioritised for studying electroweak, Higgs boson, and top-quark physics in the post-LHC era of particle physics,~\footnote{\href{https://europeanstrategy.cern/}{https://europeanstrategy.cern/}}
 we present cross-section predictions for LbL scattering at $e^-e^+$ colliders in this section.

\begin{table}[htpb!]
\centering
\tabcolsep=3.5mm
\vspace{0.2cm}
\begin{tabular}{c|c|c|c} \hline
Process: $\gamma\gamma\to \gamma\gamma$ & \multicolumn{3}{c}{\lblatnlo} \\
$\sqrt{s_{e^-e^+}}$ [GeV] & $\sigma_{e^-e^+}^{\mathrm{LO}}$ [fb] & $\sigma_{e^-e^+}^{\mathrm{NLO}\ \mathrm{QCD}}$ [fb] & $\sigma_{e^-e^+}^{(\mathrm{NLO+LP})^\prime\ \mathrm{QCD+QED}}$ [fb] \\ \hline
$10.58$ &  $0.989^{+0.036}_{-0.038}$ &  $1.063^{+0.028+0.028}_{-0.033-0.015}$ & $1.070^{+0.028+0.030}_{-0.033-0.016}$ \\
$90$ &  $19.5^{+0.5}_{-0.5}$ &  $20.6^{+0.4+0.4}_{-0.4-0.2}$ & $20.7^{+0.4+0.4}_{-0.4-0.2}$ \\
$160$ &  $25.1^{+0.6}_{-0.7}$ &  $26.4^{+0.4+0.5}_{-0.5-0.2}$ & $26.5^{+0.4+0.5}_{-0.5-0.3}$ \\
$240$ &  $28.8^{+0.7}_{-0.7}$ &  $30.3^{+0.5+0.5}_{-0.6-0.3}$ & $30.5^{+0.5+0.6}_{-0.6-0.3}$ \\
$250$ &  $29.2^{+0.7}_{-0.7}$ &  $30.7^{+0.5+0.5}_{-0.6-0.3}$ & $30.9^{+0.5+0.6}_{-0.6-0.3}$ \\
$365$ &  $32.6^{+0.8}_{-0.8}$ &  $34.2^{+0.6+0.6}_{-0.7-0.3}$ & $34.4^{+0.6+0.6}_{-0.7-0.3}$ \\
\hline
 $\sqrt{s_{e^-e^+}}$ [GeV] & & {\large $\sigma_{e^-e^+}^{\mathrm{NLO}\ \mathrm{QCD}}/\sigma_{e^-e^+}^{\mathrm{LO}}$} & $\sigma_{e^-e^+}^{(\mathrm{NLO+LP})^\prime\ \mathrm{QCD+QED}}/\sigma_{e^-e^+}^{\mathrm{LO}}$ \\ \hline
$10.58$ & & $1.075_{-0.011-0.015}^{+0.008+0.028}$ & $1.082_{-0.011-0.016}^{+0.008+0.031}$ \\
$90$ &  &  $1.054_{-0.008-0.010}^{+0.007+0.019}$ & $1.060_{-0.008-0.011}^{+0.007+0.020}$ \\
$160$ &  &  $1.052_{-0.008-0.010}^{+0.006+0.018}$ & $1.058_{-0.008-0.010}^{+0.007+0.020}$ \\
$240$ &  &  $1.051_{-0.008-0.010}^{+0.006+0.018}$ & $1.057_{-0.008-0.010}^{+0.006+0.019}$ \\
$250$ &  &  $1.051_{-0.008-0.010}^{+0.006+0.018}$ & $1.057_{-0.008-0.010}^{+0.006+0.019}$ \\
$365$ &  &  $1.051_{-0.008-0.010}^{+0.006+0.018}$ & $1.056_{-0.008-0.010}^{+0.006+0.019}$ \\
\hline
\end{tabular}
\caption{Fiducial exclusive LbL cross sections at LO, NLO QCD, and (NLO+LP)$^\prime$ QCD+QED accuracies, obtained with \lblatnlo\ using the iWW $\gamma$ flux at $e^-e^+$ colliders within the phase space defined in eq.~\eqref{eq:cuts4ee}. The first (second) quoted theoretical uncertainties stem from variations of the charm-quark mass (scale).\label{tab:eexs}
}
\end{table}

\begin{figure}[hbt!]
\includegraphics[width=0.95\columnwidth,draft=false]{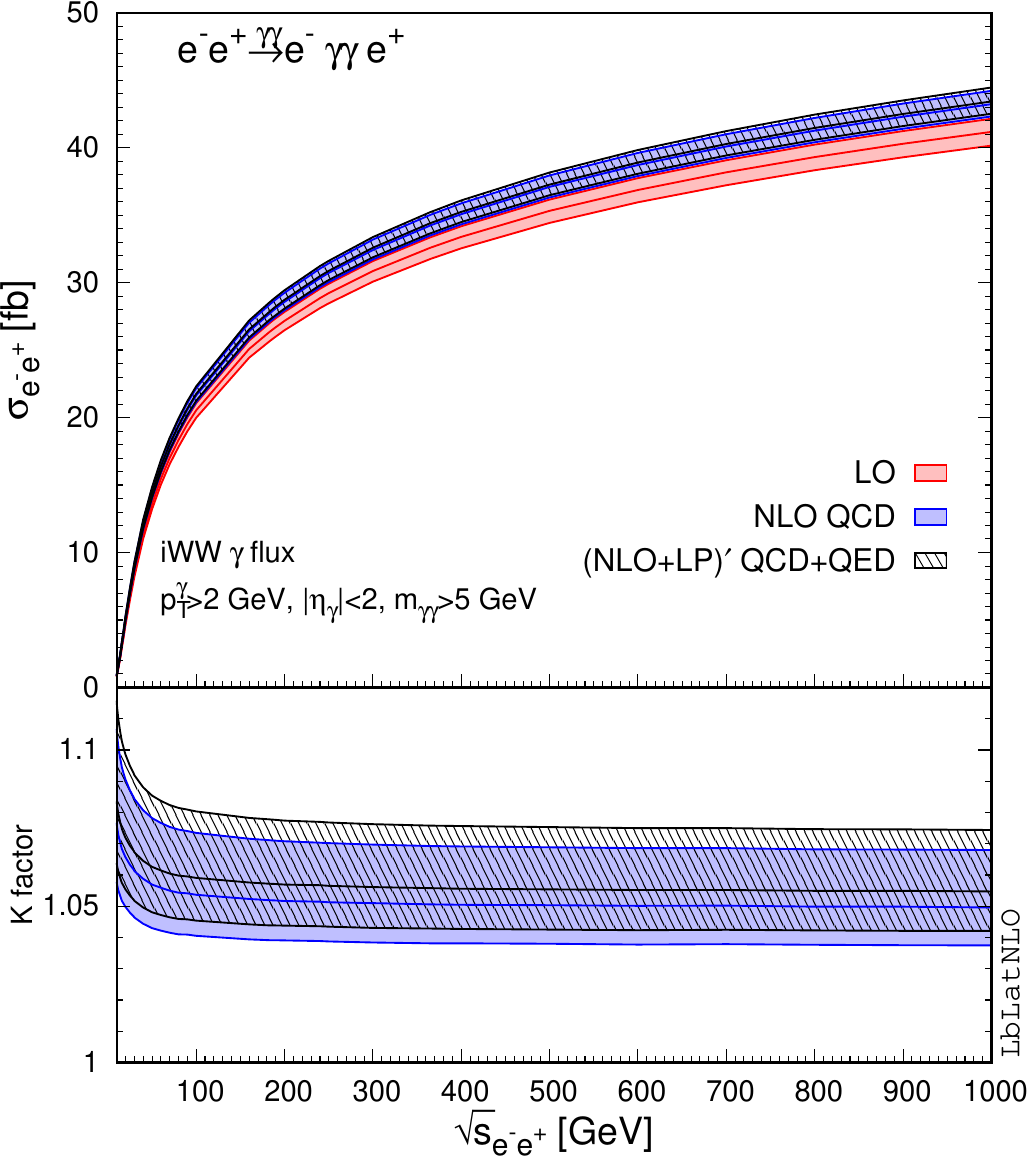}
\caption{Phase-space-integrated fiducial LbL cross section as a function of the centre-of-mass energy $\sqrt{s_{e^-e^+}}$ at $e^-e^+$ colliders, within the phase space defined in eq.~\eqref{eq:cuts4ee}. The theoretical predictions are shown at LO (red band), NLO QCD (blue band), and $(\text{NLO}+\text{LP})^\prime$ QCD+QED (black hatched) accuracy, obtained with \lblatnlo\ using the iWW $\gamma$ flux. The quoted theoretical uncertainties arise from charm-quark mass and scale variations and are combined in quadrature.}
\label{figLbLeexs} \vspace*{-0.5cm}
\end{figure}

As a showcase, we consider the following fiducial cuts for final-state photons in the centre-of-mass frame of $e^-e^+$:
\begin{equation}
m_{\gamma\gamma}>5~\mathrm{GeV}\,, \quad p_T^{\gamma}>2~\mathrm{GeV}\,,\quad |\eta_\gamma|<2\,,\label{eq:cuts4ee}
\end{equation}
where the pseudorapidity cut corresponds to restricting the photon scattering angle to the range $15.4146^\circ\lesssim\theta_\gamma\lesssim164.585^\circ$ with respect to the beam axis. We present the phase-space-integrated fiducial cross sections in table~\ref{tab:eexs} at the Belle II~\cite{Belle-II:2018jsg} energy $\sqrt{s_{e^-e^+}}=10.58$ GeV, as well as at future circular $e^-e^+$ colliders such as FCC-ee~\cite{FCC:2018evy} and CEPC~\cite{CEPCStudyGroup:2023quu}. The LbL fiducial cross section over a wider range $\sqrt{s_{e^-e^+}}\in [10,1000]$ GeV is shown in figure~\ref{figLbLeexs}. Within the fiducial region defined in eq.~\eqref{eq:cuts4ee}, the LbL cross section at $\sqrt{s_{e^-e^+}}=10.58$ GeV is of order $1$ fb, which is sizeable in view of the large integrated luminosity expected at Belle II. Owing to the increase in the photon-photon flux with $\sqrt{s_{e^-e^+}}$, the fiducial cross section becomes approximately 20 times larger at the $Z$-pole ($\sqrt{s_{e^-e^+}}\approx 90$ GeV) and about 30 times larger at energies relevant for $ZH$ associated production ($\sqrt{s_{e^-e^+}}=240$-$250$ GeV). The observability of LbL scattering at these energies depends on both the integrated luminosity and the control of backgrounds. A detailed experimental assessment, however, is beyond the scope of this work.

The NLO QCD corrections relative to LO decrease mildly with the centre-of-mass energy $\sqrt{s_{e^-e^+}}$, ranging from $+7.5\%^{+0.8\%+2.8\%}_{-1.1\%-1.5\%}$ at $\sqrt{s_{e^-e^+}}=10.58$ GeV to $+5.1\%^{+0.6\%+1.8\%}_{-0.8\%-1.0\%}$ at the top-quark pair production energy ($\sqrt{s_{e^-e^+}}=365$ GeV). The quoted uncertainties arise from charm-quark mass and scale variations. Our best predictions at (NLO+LP)$^\prime$ QCD+QED increase the LO cross sections by $8.2\%^{+0.8\%+3.1\%}_{-1.1\%-1.6\%}$ at low energies and \\
$5.6\%^{+0.6\%+1.9\%}_{-0.8\%-1.0\%}$ at higher energies. Therefore, achieving a target theoretical precision of $10\%$ requires the inclusion of these higher-order corrections.

We stress again that, since our predictions for $e^-e^+$ colliders rely on a simplified photon flux (cf. eq.~\eqref{eq:WWPDF4lepton}), the absolute cross sections presented here should be regarded as illustrative. The corresponding $K$ factors, however, are largely independent of the choice of photon flux and therefore provide state-of-the-art precision predictions. The public \lblatnlo\ code (cf. appendix~\ref{sec:lblatnlo}) also allows users to replace the default photon flux with alternative choices.

\section{Conclusions}
\label{sec:conclusions}

In this work, we have performed a detailed analysis of LbL scattering, combining analytic, numerical, and phenomenological aspects. In particular, we present asymptotic expansions of the one- and two-loop helicity amplitudes in both the low- and high-energy regimes, and demonstrate that these expansions significantly improve the numerical stability of full fermion-mass-dependent calculations. This improvement is especially important in regions where exact evaluations suffer from strong cancellations or require very high numerical precision, thereby highlighting the practical importance of compact analytic amplitudes.

In the threshold region, we have shown that fixed-order NLO predictions are affected by Coulomb singularities. By implementing Coulomb resummation, we obtain improved and well-behaved results for both two-loop helicity amplitudes and NLO cross sections, thereby removing the logarithmic enhancements that appear near threshold and ensuring reliable predictions in this regime, even though the overall impact of Coulomb resummation becomes marginal after integration over wide di-photon invariant-mass bins.

We also present phenomenological results for LbL scattering across a wide range of energies and collider setups. Our analysis shows that NLO QCD corrections can enhance the LO cross section by $5\%$ to $15\%$, while NLO QED corrections and higher-order effects beyond NLO, arising from both the squared two-loop amplitudes and Coulomb resummation, remain relatively small. The behaviour of the $K$ factor across different energy scales further illustrates the interplay between fermion thresholds, $W^\pm$ boson contributions, and asymptotic limits, with the $W^\pm$-loop contribution included only at one-loop (LO) accuracy.

Finally, we make our results available through the public event generator \lblatnlo\ (see appendix~\ref{sec:lblatnlo}), which allows for the computation of differential fiducial cross sections and the generation of helicity-dependent unweighted events. This tool enables the direct application of our calculations to experimental analyses, including studies of ultraperipheral heavy-ion collisions and future lepton colliders. 

In summary, compared to our previous studies in refs.~\cite{AH:2023ewe,AH:2023kor}, this work presents several new developments:
\begin{itemize}
\item analytic asymptotic LE and HE expansions of the one- and two-loop helicity amplitudes (see section \ref{sec:asymexpofamp} and appendix \ref{sec:oneloopHE});
\item Coulomb gluon and photon resummation in the threshold region (see section \ref{sec:Coulombres});
\item the release of a new event generator \lblatnlo\  (see appendix \ref{sec:lblatnlo}).
\end{itemize}
Taken together, these developments provide a coherent framework that improves both the theoretical understanding and the practical computation of LbL scattering, thereby reinforcing its role as a precision test of QED within the SM and as a sensitive probe of BSM physics.

Looking ahead, achieving percent-level accuracy in cross-section predictions will require including NLO electroweak and NNLO QCD corrections. NLO electroweak effects are particularly important for $\sqrt{s} \gtrsim 2 m_W$, where only LO predictions exist for the $W^\pm$ loop contribution. This regime is phenomenologically relevant for ongoing LbL measurements in \pp\ UPCs at the LHC. Three-loop NNLO QCD corrections are also of interest, motivated by the observed enhancement from singlet contributions in the massless case~\cite{Bargiela:2026tcn}, while the size of the singlet contribution in the massive case remains to be assessed. Additionally, pursuing the resummation of subleading-power Sudakov logarithms in LbL scattering may enhance our understanding of such logarithms in loop-induced processes. Finally, it would be intriguing to construct the LE Euler-Heisenberg effective Lagrangian at higher orders in both $m_f^{-2}$ and coupling expansions based on our LE-expanded results, which may be relevant for laser-based experiments. Beyond these perturbative results, a reliable theoretical description of LbL scattering in the non-perturbative hadronic regime ($134~\mathrm{MeV} \lesssim \sqrt{s} \lesssim 4~\mathrm{GeV}$) remains an open question.

\section*{Acknowledgements}
We are grateful to Mathijs Fraaije and Valentin Hirschi for their valuable contributions in the early stages of this project. HSS gratefully acknowledges valuable discussions with Vasily Sotnikov on \texttt{PentagonFunctions++}, with Guoxing Wang on Coulomb resummation, and with Zeno Capatti and Valentin Hirschi on the Local Unitarity method. This work is supported by grants from the ERC (grant 101041109 ``BOSON", grant 101043686 ``LoCoMotive"), the French ANR (grant ANR-20-CE31-0015 ``PrecisOnium"), and the UK Royal Society University Research Fellowship (URF/R1/201268, URF/R/251034). Views and opinions expressed are however those of the authors only and do not necessarily reflect those of the European Union or the European Research Council Executive Agency. Neither the European Union nor the granting authority can be held responsible for them. For the purpose of Open Access, a CC-BY public copyright license has been applied by the authors to the present document and will be applied to all subsequent versions up to the Author Accepted Manuscript arising from this submission.

\newpage
\appendix

\section{High-energy expansion of one-loop amplitudes\label{sec:oneloopHE}}

In this section, we present the explicit coefficients of the HE expansion of the one-loop amplitudes in eqs.~\eqref{eq:Ml1HE} and \eqref{eq:Ml2HE} up to $\mathcal{O}(m_f^{10})$. Higher-order terms in the HE expansion are provided in the ancillary files.

For the one-loop fermionic contribution, the non-vanishing coefficients of the all-plus amplitude are given by
\begin{align}
\mathcal{M }^{(0,0,f,0,0)}_{++++}(x,y) =&\frac{1}{2}\,,\quad \mathcal{M}^{(0,0,f,2,1)}_{++++}(x,y)=-\frac{4 \Xb}{y}\,,\quad \mathcal{M}^{(0,0,f,2,0)}_{++++}(x,y)=-\frac{2 \Xb \Yb}{x y}\, \nonumber \\

\mathcal{M}^{(0,0,f,3,2)}_{++++}(x,y) =&-\frac{12}{x y}\,,\quad \mathcal{M}^{(0,0,f,3,1)}_{++++}(x,y)=\frac{8 \left(y^3+1\right) \Xb}{x^2 y^2}-\frac{36}{x y}\,,\nonumber\\
\mathcal{M}^{(0,0,f,3,0)}_{++++}(x,y)&=-\frac{4 \Xb\Yb}{x^2 y^2}
    -\frac{8 (y-1)^2\Xb}{xy^2}
    +\frac{36 \zeta_2+24}{x y}\,,\nonumber\\

\mathcal{M }^{(0,0,f,4,2)}_{++++}(x,y)=&-\frac{12 \left(2 x^5+1\right)}{x^3 y^3}\,,\quad \mathcal{M }^{(0,0,f,4,1)}_{++++}(x,y)=\frac{24 \left(y^5+1\right) \Xb}{x^3 y^3}-\frac{206 \left(1-x y\right)}{x^2 y^2}\,,
\nonumber\\

\mathcal{M }^{(0,0,f,4,0)}_{++++}(x,y)=&-\frac{12\Xb\Yb}{x^3 y^3}-\frac{4 \Xb}{x^2 y}\left(\frac{7 (1-x y)^2}{y^2}-\frac{27 (1-x y)}{y}+22\right)
   +\frac{72 \zeta_2 \left(x^4-1\right)}{x^2 y^3}\nonumber\\
   &+\frac{96 \left(1-x y\right)}{x^2 y^2}\,,
\nonumber\\

\mathcal{M }^{(0,0,f,5,2)}_{++++}(x,y)=&-\frac{40 \left(2 x^7+1\right)}{x^4 y^4}\,,\quad \mathcal{M }^{(0,0,f,5,1)}_{++++}(x,y)=\frac{80 \left(y^7+1\right) \overline{X}}{x^4 y^4}-\frac{1044 \left(1-xy\right)^2}{x^3 y^3}\,,\nonumber\\

\mathcal{M }^{(0,0,f,5,0)}_{++++}(x,y)=&\frac{240 \zeta _2 \left(x^5-2 y+1\right)}{x^2 y^4}+\frac{3248\left(1-xy \right)^2}{9 x^3 y^3}-\frac{40\overline{X}\overline{Y}}{x^4 y^4}\nonumber\\
&-\frac{8\overline{X} \left(37 y^6-67 y^5+82 y^4-92 y^3+82 y^2-67 y+37\right)}{3 x^3 y^4}\,.
\end{align}
The non-vanishing coefficients of the single-minus amplitude read
\begin{align}
\mathcal{M}^{(0,0,f,0,0)}_{-+++}(x,y) =&\frac{1}{2}\,,\quad \mathcal{M }^{(0,0,f,1,2)}_{-+++}(x,y)=\frac{1-x y}{2x y}\,,\quad \mathcal{M }^{(0,0,f,1,1)}_{-+++}(x,y)=\frac{2 \overline{X}}{x}\,,\nonumber\\

\mathcal{M }^{(0,0,f,1,0)}_{-+++}(x,y) =&\frac{\left(1-x y\right) \overline{X}^2}{x y}
    +\overline{X} \overline{Y}
    +\frac{3 \zeta _2 (1-xy)}{xy}\,,\nonumber\\

\mathcal{M }^{(0,0,f,2,1)}_{-+++}(x,y)=&\frac{2 \left(1-x y\right)^2}{x^2 y^2}\,,\quad \mathcal{M }^{(0,0,f,2,0)}_{-+++}(x,y)=\frac{4 \left(1-x y\right) \overline{X}}{x^2 y}\,,\nonumber\\ 

\mathcal{M }^{(0,0,f,3,2)}_{-+++}(x,y)=&\frac{6}{x y}\,,\quad \mathcal{M }^{(0,0,f,3,1)}_{-+++}(x,y)=\frac{3 (1-x y)^3}{x^3 y^3}+\frac{15}{x y}-\frac{4\left(y^3+1\right) \overline{X}}{x^2y^2}\,,\nonumber\\

\mathcal{M }^{(0,0,f,3,0)}_{-+++}(x,y)=&\frac{2 \overline{X}\overline{Y}}{x^2y^2}
     +\frac{2\overline{X} \left(3 y^4+y^3-y^2+y+3\right)}{x^3 y^2}-\frac{2 \left(x^3+y^3\right)}{x^3 y^3}-\frac{2 (x y+3)}{x y}
     -\frac{18 \zeta _2}{x y}\,,\nonumber\\

\mathcal{M }^{(0,0,f,4,2)}_{-+++}(x,y)=&\frac{8+16 x^5}{x^3 y^3}\,,\quad \mathcal{M }^{(0,0,f,4,1)}_{-+++}(x,y)=\frac{20 (1-x y)^4}{3 x^4 y^4}
   +\frac{124 (1-x y)}{x^2 y^2}
   -\frac{16 \left(y^5+1\right) \overline{X}}{x^3 y^3}\,,\nonumber\\

\mathcal{M }^{(0,0,f,4,0)}_{-+++}(x,y)=&\frac{8 \overline{X}\overline{Y}}{x^3 y^3}
    +\frac{6 (1-x y)^4}{x^4 y^4}+\frac{8  \overline{X}\left(8 y^6+2 y^5-3
   y^4+y^3-3 y^2+2 y+8\right)}{3 x^4 y^3}\nonumber\\
   &-\frac{2(1-xy)}{x^2 y^2}\left(60 \zeta_2+\frac{313}{9}\right)\,,\nonumber\\

\mathcal{M }^{(0,0,f,5,2)}_{-+++}(x,y)=&\frac{30+60 x^7}{x^4 y^4}\,,\quad \mathcal{M }^{(0,0,f,5,1)}_{-+++}(x,y)=-\frac{60 \left(y^7+1\right) \overline{X}}{x^4 y^4}
   +\frac{35 (1-x y)^5}{2 x^5 y^5}
   +\frac{1461 (1-x y)^2}{2 x^3 y^3}\,,\nonumber\\

\mathcal{M }^{(0,0,f,5,0)}_{-+++}(x,y)=&\frac{30\overline{X} \overline{Y}}{x^4 y^4}+\frac{\overline{X} \left(79 y^8+19 y^7-30y^6+10 y^5-5 y^4+10 y^3-30 y^2+19 y+79\right)}{x^5 y^4}\nonumber\\
&-\frac{70\left(1-x y\right)^2}{x^3 y^3}\left(9\zeta_2+\frac{13}{3}\right)+\frac{107 (1-x y)^5}{6 x^5 y^5}\,.

\end{align}
The non-vanishing coefficients of $\mathcal{M}_{--++}^{(0,0,f)}$ are
\begin{align}
\mathcal{M}^{(0,0,f,0,0)}_{--++}(x,y)=&
    {1\over 2}(1-2x y)\left(\overline{X}\overline{Y}
    -\overline{X}^2-3\zeta_2\right)
    -\overline{X}(x-y) -\frac{1}{2}\,,\nonumber\\

\mathcal{M}^{(0,0,f,1,2)}_{--++}(x,y)=&\frac{1}{xy}\,,\quad \mathcal{M}^{(0,0,f,1,1)}_{--++}(x,y)=\frac{2 \overline{X}}{x}+\frac{1}{x y}\,,\nonumber\\
\mathcal{M}^{(0,0,f,1,0)}_{--++}(x,y)=&2 \left(\overline{X}\overline{Y}-\overline{X}^2\right)
    +\frac{2 \overline{X}}{y}
    -\frac{3\zeta_2\left(1+2 xy\right)}{xy}
    -\frac{1}{x y}\,,\nonumber\\

\mathcal{M}^{(0,0,f,2,2)}_{--++}(x,y)=&\frac{3-4 x y}{x^2 y^2}\,,\quad \mathcal{M}^{(0,0,f,2,1)}_{--++}(x,y)=-\frac{\left(4 y^2+2\right) \overline{X}}{x^2 y^2}-\frac{8}{x}+\frac{18}{y^2}\,,\nonumber\\

\mathcal{M}^{(0,0,f,2,0)}_{--++}(x,y)=&\frac{\overline{X}\overline{Y}}{x^2 y^2}+\frac{\overline{X} \left(4 y^3+4y-3\right)}{x^2 y^2}+\frac{16 x y-11}{2 x^2 y^2}+\frac{6 \zeta_2 (2 y-3)}{y^2}\,,\nonumber\\

\mathcal{M}^{(0,0,f,3,2)}_{--++}(x,y)=&\frac{4}{x^3 y^3}+\frac{4 (2-x) x}{y^3}\,,\nonumber\\ \mathcal{M}^{(0,0,f,3,1)}_{--++}=&-\overline{X}\left(
    \frac{8 }{x^3 y^3}+\frac{4 (2-y) y}{x^3}\right)
    +\frac{2 \left(383 y^4+474 y^3+148 y^2-116 y-113\right)}{3 x^2 y^3}\,,\nonumber\\

\mathcal{M}^{(0,0,f,3,0)}_{--++}(x,y)=&\frac{4 \overline{X} \overline{Y}}{x^3 y^3}
    +\frac{2 \overline{X}\left(9 y^5-18 y^4+15 y^3-8 y^2+10 y-16\right)}{3 x^3 y^3}\nonumber\\
    &-\frac{12 \zeta_2 \left(5 x+\left(19 y^2+24 y+8\right)
   y^2\right)}{x^2 y^3}-\frac{2 \left(536 y^4+696 y^3+253 y^2-113 y-116\right)}{9 x^2 y^3}\,,\nonumber\\

\mathcal{M}^{(0,0,f,4,2)}_{--++}(x,y)=&\frac{15-2 x y}{x^4 y^4}+\frac{8 (3-2 x) x^2}{y^4}\,,\nonumber\\

\mathcal{M}^{(0,0,f,4,1)}_{--++}(x,y)=&\frac{2 \left(120 y^5+654 y^4+1642 y^3+1781y^2+724 y+444\right)}{3 x^2 y^4}+\overline{X}\left(\frac{4 x y-30}{x^4 y^4}+\frac{8 (2 y-3) y^2}{x^4}\right)\,,\nonumber\\

\mathcal{M}^{(0,0,f,4,0)}_{--++}(x,y)=&\frac{(15-2 x y) \overline{X}\overline{Y}}{x^4 y^4}+\frac{\overline{X}\left(128 y^7-264 y^6+192 y^5-104 y^4+72 y^3-116 y^2+130 y-237\right)}{6 x^4 y^4}\nonumber\\
&-\frac{6 \zeta_2 \left(8 y^5+52 y^4+148 y^3+162 y^2+60 y+35\right)}{x^2 y^4}\nonumber\\
&-\frac{16 y^5+1544 y^4+7022 y^3+7779 y^2+2256 y+1042}{18 x^2 y^4}\,,\nonumber\\

\mathcal{M}^{(0,0,f,5,2)}_{--++}(x,y)=&\frac{56-12 x y}{x^5 y^5}+\frac{20 (4-3 x) x^3}{y^5}\,,\nonumber\\

\mathcal{M}^{(0,0,f,5,1)}_{--++}(x,y)=&\overline{X}\left(\frac{8 (3 x y-14) }{x^5 y^5}-\frac{20 (3 x+7) y^3}{x^5}\right)\nonumber\\
&+\frac{1515 y^7+11025 y^6+91873 y^5+148707y^4+126247 y^3+61789 y^2+15405 y+5739}{5 x^3 y^5}\,,\nonumber\\

\mathcal{M}^{(0,0,f,5,0)}_{--++}(x,y)=&\frac{(56-12 x y)\overline{X} \overline{Y}}{x^5y^5}\nonumber\\
&-\frac{12 \zeta_2 \left(15 y^7+125 y^6+1191 y^5+1959 y^4+1669 y^3+798 y^2+175 y+63\right)}{x^3 y^5}\nonumber\\
&+\frac{\overline{X}\left(1185 y^9-2380 y^8+1650 y^7-900 y^6+625 y^5-480
   y^4\right)}{15 x^5 y^5}\nonumber\\
&+\frac{\overline{X}\left(740 y^3-1062 y^2+1098 y-2216\right)}{15 x^5 y^5}-\frac{-1200
   y^7+62800 y^6+1162722 y^5}{225 x^3 y^5}\nonumber\\
&+\frac{2019108 y^4+1751513 y^3+785661 y^2+99005 y+25896}{225 x^3 y^5}\,.
\end{align}
For $\mathcal{M}_{+-+-}^{(0,0,f)}$, we explicitly list the non-vanishing coefficients up to $\mathcal{O}(m_f^8)$,
\begin{align}
\mathcal{M}^{(0,0,f,0,0)}_{+-+-}(x,y)=&\left(\frac{x}{y^2}-\frac{1}{2}\right) \overline{X}^2+\frac{(y+2) \overline{X}}{y}+\zeta_2 \left(\frac{6 x}{y^2}-3\right)-1\,,\nonumber\\

\mathcal{M}^{(0,0,f,1,2)}_{+-+-}(x,y)=&\frac{2 y}{x}\,,\quad \mathcal{M}^{(0,0,f,1,1)}_{+-+-}(x,y)=\frac{2 \overline{X}}{x}-\frac{2 y }{x}\left(\overline{Y}-1\right)\,,\nonumber\\

\mathcal{M}^{(0,0,f,1,0)}_{+-+-}(x,y)=&\overline{X} \left(2-\frac{2\overline{Y}}{x}\right)-\frac{2 \overline{X}^2}{y}-\frac{6 \zeta_2\left(y^2+2 x\right)}{x y}-\frac{2 y}{x}\,,\nonumber\\

\mathcal{M}^{(0,0,f,2,2)}_{+-+-}(x,y)=&\frac{6 y^2-8 x}{x^2}\,,\quad \mathcal{M}^{(0,0,f,2,1)}_{+-+-}(x,y)=-\frac{2 \left(y^2+2\right) \overline{X}}{x^2}-\frac{4(1+x^2)\overline{Y}}{x^2}+\frac{18 y^2-28x}{x^2}\,,\nonumber\\

\mathcal{M}^{(0,0,f,2,0)}_{+-+-}(x,y)=&\frac{4 \overline{X} \overline{Y}}{x^2}+\frac{\overline{X} (4-3 y) y^2+4\overline{X}}{x^2y}-\frac{4 (y+2)^2 \overline{Y}}{x^2}+\frac{6 \zeta _2 \left(4 x-3y^2\right)}{x^2}+\frac{16 x-11 y^2}{x^2}\,,\nonumber\\

\mathcal{M}^{(0,0,f,3,2)}_{+-+-}(x,y)=&-\frac{4 \left(3 x^5+2 x^4+2 x-2 y^5+3\right)}{x^3 y^2}\,,\nonumber\\

\mathcal{M}^{(0,0,f,3,1)}_{+-+-}(x,y)=&\frac{\left(4-8 y^5-8 y\right) \overline{X}}{x^3 y^2}+\frac{4 \left(3 x^5+2 x^4+2 x+3\right) \overline{Y}}{x^3 y^2}\nonumber\\
&+\frac{226 y^4+578 y^3+980 y^2+804 y+402}{3 x^3 y}\,,\nonumber\\

\mathcal{M}^{(0,0,f,3,0)}_{+-+-}(x,y)=&\frac{(8 y-4) \overline{X} \overline{Y}}{x^3 y^2}-\frac{12 \zeta _2 \left(5 y^4+13 y^3+22 y^2+18 y+9\right)}{x^3y}\nonumber\\
&-\frac{2 \overline{X} \left(16 y^5-10 y^4+8 y^3-15 y^2+18y-9\right)}{3 x^3 y^2}-\frac{2 \overline{Y}\left(7 y^4+37 y^3+77 y^2+80 y+40\right) }{x^3 y}\nonumber\\
&-\frac{2 \left(116 y^4+319 y^3+535 y^2+432 y+216\right)}{9 x^3 y}\,,\nonumber \\

\mathcal{M}^{(0,0,f,4,2)}_{+-+-}(x,y)=&-\frac{2 \left(20 x^7+12 x^6+2 x \left(y^5+6\right)-15 y^7+20\right)}{x^4y^3}\,,\nonumber\\

\mathcal{M}^{(0,0,f,4,1)}_{+-+-}(x,y)=&\frac{\left(4 x\left(6+y^5\right)-30 y^7+40\right) \overline{X}}{x^4y^3}+\frac{8 \left(5 x^7+3 x^6+3 x+5\right) \overline{Y}}{x^4 y^3}\nonumber\\
&+\frac{4 \left(222y^6+806 y^5+2033 y^4+2928 y^3+2649 y^2+1422 y+474\right)}{3 x^4 y^2}\,,\nonumber\\

\mathcal{M}^{(0,0,f,4,0)}_{+-+-}(x,y)=&\frac{8 (3 y-2) \overline{X}\overline{Y}}{x^4 y^3}+\frac{\overline{X}\left(-237 y^7+130 y^6-116 y^5+72y^4-104 y^3+192 y^2-264 y+128\right)}{6 x^4 y^3}\nonumber\\
&-\frac{4 \left(37 y^6+268 y^5+828
   y^4+1416 y^3+1448 y^2+888 y+296\right) \overline{Y}}{3 x^4 y^2}\nonumber\\
&-\frac{30 \zeta_2
   \left(7 y^6+26 y^5+70 y^4+104 y^3+92 y^2+48 y+16\right)}{x^4 y^2}\nonumber\\
&-\frac{521 y^6+2170y^5+7202 y^4+11584 y^3+9592 y^2+4560 y+1520}{9 x^4 y^2}\,.

\end{align}

For the one-loop $W^\pm$ boson contribution, a decomposition analogous to eqs.~\eqref{eq:Ml1HE} and \eqref{eq:Ml2HE} applies, with $m_f$ replaced by $m_W$ and the prefactor
\begin{equation}
C^{(0,0,W)}=12\alpha^2\,.
\end{equation}
For the all-plus and single-minus helicity configurations, the $W^\pm$ amplitudes are simply related to the corresponding fermion-loop contributions:
\begin{equation}
\mathcal{M}_{++++}^{(0,0,W),\mathrm{HE}}=-\frac{3}{2N_{c,f}Q_f^4}
\mathcal{M}_{++++}^{(0,0,f),\mathrm{HE}}\,,\quad
\mathcal{M}_{-+++}^{(0,0,W),\mathrm{HE}}=-\frac{3}{2N_{c,f}Q_f^4}\mathcal{M}_{-+++}^{(0,0,f),\mathrm{HE}}\,.
\end{equation}
Consequently,
\begin{equation}
\mathcal{M}_{++++}^{(0,0,W,p,q)}(x,y)=\mathcal{M}_{++++}^{(0,0,f,p,q)}(x,y)\,,\quad \mathcal{M}_{-+++}^{(0,0,W,p,q)}(x,y)=\mathcal{M}_{-+++}^{(0,0,f,p,q)}(x,y)\,.
\end{equation}
For the two-minus-two-plus helicity configurations, the non-vanishing coefficients up to $\mathcal{O}(m_W^8)$ are given by
\begin{align}

\mathcal{M}^{(0,0,W,0,1)}_{--++}(x,y)=&-\frac{4 \overline{X}}{3 y}\,,\nonumber\\
\mathcal{M}^{(0,0,W,0,0)}_{--++}(x,y)=&-\frac{2\overline{X} \overline{Y}}{3 x y}+\overline{X} \left(y-x\right)+\frac{3 xy-4}{3}\left(\overline{X}^2-\overline{X} \overline{Y}+3\zeta_2\right)-\frac{1}{2}\,,\nonumber\\

\mathcal{M}^{(0,0,W,1,2)}_{--++}(x,y)=&-\frac{4}{3 x y}\,,\quad 
\mathcal{M}^{(0,0,W,1,1)}_{--++}(x,y)=\frac{8\left(1-y^2(1-x)\right) \overline{  X}}{3 x^2 y^2}-\frac{23}{3 x y}\,,\nonumber\\

\mathcal{M}^{(0,0,W,1,0)}_{--++}(x,y)=&-\frac{4}{3xy}\left(
    \frac{\overline{X}\overline{Y}}{xy}
    +\frac{2 \overline{X}}{y}
    -3\zeta_2
    -\frac{11}{4}
    \right)
    -2\left(\overline{X}^2- \overline{X}\overline{Y} + 3\zeta_2\right)-\frac{14\overline{X}}{3x}\,,\nonumber\\

\mathcal{M}^{(0,0,W,2,2)}_{--++}(x,y)=&\frac{6 (x y-2)}{x^2 y^2}\,,\quad \mathcal{M}^{(0,0,W,2,1)}_{--++}(x,y)=\frac{74 x y-129}{3 x^2y^2}
   +\frac{4 \left(-4 x y+y^5-2 y^4+3 y^3+6\right) \overline{  X}}{3 x^3y^3}\,,\nonumber\\

\mathcal{M}^{(0,0,W,2,0)}_{--++}(x,y)=&\frac{4(2 x y-3)\overline{X}\overline{Y}}{3 x^3y^3}+\frac{\overline{X} \left(
   4 y^4+32y^3-42 y^2+28 y-28\right)}{3 x^2 y^3}+\frac{18 \zeta_2 \left(2-x y\right)}{x^2y^2}-\frac{24 x y-109}{6 x^2 y^2}\,,\nonumber\\

\mathcal{M}^{(0,0,W,3,2)}_{--++}(x,y)=&-\frac{4 \left(19 x^2 y^2-83 x y+50\right)}{3 x^3 y^3}\,,\nonumber\\
\mathcal{M}^{(0,0,W,3,1)}_{--++}(x,y)=&\frac{8 \left(-8 x y+y^7-6 y^6+3 y^5+10\right) \overline{  X}}{3 x^4 y^4}-\frac{2 \left(1099-1851 x y+492 x^2 y^2\right)}{9x^3y^3}\,,\nonumber\\
\mathcal{M}^{(0,0,W,3,0)}_{--++}(x,y)=&-\frac{4\overline{X} \left(8 y^6-62 y^5+119 y^4-157 y^3+119 y^2-70 y+74\right)}{9 x^3 y^4}\nonumber\\
&+\frac{8 (4 x y-5)\overline{X}\overline{Y}}{3 x^4y^4}
   +\frac{4\zeta_2 \left(19 x^2 y^2-83 x y+50\right)}{x^3 y^3}
  +\frac{440 x^2 y^2-3709 x y+2395}{27 x^3 y^3}\,,\nonumber\\
\mathcal{M}^{(0,0,W,4,2)}_{--++}(x,y)=&\frac{10 \left(32 x^3 y^3-214 x^2 y^2+269 x y-98\right)}{3 x^4 y^4}\,,\nonumber\\
\mathcal{M}^{(0,0,W,4,1)}_{--++}(x,y)=&-\frac{-475 x^3 y^3+2722 x^2 y^2-3459 x y+1258}{x^4 y^4}\nonumber\\
&+\frac{4 \left(3x^2 \left(6 y^7+y^2\right)
    +20 x y \left(4 y^7-3\right)
    +70 \left(y^9+1\right)\right) \overline{X}}{3 x^5 y^5}\,,\nonumber\\

\mathcal{M}^{(0,0,W,4,0)}_{--++}(x,y)=&-\frac{2 \left(3 x^2 y^2-60 x y+70\right)\overline{X} \overline{Y}}{3 x^5 y^5}-\frac{10\zeta_2 \left(32 x^3 y^3-214 x^2 y^2+269 x y-98\right)}{x^4 y^4}\nonumber\\
&-\frac{134 y^8-910 y^7+1738 y^6-2250 y^5+2558 y^4-2207 y^3+1681 y^2-958 y+1066}{9 x^4 y^5}\overline{X}\nonumber\\
&-\frac{5440 x^3 y^3-85976 x^2 y^2+106360 x y-39589}{108 x^4 y^4}\,,
\end{align}
and
\begin{align}
\mathcal{M }^{(0,0,W,0,1)}_{+-+-}(x,y)=&-\frac{4 y}{3x}
    \left(y\overline{Y}+ x\overline{X}\right)\,,\nonumber\\
\mathcal{M }^{(0,0,W,0,0)}_{+-+-}(x,y)=&-\frac{4 y \overline{X}\overline{Y}}{3x}-\frac{\overline{X} (x-1)}{y}-\frac{4 y^2-3x}{3y^2}\left(\overline{X}^2+6\zeta_2\right)-1\,,\nonumber\\

\mathcal{M }^{(0,0,W,1,2)}_{+-+-}(x,y)=&-\frac{8 y}{3 x}\,,\quad \mathcal{M }^{(0,0,W,1,1)}_{+-+-}(x,y)=\frac{8 \left(x-y+y^3\right) \overline{X}}{3 x^2}-\frac{8 \left(1-x y\right) y \overline{Y}}{3 x^2}-\frac{46 y}{3 x}\,,\nonumber\\

\mathcal{M }^{(0,0,W,1,0)}_{+-+-}(x,y)=&-\frac{2\overline{X}\left(4 y^2+7\right)}{3 x}-\frac{8\overline{X} (x-y) \overline{Y}}{3 x^2}-\frac{8 y (2+y)^2 \overline{Y}}{3 x^2}-\frac{2 \overline{  X}^2}{y}-\frac{4\zeta_2\left(3x-2y^2\right)}{xy}+\frac{22 y}{3 x}\,,\nonumber\\

\mathcal{M }^{(0,0,W,2,2)}_{+-+-}(x,y)=&\frac{12 \left(x-2 y^2\right)}{x^2}\,,\nonumber\\
\mathcal{M}^{(0,0,W,2,1)}_{+-+-}(x,y)=&\frac{2 \left(74 x-129 y^2\right)}{3 x^2}-\frac{4 \left(6 x^4+2 x^3+x^2+2 x+6\right)\overline{Y}}{3 x^3}-\frac{4 \left(6 y^4-2 y^3+6 y^2-3 y+1\right) \overline{X}}{3 x^2 y}\,,\nonumber\\
\mathcal{M}^{(0,0,W,2,0)}_{+-+-}(x,y)=&-\frac{4 \left(7 x^4-5 x^3-5 x+7\right) \overline{Y}}{3 x^3}-\frac{4 \left(1-2y+3 y^2\right)\overline{X} \overline{Y}}{3x^3 y }-\frac{24 x-109 y^2}{3 x^2}\nonumber\\
&-\frac{28 y^4-28 y^3+42 y^2-32 y-4}{3 x^2 y}\overline{X} 
    -\frac{36 \zeta_2 \left(x-2 y^2\right)}{x^2}\,,\nonumber\\

\mathcal{M}^{(0,0,W,3,2)}_{+-+-}(x,y)=&-\frac{8 \left(50 x^2+17 x+69\right) y}{3 x^3}+\frac{152 (x-y)}{3 x^3 y}\,,\nonumber\\

\mathcal{M}^{(0,0,W,3,1)}_{+-+-}(x,y)=&-\frac{8 \left(8 x y^5-3 x (x+4)-10 y^7-10\right) \overline{X}}{3 x^4 y^2}-\frac{656 (y-x)}{3 x^3 y}\nonumber\\
&-\frac{4 \left(1099 x^2+347 x+1591\right) y}{9 x^3}+\frac{8\left(10 x^7+12 x^6+3 x^5+3 x^2+12 x+10\right) \overline{Y}}{3 x^4 y^2}\,,\nonumber\\

\mathcal{M}^{(0,0,W,3,0)}_{+-+-}(x,y)=&-\frac{4\overline{X}\left(74 y^6-70 y^5+119 y^4-157 y^3+119 y^2-62 y+8\right)}{9 x^3 y^2}-\frac{8 \left(1-6y+3 y^2\right)
\overline{X} \overline{Y}}{3 x^4 y^2}\nonumber\\
&-\frac{8 \left(37 x^6-25 x^5+13 x^4-14 x^3+13 x^2-25 x+37\right) \overline{Y}}{9 x^4 y}+\frac{8 \zeta_2 \left(50 x^2+17 x+69\right) y}{x^3}\nonumber\\
&-\frac{8 (x-y)}{27 x^3 y}\left(110+513 \zeta_2\right)+\frac{2 \left(2395 x^2+1081 x+2835\right) y}{27 x^3}\,,\nonumber\\

\mathcal{M}^{(0,0,W,4,2)}_{+-+-}(x,y)=&-\frac{20 \left(98 y^6+269 y^5+483 y^4+460 y^3+310 y^2+96 y+32\right)}{3 x^4 y^2}\,,\nonumber\\

\mathcal{M}^{(0,0,W,4,1)}_{+-+-}(x,y)=&-\frac{4\overline{X}}{x^3 y^3}
   \left(\frac{20 \left(3 y^7-4\right)}{3 x}-\frac{70 \left(y^9+1\right)}{3 x^2}-y^5-6\right)-\frac{950 (1+3y)}{x^4 y^2}\nonumber\\
&+\frac{8 \left(35 x^9+40 x^8+9 x^7+9 x^2+40 x+35\right) \overline{Y}}{3 x^5 y^3}\nonumber\\
&-\frac{2516 x^4+3146 x^3+6704 x^2+2196 x+4416}{x^4}\,,\nonumber\\

\mathcal{M}^{(0,0,W,4,0)}_{+-+-}(x,y)=&-\frac{2 \left(533 x^8-361 x^7+217 x^6-225 x^5+224 x^4-225 x^3+217 x^2-361 x+533\right) \overline{Y}}{9 x^5 y^2}\nonumber\\
&-\frac{8 \left(9 y^2-22 y+4\right) \overline{X}\overline{Y}}{3 x^5 y^3}+\frac{640 \zeta_2 (1+3 y)}{x^4 y^2}+\frac{2720 (1+3 y)}{27 x^4 y^2}\nonumber\\
&-\frac{1066 y^8-958 y^7+1681 y^6-2207 y^5+2558y^4-2250 y^3+1738 y^2-910 y+134}{9 x^4 y^3}\overline{X}\nonumber\\
&+\frac{20 \zeta_2 \left(98 x^4+123 x^3+264 x^2+91 x+162\right)}{x^4}\nonumber\\
&+\frac{39589 x^4+51996 x^3+110790 x^2+46556 x+50469}{54 x^4}\,.
\end{align}
In contrast to the fermion case, the $W^\pm$ loop exhibits non-vanishing Sudakov logarithms at leading power in $m_W$ for the two-minus-two-plus helicity amplitudes.
This also implies that a calculation performed with massless $W^\pm$ bosons from the outset leads to infrared divergences already at the one-loop level.

\section{Effective QED coupling $\alpha$ in the on-shell scheme\label{sec:alphaRGrun}}

For completeness, we describe in this appendix how we evaluate the effective electromagnetic coupling $\alpha$ in the on-shell scheme, as implemented in the \lblatnlo\ code (see appendix \ref{sec:lblatnlo}). The effective QED coupling at a given energy scale $\mu$ is defined as
\begin{eqnarray}
\alpha(\mu^2)&=&\frac{\alpha(0)}{1-\Delta \alpha(\mu^2)}\,,
\end{eqnarray}
where
\begin{eqnarray}
\Delta\alpha(\mu^2)&=&\sum_{q\neq t}{\Delta\alpha_{q}(\mu^2)}+\Delta \alpha_{t}(\mu^2)+\sum_{\ell}{\Delta\alpha_\ell(\mu^2)}\,.\label{eq:deltaalphaOSdecomp}
\end{eqnarray}
The sums in eq.~\eqref{eq:deltaalphaOSdecomp} run over all quark flavours except the top quark and over the three charged leptons. Due to gauge invariance, contributions from weak gauge bosons are excluded, and only charged fermion loops are included within the SM. The evolution of $\alpha$ has been experimentally observed by the OPAL~\cite{OPAL:2005xqs} and L3~\cite{L3:2005tsb} collaborations at LEP in small- and large-angle Bhabha scattering, $e^-e^+\to e^-e^+$, respectively. 

For a given fermion species $f$, $\Delta\alpha_f(\mu^2)$ is directly related to the contribution of $f$ to the photon vacuum polarisation function, $\Pi^{\gamma\gamma}_f\left(\mu^2\right)$, via
\begin{eqnarray}
\Delta \alpha_f\left(\mu^2\right)&=&-\left(\Re{\Pi^{\gamma\gamma}_f\left(\mu^2\right)}-\Pi^{\gamma\gamma}_f\left(0\right)\right)\,.
\end{eqnarray}
The perturbative expression for $\Pi^{\gamma\gamma}_f\left(\mu^2\right)$ is known analytically up to two loops in QCD and QED~\cite{Kallen:1955fb,Sabry:1962rge}, and up to three loops in QED for $f=e^-$~\cite{Baikov:1995ui,Forner:2024ojj}. However, for quark contributions other than the top quark, $\Delta \alpha_q(\mu^2)$ is affected by hadronic resonances, which prevents a reliable direct use of perturbation theory. Although the hadronic contribution can be evaluated through
a dispersion integral over the measured cross section of $e^-e^+\to\mathrm{hadrons}$, supplemented with perturbative QCD at energies above the resonance region~\cite{Eidelman:1995ny,Burkhardt:2001xp}, we adopt a much simpler yet pragmatic approach here. In practice, we use the following equation:
\begin{equation}
\alpha(m_Z^2)=\frac{\alpha(0)}{1-\Delta \alpha_{\rm had}^{(5)}(m_Z^2)-\Delta \alpha_t(m_Z^2)-\sum_{\ell}{\Delta \alpha_\ell(m_Z^2)}}
\end{equation}
to extract the non-perturbative hadronic contribution $\Delta \alpha_{\rm had}^{(5)}(m_Z^2)$ from the inputs $\alpha(0)$ and $\alpha(m_Z^2)$~\cite{ParticleDataGroup:2024cfk}. For scales $\mu^2 \neq m_Z^2$, we then use
\begin{equation}
\alpha(\mu^2)=\frac{\alpha(0)}{1-\sum_{q\neq t}{\left[\Delta \alpha_q(\mu^2)-\Delta \alpha_q(m_Z^2)\right]}-\Delta \alpha_{\rm had}^{(5)}(m_Z^2)-\Delta \alpha_t(\mu^2)-\sum_{\ell}{\Delta \alpha_\ell(\mu^2)}}\,,\label{eq:alphaOSrun2}
\end{equation}
where two-loop perturbative expressions for $\Delta \alpha_q$, $\Delta \alpha_t$, and $\Delta \alpha_\ell$ in QCD and QED are employed.
With eq.~\eqref{eq:alphaOSrun2}, the hadronic contribution is effectively included, and the result becomes largely insensitive to the light-quark masses $m_q$ in the regime $|\mu| \gg m_q$.

\section{\lblatnlo: Light-by-Light scattering at NLO\label{sec:lblatnlo}}

In this appendix, we present the main features and usage of \lblatnlo, which is written in \texttt{Fortran90}. The code is publicly available in the following git repository:

~~~~~~~~~~~~~~~~~~~~~~\url{https://github.com/huashengshao/LbLatNLO}\\
It can be cloned using:
\vskip 0.25truecm
\noindent
~~~{\tt git clone https://github.com/huashengshao/LbLatNLO.git}

\vskip 0.25truecm
\noindent 
The implementation of master integrals and scattering amplitudes in the code is described in sections~\ref{sec:MIinLbLatNLO} and~\ref{sec:AmpinLbLatNLO}, respectively. A brief guide to using \lblatnlo\ for event simulation is provided in section~\ref{sec:useLbLatNLO}.

\subsection{Implementation of master integrals\label{sec:MIinLbLatNLO}}

All one- and two-loop master integrals necessary for evaluating the two-loop LbL amplitudes, as implemented in the \lblatnlo\ code, follow the procedure explained in section 3 of ref.~\cite{AH:2023ewe}. We briefly recall the main idea using the same notation and explain how readers can call them from the \lblatnlo\ code.

For the numerical evaluation of the five two-loop master integrals with square roots, $f_{18}^{(2,3)}, f_{25}^{(2,4)}, f_{26}^{(2,4)}, f_{28}^{(2,4)}, f_{29}^{(2,4)}$, these integrals cannot be expressed in terms of multiple polylogarithms but can instead be written in terms of Chen's iterated path integrals~\cite{Chen:1977oja} with logarithmic one-forms. To evaluate these iterated integrals, we adopt the technique described in refs.~\cite{Caron-Huot:2014lda,Chicherin:2020oor}. In particular, the kernels in the first two-fold integration can be rationalised and thus expressed in terms of logarithms and classical polylogarithms $\mathrm{Li}_n$ of the kinematic variables $x_{ij}=s_{ij}/m_f^2, x_{jk}=s_{jk}/m_f^2, x_{ik}=s_{ik}/m_f^2$.~\footnote{The tuple $(x_{ij}, x_{jk}, x_{ik})$ is a permutation of $(x_s, x_t, x_u)$.} For the weight-$3$ integral $f_{18}^{(2,3)}$, only a single integration remains to be performed numerically, while for the remaining four weight-$4$ integrals, two-fold integrations remain. For the latter, we reduce the final two-fold integration to a one-fold integration and evaluate it numerically using the double-exponential (tanh-sinh) quadrature method~\cite{Mori:1973}. We further improve the convergence of the numerical integration by choosing the following path parametrisation from the starting point $(x_{ij,0}, x_{jk,0})$ to the endpoint $(x_{ij}, x_{jk})$:
\begin{eqnarray}
x_{ij}(r)&=&x_{ij,0}(1-r)^5\left(1+5r+15r^2+35r^3+70r^4\right)\nonumber\\
&&+x_{ij}r^5\left(126-420r+540r^2-315r^3+70r^4\right)\,,\nonumber\\
x_{jk}(r)&=&x_{jk,0}(1-r)^5\left(1+5r+15r^2+35r^3+70r^4\right)\nonumber\\
&&+x_{jk}r^5\left(126-420r+540r^2-315r^3+70r^4\right)\,,\label{eq:path4int}
\end{eqnarray}
where $r \in [0,1]$, $(x_{ij}(0), x_{jk}(0)) = (x_{ij,0}, x_{jk,0})$, and $(x_{ij}(1), x_{jk}(1)) = (x_{ij}, x_{jk})$. We choose the starting point as
\begin{eqnarray}
\left(x_{ij,0},x_{jk,0}\right)&=&\left\{\begin{array}{ll}(0,0), & \quad {\rm if}~x_{ij},x_{jk}<0\\
(x_{ij},0), & \quad {\rm if}~x_{ij}>0,x_{jk}<0\\
(0,x_{jk}), & \quad {\rm if}~x_{ij}<0,x_{jk}>0\,,\end{array}\right.\,
\end{eqnarray}
where the boundary conditions of these master integrals are known analytically in terms of GPLs (see appendix B of ref.~\cite{AH:2023ewe}). The implementation of these five master integrals with square roots can be found in the \texttt{Fortran} file \texttt{calc\_IterInts\_RootMIs.f} in the \texttt{src/} subdirectory of \lblatnlo. Their correspondence is collected in table~\ref{tab:master_integrals_fortran}.

The remaining master integrals can be expressed in terms of multiple polylogarithms. Their numerical evaluation is straightforward using the algorithm outlined in ref.~\cite{Vollinga:2004sn}. In \lblatnlo, both \fastgpl~\cite{Wang:2021imw} and \handyg~\cite{Naterop:2019xaf} can be linked to evaluate GPLs. The corresponding master integrals are implemented in the \texttt{Fortran} file \texttt{calc\_IterInts\_GPLMIs.f} in the \texttt{src/} subdirectory. The one-to-one correspondence between master integrals and \texttt{Fortran} functions in double precision is summarised in table~\ref{tab:master_integrals_fortran}. Their evaluation in both double and quadruple precision is handled by the \texttt{Fortran} subroutine \texttt{UTtwoloopbasis}, defined in the file \texttt{src/twoloopbasis.f}. In the case of quadruple precision, only \handyg\ is available, as \fastgpl\ currently provides only a double-precision implementation.

\begin{table}[h!]
\centering
\begin{tabular}{| c c | c c |}
\hline
Master integral & \texttt{Fortran} function & Master integral & \texttt{Fortran} function \\
\hline
$f_{3}^{(1,2)}$ & \texttt{f1E3\_w2} & $f_{5}^{(1,3)}$ & \texttt{f1E5\_w3} \\
$f_{13}^{(2,4)}$ & \texttt{fE13\_w4} & $f_{14}^{(2,3)}$ & \texttt{fE14\_w3} \\
$f_{15}^{(2,2)}$ & \texttt{fE15\_w2} & $f_{17}^{(2,4)}$ & \texttt{fE17\_w4} \\
$f_{18}^{(2,3)}$ & \texttt{fE18\_w3} & $f_{19}^{(2,2)}$ & \texttt{fE19\_w2} \\
$f_{19}^{(2,3)}$ & \texttt{fE19\_w3} & $f_{20}^{(2,3)}$ & \texttt{fE20\_w3} \\
$f_{21}^{(2,4)}$ & \texttt{fE21\_w4} & $f_{22}^{(2,3)}$ & \texttt{fE22\_w3} \\
$f_{25}^{(2,4)}$ & \texttt{fE25\_w4} & $f_{26}^{(2,4)}$ & \texttt{fE26\_w4} \\
$f_{27}^{(2,4)}$ & \texttt{fE27\_w4} & $f_{28}^{(2,4)}$ & \texttt{fE28\_w4} \\
$f_{29}^{(2,4)}$ & \texttt{fE29\_w4} & & \\
\hline
\end{tabular}
\caption{Correspondence between the master integrals $f_n^{(\ell,\omega)}$ and their double-precision \texttt{Fortran} implementations in \lblatnlo. The quadruple-precision versions are denoted by names ending with \texttt{\_qp}.}
\label{tab:master_integrals_fortran}
\end{table}

Finally, we note that many other master integrals appearing in the two-loop LbL scattering amplitudes can be related to those listed in table~\ref{tab:master_integrals_fortran} using the relations given in eqs.~(3.13) and (3.14) of ref.~\cite{AH:2023ewe}.

\subsection{Implementation of scattering amplitudes\label{sec:AmpinLbLatNLO}}

We provide a set of helicity-amplitude implementations in the \lblatnlo\ code. The \texttt{Fortran} functions and subroutines corresponding to the amplitudes presented in this work are listed in table~\ref{tab:amplitudes_fortran}. The following quantities are defined after factoring out global prefactors:
\begin{eqnarray}
\mathcal{M}_{\lambda_1\lambda_2\lambda_3\lambda_4}^{(0,0,f)}&=&8iN_{c,f}Q_f^4\alpha^2\hat{\mathcal{M}}_{\lambda_1\lambda_2\lambda_3\lambda_4}^{(0,f)}\,,\quad \mathcal{M}_{\lambda_1\lambda_2\lambda_3\lambda_4}^{(0,0,W)}=-12i\alpha^2\hat{\mathcal{M}}_{\lambda_1\lambda_2\lambda_3\lambda_4}^{(0,W)}\,,\nonumber\\
\mathcal{M}_{\lambda_1\lambda_2\lambda_3\lambda_4}^{(0,0,f),\mathrm{LE}}&=&iN_{c,f}Q_f^4\alpha^2\hat{\mathcal{M}}_{\lambda_1\lambda_2\lambda_3\lambda_4}^{(0,f),\mathrm{LE}}\,,\quad
\mathcal{M}_{\lambda_1\lambda_2\lambda_3\lambda_4}^{(0,0,f),\mathrm{HE}}=8iN_{c,f}Q_f^4\alpha^2\hat{\mathcal{M}}_{\lambda_1\lambda_2\lambda_3\lambda_4}^{(0,f),\mathrm{HE}}\,,\nonumber\\
\mathcal{M}_{\lambda_1\lambda_2\lambda_3\lambda_4}^{(0,0,W),\mathrm{LE}}&=&-12i\alpha^2\hat{\mathcal{M}}_{\lambda_1\lambda_2\lambda_3\lambda_4}^{(0,W),\mathrm{LE}}\,,\quad
\mathcal{M}_{\lambda_1\lambda_2\lambda_3\lambda_4}^{(0,0,W),\mathrm{HE}}=-12i\alpha^2\hat{\mathcal{M}}_{\lambda_1\lambda_2\lambda_3\lambda_4}^{(0,W),\mathrm{HE}}\,,\nonumber\\
\hat{\mathcal{M}}_{\lambda_1\lambda_2\lambda_3\lambda_4}^{(1,f)}&=&\frac{i\pi\mathcal{M}_{\lambda_1\lambda_2\lambda_3\lambda_4}^{(1,0,f)}}{N_{c,f}Q_f^4\alpha^2C_{F,f}\alpha_s}=\frac{i\pi\mathcal{M}_{\lambda_1\lambda_2\lambda_3\lambda_4}^{(0,1,f)}}{N_{c,f}Q_f^6\alpha^3}\,,\nonumber\\
\hat{\mathcal{M}}_{\lambda_1\lambda_2\lambda_3\lambda_4}^{(1,f),\mathrm{LE}}&=&\frac{i\pi\mathcal{M}_{\lambda_1\lambda_2\lambda_3\lambda_4}^{(1,0,f),\mathrm{LE}}}{N_{c,f}Q_f^4\alpha^2C_{F,f}\alpha_s}=\frac{i\pi\mathcal{M}_{\lambda_1\lambda_2\lambda_3\lambda_4}^{(0,1,f),\mathrm{LE}}}{N_{c,f}Q_f^6\alpha^3}\,,\nonumber\\
\hat{\mathcal{M}}_{\lambda_1\lambda_2\lambda_3\lambda_4}^{(1,f),\mathrm{HE}}&=&\frac{i\pi\mathcal{M}_{\lambda_1\lambda_2\lambda_3\lambda_4}^{(1,0,f),\mathrm{HE}}}{N_{c,f}Q_f^4\alpha^2C_{F,f}\alpha_s}=\frac{i\pi\mathcal{M}_{\lambda_1\lambda_2\lambda_3\lambda_4}^{(0,1,f),\mathrm{HE}}}{N_{c,f}Q_f^6\alpha^3}\,,\nonumber\\
\hat{\mathcal{M}}_{\lambda_1\lambda_2\lambda_3\lambda_4}^{(1,f),\mathrm{LP}}&=&\frac{i\pi\mathcal{M}_{\lambda_1\lambda_2\lambda_3\lambda_4}^{(1,0,f),\mathrm{LP}}}{N_{c,f}Q_f^4\alpha^2C_{F,f}\alpha_s}=\frac{i\pi\mathcal{M}_{\lambda_1\lambda_2\lambda_3\lambda_4}^{(0,1,f),\mathrm{LP}}}{N_{c,f}Q_f^6\alpha^3}\,,\nonumber\\
\hat{\mathcal{M}}_{\lambda_1\lambda_2\lambda_3\lambda_4}^{(1,f),\mathrm{Coul~approx}}&=&\frac{i\pi\mathcal{M}_{\lambda_1\lambda_2\lambda_3\lambda_4}^{(1,0,f),\mathrm{Coul~approx}}}{N_{c,f}Q_f^4\alpha^2C_{F,f}\alpha_s}=\frac{i\pi\mathcal{M}_{\lambda_1\lambda_2\lambda_3\lambda_4}^{(0,1,f),\mathrm{Coul~approx}}}{N_{c,f}Q_f^6\alpha^3}\,,\nonumber\\
\left.\hat{\mathcal{M}}_{\lambda_1\lambda_2\lambda_3\lambda_4}^{(1,f)}\right|_{m_f=0}&=&\left.\frac{-i\pi\mathcal{M}_{\lambda_1\lambda_2\lambda_3\lambda_4}^{(1,0,f)}}{8N_{c,f}Q_f^4\alpha^2C_{F,f}\alpha_s}\right|_{m_f=0}=\left.\frac{-i\pi\mathcal{M}_{\lambda_1\lambda_2\lambda_3\lambda_4}^{(0,1,f)}}{8N_{c,f}Q_f^6\alpha^3}\right|_{m_f=0}\,.\label{eq:Mhatdef1}
\end{eqnarray}
The functions for evaluating one-loop massive and massless helicity amplitudes are implemented in \texttt{src/LOME.f}, while their HE and LE expansions are provided in\\ \texttt{src/OneLoopHighEnergyME.f} and \texttt{src/OneLoopLowEnergyME.f}, respectively. 
The \texttt{Fortran} subroutines and functions for computing the full fermion-mass-dependent and massless two-loop amplitudes are implemented in \texttt{src/TwoLoopMassiveME.f} and \texttt{src/TwoLoopMasslessME.f}. The corresponding HE and LE expansions are available in \texttt{src/TwoLoopHighEnergyME.f} and \texttt{src/TwoLoopLowEnergyME.f}, respectively. The two-loop Coulomb-approximation and LP Coulomb-resummed results are implemented in \texttt{src/TwoLoopCoulombME.f}. All these \texttt{Fortran} functions and subroutines are called within the subroutine \texttt{stwoloopmatrix\_LbL} defined in \texttt{src/NLOME.f90}. For phenomenological applications, a grid \texttt{Amp2L.grid} containing two-loop helicity amplitudes is provided in the \texttt{grid/} subdirectory. It is generated using the subroutine \texttt{eval\_twoloop\_massive\_amplitudes} defined in \texttt{src/NLOME.f90}. This grid combines double- and quadruple-precision implementations of $\hat{\mathcal{M}}_{\lambda_1\lambda_2\lambda_3\lambda_4}^{(1,f)}$, together with the corresponding LE and HE expansions $\hat{\mathcal{M}}_{\lambda_1\lambda_2\lambda_3\lambda_4}^{(1,f),\mathrm{LE}}$ and $\hat{\mathcal{M}}_{\lambda_1\lambda_2\lambda_3\lambda_4}^{(1,f),\mathrm{HE}}$. This avoids the need to interface with external \fastgpl\ and \handyg\ packages, thereby improving the usability of the programme.

\begin{table}[h!]
\hspace{-1cm}
\small
\begin{tabular}{| c c | c c |}
\hline
Amplitude & \texttt{Fortran} & Amplitude & \texttt{Fortran} \\
\hline
$\hat{\mathcal{M}}_{\lambda_1\lambda_2\lambda_3\lambda_4}^{(0,f)}$ & \texttt{\footnotesize OneLoop\_HelAmp\_Massive} & $\hat{\mathcal{M}}_{\lambda_1\lambda_2\lambda_3\lambda_4}^{(0,W)}$ & \texttt{\footnotesize OneLoop\_HelAmp\_MassiveW} \\
$\hat{\mathcal{M}}_{\lambda_1\lambda_2\lambda_3\lambda_4}^{(0,f),\mathrm{LE}}$ & \texttt{\footnotesize OneLoop\_HelAmp\_LE} & $\hat{\mathcal{M}}_{\lambda_1\lambda_2\lambda_3\lambda_4}^{(0,W),\mathrm{LE}}$ & \texttt{\footnotesize OneLoopW\_HelAmp\_LE} \\
$\hat{\mathcal{M}}_{\lambda_1\lambda_2\lambda_3\lambda_4}^{(0,f),\mathrm{HE}}$ & \texttt{\footnotesize OneLoop\_HelAmp\_HE} & $\hat{\mathcal{M}}_{\lambda_1\lambda_2\lambda_3\lambda_4}^{(0,W),\mathrm{HE}}$ & \texttt{\footnotesize OneLoopW\_HelAmp\_HE} \\
$\hat{\mathcal{M}}_{\lambda_1\lambda_2\lambda_3\lambda_4}^{(1,f)}$ & \texttt{\footnotesize Get\_TwoLoop\_HelAmp\_Massive} & $\hat{\mathcal{M}}_{\lambda_1\lambda_2\lambda_3\lambda_4}^{(1,f),\mathrm{LP}}$ & \texttt{\footnotesize Get\_TwoLoop\_HelAmp\_LPCoulImproved} \\
$\hat{\mathcal{M}}_{\lambda_1\lambda_2\lambda_3\lambda_4}^{(1,f),\mathrm{LE}}$ & \texttt{\footnotesize TwoLoop\_HelAmp\_LE} & $\hat{\mathcal{M}}_{\lambda_1\lambda_2\lambda_3\lambda_4}^{(1,f),\mathrm{Coul~approx}}$ & \texttt{\footnotesize Get\_TwoLoop\_HelAmp\_LPCoulombApprox} \\
$\hat{\mathcal{M}}_{\lambda_1\lambda_2\lambda_3\lambda_4}^{(1,f),\mathrm{HE}}$ & \texttt{\footnotesize TwoLoop\_HelAmp\_HE} & $\left.\hat{\mathcal{M}}_{\lambda_1\lambda_2\lambda_3\lambda_4}^{(0,f)}\right|_{m_f=0}$ & \texttt{\footnotesize OneLoop\_HelAmp\_Massless} \\
$\Delta \hat{\mathcal{M}}_{\lambda_1\lambda_2\lambda_3\lambda_4}^{(1,f),\MSbar}$ & \texttt{\footnotesize TwoLoop\_HelAmp\_Massive\_MSbarOS} & $\left.\hat{\mathcal{M}}_{\lambda_1\lambda_2\lambda_3\lambda_4}^{(1,f)}\right|_{m_f=0}$ & \texttt{\footnotesize TwoLoop\_HelAmp\_Massless} \\
\hline
\end{tabular}
\caption{Correspondence between the amplitudes and their double-precision \texttt{Fortran} implementations in \lblatnlo. The quadruple-precision versions are denoted by names ending with \texttt{\_qp}.}
\label{tab:amplitudes_fortran}
\end{table}

As an aside in the present context, which does not affect the numerical results presented so far, it is worth mentioning that the difference between two-loop scattering amplitudes in the $\MSbar$ and on-shell mass schemes can be straightforwardly obtained. The explicit expressions for the difference are
\begin{eqnarray}
i\Delta \mathcal{M}^{(1,0,f),\MSbar}_{++++}&=&16N_{c,f}Q_f^4\alpha^2\frac{\alpha_s}{\pi}C_{F,f}\left(3\log{\left(\frac{m_f^2}{\mu_R^2}\right)}-4\right)\times\nonumber\\
&&\mathop{\sum_{(i,j,k)=(s,t,u),}}_{(t,u,s),(u,s,t)}{\left[\frac{x_j^2\left(4+x_j\right)}{\left(x_ix_j+4x_k\right)\left(x_jx_k+4x_i\right)}\frac{f_3^{(1,1)}(x_i,x_j,x_k)}{\sqrt{x_j\left(x_j-4\right)}}\right.}\nonumber\\
&&\left.-2\left(1-\frac{x_k}{x_ix_j+4x_k}\right)\frac{f_6^{(1,2)}(x_i,x_j,x_k)}{\sqrt{x_ix_j(x_ix_j+4x_k)}}\right]\,,\\
i\Delta \mathcal{M}^{(1,0,f),\MSbar}_{-+++}&=&8N_{c,f}Q_f^4\alpha^2\frac{\alpha_s}{\pi}C_{F,f}\left(3\log{\left(\frac{m_f^2}{\mu_R^2}\right)}-4\right)\times\nonumber\\
&&\mathop{\sum_{(i,j,k)=(s,t,u),}}_{(t,u,s),(u,s,t)}{\left[\frac{\left(4-x_j\right)\left(x_ix_k(4-x_j)-2x_j^2\right)}{\left(x_ix_j+4x_k\right)\left(x_jx_k+4x_i\right)}\frac{f_3^{(1,1)}(x_i,x_j,x_k)}{\sqrt{x_j\left(x_j-4\right)}}\right.}\\
&&\left.+\left(\frac{1}{x_k}-\frac{x_k}{x_ix_j}\right)f_5^{(1,2)}(x_i,x_j,x_k)-\left(2+\frac{x_ix_j}{x_k}+\frac{4x_k}{x_ix_j+4x_k}\right)\frac{f_6^{(1,2)}(x_i,x_j,x_k)}{\sqrt{x_ix_j(x_ix_j+4x_k)}}\right]\,,\nonumber\\
i\Delta \mathcal{M}^{(1,0,f),\MSbar}_{--++}&=&8N_{c,f}Q_f^4\alpha^2\frac{\alpha_s}{\pi}C_{F,f}\left(3\log{\left(\frac{m_f^2}{\mu_R^2}\right)}-4\right)\times\nonumber\\
&&\left\{\frac{x_s^2\left(2-x_s\right)\left(4+x_s\right)}{\left(x_sx_t+4x_u\right)\left(x_sx_u+4x_t\right)}\frac{f_3^{(1,1)}(x_t,x_s,x_u)}{\sqrt{x_s\left(x_s-4\right)}}\right.\nonumber\\
&&-2\left(2+\frac{x_tx_u}{x_s}-\frac{x_s(2+x_s)}{x_tx_u+4x_s}\right)\frac{f_6^{(1,2)}(x_t,x_u,x_s)}{\sqrt{x_tx_u(x_tx_u+4x_s)}}\nonumber\\
&&+\mathop{\sum_{(i,j,k)=(s,t,u),}}_{(s,u,t)}{\left[\frac{\left(4-x_j\right)\left(x_i^2\left(2+x_j\right)+2x_k^2\right)}{\left(x_ix_j+4x_k\right)\left(x_jx_k+4x_i\right)}\frac{f_3^{(1,1)}(x_i,x_j,x_k)}{\sqrt{x_j\left(x_j-4\right)}}\right.}\nonumber\\
&&\left.\left.+\frac{2}{x_i}f_5^{(1,2)}(x_i,x_j,x_k)+\frac{x_i^2\left(x_j-2\right)-6\left(x_ix_j+2x_k\right)}{x_ix_j+4x_k}\frac{f_6^{(1,2)}(x_i,x_j,x_k)}{\sqrt{x_ix_j(x_ix_j+4x_k)}}\right]\right\}\,,\\
\Delta\mathcal{M}_{+-+-}^{(1,0,f),\MSbar}&=&\left.\Delta\mathcal{M}_{--++}^{(1,0,f),\MSbar}\right|_{x_s\leftrightarrow x_u}\,,\quad \Delta\mathcal{M}_{+--+}^{(1,0,f),\MSbar}=\left.\Delta\mathcal{M}_{--++}^{(1,0,f),\MSbar}\right|_{x_s\leftrightarrow x_t}\,.
\end{eqnarray}
Analogous to the notation defined in eq.~\eqref{eq:Mhatdef1}, we define the hatted quantity
\begin{equation}
\Delta \hat{\mathcal{M}}_{\lambda_1\lambda_2\lambda_3\lambda_4}^{(1,f),\MSbar}=\frac{i\pi\mathcal{M}_{\lambda_1\lambda_2\lambda_3\lambda_4}^{(1,0,f),\MSbar}}{N_{c,f}Q_f^4\alpha^2C_{F,f}\alpha_s}=\frac{i\pi\mathcal{M}_{\lambda_1\lambda_2\lambda_3\lambda_4}^{(0,1,f),\MSbar}}{N_{c,f}Q_f^6\alpha^3}\,.
\end{equation}
The corresponding \texttt{Fortran} implementation is provided as \texttt{TwoLoop\_HelAmp\_Massive\_MSbarOS} in the file \texttt{src/TwoLoopMassiveME.f}. We have not yet enabled the option of using the $\MSbar$ mass scheme in \lblatnlo. However, users who wish to obtain two-loop amplitudes in the $\MSbar$ mass scheme can do so by combining $\Delta \hat{\mathcal{M}}_{\lambda_1\lambda_2\lambda_3\lambda_4}^{(1,f),\MSbar}$ with $\hat{\mathcal{M}}_{\lambda_1\lambda_2\lambda_3\lambda_4}^{(1,f)}$.

\subsection{Basic usage\label{sec:useLbLatNLO}}

The \lblatnlo\ code can be compiled as follows:
\vskip 0.25truecm
\noindent
~~~{\tt cd LbLatNLO; ./config}

\vskip 0.25truecm
\noindent
The file {\tt config} is a {\tt bash} script that generates a {\tt makefile}, creates several subdirectories, and runs {\tt make} to compile all {\tt Fortran} and {\tt C++} source files using the {\tt gfortran} and {\tt gcc} compilers. The third-party \gammaUPC\ code~\cite{Shao:2022cly} is included in \lblatnlo\ to enable calculations in UPCs. The main code resides in the {\tt src/} subdirectory. The files related to $\alpha_s$ renormalisation group evolution are adapted from the {\tt HOPPET} programme~\cite{Salam:2008qg}, with extensions to include five-loop QCD running and four-loop decoupling relations. After compilation, an executable {\tt LbLatNLO} is created in the {\tt bin/} subdirectory. A symbolic link to {\tt bin/LbLatNLO} is also created in the main directory. Although not required, users may link the \fastgpl\ and/or \handyg\ external libraries for the evaluation of two-loop master integrals and full fermion-mass-dependent two-loop scattering amplitudes. To do so, the paths to these libraries must be specified in \texttt{input/configuration.txt} before compilation.

The use of the programme is straightforward once the run setup has been specified in the {\tt input/run.inp} file. The code can be executed simply via
\vskip 0.25truecm
\noindent
~~~{\tt ./LbLatNLO}

\vskip 0.25truecm

\noindent
The phase-space-integrated fiducial cross section is displayed on the screen, and several output files, including histograms and a Les Houches event file~\cite{Alwall:2006yp}, are written to the {\tt output/} subdirectory. As a validation of the installation, we consider LbL scattering in \PbPb\ UPCs at $\sqrtsnn=5.02$ TeV with the default setup in {\tt input/run.inp}. The integrated fiducial cross section, with the cuts $p_T^\gamma>2$ GeV, $|\eta_\gamma|<5$, and $m_{\gamma\gamma}>5$ GeV at (NLO+LP)$^\prime$ QCD+QED accuracy, should be reported by \lblatnlo\ as
\begin{equation}
\sigma^{\mathrm{(NLO+LP)}^\prime~\mathrm{QCD+QED}}_{\mathrm{PbPb}}=1.50187(86)\times 10^{5}~\mathrm{pb}\,.
\end{equation}

We now discuss the parameters in the {\tt input/run.inp} file. The format of the file is analogous to the \texttt{user.inp} file in \helaconia~\cite{Shao:2012iz,Shao:2015vga} and the \texttt{run.inp} file in \phique~\cite{Capatti:2025khs}. The beam configuration includes the following parameters:
\begin{itemize}
\item {\tt colpar} is an integer specifying the type of collision: $1$ (UPC at hadron colliders, cf. eq.~\eqref{eq:LgginUPC}), $2$ ($e^\pm$-proton two-photon collisions), $3$ ($e^-e^+$ colliders, cf. eq.~\eqref{eq:Lgginee}), and $4$ ($\gamma\gamma$ colliders, cf. eq~\eqref{eq:Lggingg}).
\item In the case of hadron colliders ({\tt colpar=1}), one must specify the atomic mass and atomic numbers of the incoming hadrons via the parameters {\tt nuclearA\_beam1}, \\
{\tt nuclearA\_beam2}, {\tt nuclearZ\_beam1}, and {\tt nuclearZ\_beam2}.
\item The energies of the two beam particles, in units of GeV, are specified by the parameters {\tt energy\_beam1} and {\tt energy\_beam2}. If a beam particle is a nucleus, the corresponding value denotes the energy per nucleon.
\end{itemize}
In ultraperipheral hadron collisions ({\tt colpar=1}), the two-photon flux can be selected via the parameter {\tt UPC\_photon\_flux}, which can take the values $1$ (ChFF), $2$ (EDFF), or $3$ (iWW, available only for protons). Furthermore, the bool parameters {\tt use\_MC\_Glauber} and\\
{\tt intrinsic\_kt} control whether Monte Carlo ({\tt T}) or optical ({\tt F}) Glauber modelling is used for the nuclear thickness $T_{\mathrm{A}}(b)$ and overlap $T_{\mathrm{AB}}(b)$ functions in the calculation of the soft survival probability, and whether the initial $k_\perp$ smearing due to non-zero photon virtualities (see appendix A of ref.~\cite{Shao:2024dmk}) is included in UPCs, respectively. When the iWW approximation ({\tt UPC\_photon\_flux=3}) is used for photon fluxes of $e^\pm$ or protons, the value of $Q_{\mathrm{max}}^2$ must be specified via the parameter {\tt q2max}. 

The perturbative order is controlled by the integer parameters {\tt order} and {\tt coulomb}. The former can take the values $0$ (LO), $\pm1$ (NLO QED or NLO$^\prime$ QED), $\pm2$ (NLO QCD or NLO$^\prime$ QCD), and $\pm3$ (NLO QCD+QED or NLO$^\prime$ QCD+QED), while the latter can be $0$ (no Coulomb resummation) or $1$ (LP Coulomb resummation). The on-shell masses of charged fermions and the $W^\pm$ boson can be set to positive or negative values; a negative value indicates that the corresponding particle is excluded from the loop. The fermion masses may also be set to zero, while this is not allowed for the $W^\pm$ boson due to the presence of infrared poles in the massless limit.
Additionally, the following parameters control the coupling constants:
\begin{itemize}
\item {\tt alphaemm1} denotes $\alpha^{-1}(0)$, {\tt alphaMZm1} denotes $\alpha^{-1}(m_Z^2)$, {\tt Gfermi} is the Fermi constant, and {\tt alphasMZ} is $\alpha_s(m_Z^2)$.
\item {\tt alphas\_nloop} specifies the perturbative order of the QCD beta function.
\item {\tt alpha\_scheme} is an integer parameter specifying the electroweak scheme for $\alpha$ in QED corrections: 0 ($\alpha(0)$ scheme), 1 ($G_\mu$ scheme), or 2 (on-shell scheme). In the on-shell scheme, the number of loops used for the evaluation of $\alpha(\mu^2)$ is set by {\tt alpha\_nloop}, which can be either 1 or 2.
\end{itemize}
The fiducial cuts in the laboratory frame can be specified in the corresponding kinematic cuts block. For instance, the minimum transverse momenta of the final-state photons can be set via {\tt minptp}, and the minimum invariant mass of the di-photon system can be controlled by {\tt minmpp}.

The remaining parameters are less essential and can be readily understood from the comments in the {\tt input/run.inp} file, which reads:

\begingroup
\fontsize{10pt}{10pt}\selectfont

\begin{Verbatim}[breaklines=true, breakanywhere=true]
# beam configuration
colpar 1                   # colliding particles: 1=hadron-hadron UPC,
                           #  2=electron/positron-proton, 3=e-e+, 
                           # 4=gamma-gamma
energy_beam1 2510d0        # beam 1 energy [in units of GeV]
                           # per nucleon/electron/positron/gamma
energy_beam2 2510d0        # beam 2 energy [in units of GeV]
                           # per nucleon/electron/positron/gamma
nuclearA_beam1  208        # A of nuclear of beam 1, 0: it is a proton
nuclearA_beam2  208        # A of nuclear of beam 2, 0: it is a proton
nuclearZ_beam1  82         # Z of nuclear of beam 1, 0: it is a proton
nuclearZ_beam2  82         # Z of nuclear of beam 2, 0: it is a proton

# gamma-UPC (arXiv:2207.03012) options
UPC_photon_flux 1          # only colpar=1. 1: ChFF (2207.03012); 
                           # 2: EDFF (2207.03012); 
                           # 3: improved Weizsacker-Williams 
                           #    (only for proton beams)
use_MC_Glauber F           # whether or not to use MC-Glauber TAA for 
                           # the survival probability when colpar=1 and 
                           # UPC_photon_flux=1 or 2.
intrinsic_kt T             # whether to include the intrinsic kT due to 
                           # non-zero virtuality of the initial photon in UPCs
                           # see appendix A of arXiv:2407.13610
# q2max in electron or proton iWW photon fluxes
q2max 1d0                  # maximal q2 cut for iWW photon flux 
                           # [in units of GeV**2]

# order
# W loop we only have LO
# for fermion loops, we have
# A(0,0) one-loop amplitude
# A(1,0) two-loop QCD amplitude
# A(0,1) two-loop QED amplitude
# 0: LO |A(0,0)|**2
# 1: NLO QED |A(0,0)|**2+2*Re(DCONJG(A(0,0))*A(0,1))
# 2: NLO QCD |A(0,0)|**2+2*Re(DCONJG(A(0,0))*A(1,0))
# 3: NLO QED+QCD |A(0,0)|**2+2*Re(DCONJG(A(0,0))*(A(1,0)+A(0,1)))
# -1: NLO QED |A(0,0)+A(0,1)|**2
# -2: NLO QCD |A(0,0)+A(1,0)|**2
# -3: NLO QED+QCD |A(0,0)+A(0,1)+A(1,0)|**2
order -3

# Coulomb resummation
# 0: no Coulomb resummation
# 1: Leading power Coulomb resummation when order =/= 0 for massive fermion loops
coulomb 1

# masses (negative means we will not include such a contribution)
# in units of GeV
emass  0.51099895d-3       # electron mass (pole mass)
mumass  0.1056583755d0     # muon mass (pole mass)
taumass 1.77693d0          # tau mass (pole mass)
umass 0.335d0              # up quark mass (pole mass)
dmass 0.340d0              # down quark mass (pole mass)
smass 0.490d0              # strange quark mass (pole mass)
cmass 1.5d0                # charm quark mass (pole mass)
bmass 4.75d0               # bottom quark mass (pole mass)
tmass 172.56d0             # top quark mass (pole mass)
wmass 80.3692d0            # W mass

# couplings
alphaemm1     137.036d0    # alpha(0)**(-1)
alphasMZ      0.118d0      # alpha_s(mZ**2) for the QCD corrections
alphas_nloop  5            # how many loops run of alpha_s
                           # in the MSbar scheme (maximal 5)
Gfermi        1.1663787d-5 # Gfermi only alpha_scheme=1
alphaMZm1     128.94d0     # alpha(mZ**2)**(-1) with alpha_scheme=2
alpha_scheme  2            # 0: use alpha(0) for the QED corrections
                           # 1: use Gmu scheme for the QED corrections
                           # 2: use on-shell alpha(mu**2) scheme 
                           #    from alpha(mZ**2)
                           # for the QED corrections
alpha_nloop   2            # how many fermion loops (1 or 2) run of alpha
                           # when alpha_scheme=2
Scale         2            # renormalisation scale.
                           #  0: fixed scale;
                           #  1: pT of photon;
                           #  2: invariant mass of the final photons
FScaleValue   10d0         # the scale value in the fixed scale (Scale = 0)
muR_over_ref  0.5d0        # the true renorm central scale is muR_over_ref*scale
reweight_Scale T           # reweight to get (renorm) scale dependence
rw_RScale_down 0.5d0       # lower bound for renormalisation scale variations
rw_RScale_up   2.0d0       # upper bound for renormalisation scale variations

# kinematic cuts in the lab frame
minptp  2.0d0              # minimum photon pt
maxptp -1.0d0              # maximum photon pt, negative no such cuts
maxptpp -1.0d0             # maximum photon-pair pt, negative no such cuts
maxaco  -1.0d0             # maximum acoplanarity, negative no such cuts
maxyrapp 5d0               # maximum photon (pseudo-)rapidity
minyrapp -5d0              # minimum photon (pseudo-)rapidity
minmpp   5d0               # minimum invariant mass of photon-photon
maxmpp   -1d0              # maximum invariant mass of photon-photon, negative no such cuts

# event generation setup
gener    0                 # integrator: 0 VEGAS; 1 Romberg
unwgt    T                 # unweighting on/off (only gener=0)
lhewgtup T                 # xwgtup=1 (F) or not (T) in unweighted lhe file
nmc      1000000           # number of Monte Carlo points or integrating points
seed     0                 # seed for MC integrator (only works when gener = 0)
                           # the range should be [0, 20000] (mod 20001)
                           # <0 means it will update automatically

# histogram output
topdrawer_output F         # topdrawer output file (T) or not (F)
gnuplot_output   F         # gnuplot output file (T) or not (F)
root_output      F         # root output file (T) or not (F)
hwu_output       T         # hwu output file (T) or not (F)

# technical stuff (do not touch them unless you understand)
grid_gen   0               # 0: check whether there is a grid. 
                           #    If yes, just use the grid.
                           #    Otherwise, generate a grid first and 
                           #    write to the disk and then use it.
                           #    (the only option if there is 
                           #      neither FastGPL nor handyG)
                           # 1: force to generate a grid and 
                           #    write to the disk and then use it
                           # 2: just evaluate it directly without using the grid
                           
improve_w_LE -1            # -1: We need to use the low/high-energy expansion
                           #     results to improve the massive amplitudes
                           #  0: We do not use the low/high-energy expansion
                           #     results to improve the massive amplitudes
                           #  1: We simply use the low-energy expansion
                           #     results for the massive amplitudes
\end{Verbatim}

\endgroup

\vskip 0.5truecm
\noindent

\bibliography{literature} 
\bibliographystyle{utphysM}
\end{document}